\documentclass[12pt]{article}
\usepackage{cite}
\usepackage{amsmath, amsthm, amssymb,slashed,url
}
\usepackage{color}
\usepackage{ifpdf}
\ifpdf
  \usepackage[pdftex]{graphicx}
  \usepackage{epstopdf}
\else  
  \usepackage[dvips]{graphicx}
\fi
\usepackage[utf8]{inputenc}
\usepackage[T1]{fontenc}
\parskip=\baselineskip
\def\Bbb{\mathbb}
\def\Tr{{\rm Tr}}
\def\16{{\bf 16}}
\def\bY{{\bar Y}}
\def\S{S}
\def\sR{{\sf R}}
\def\Pin{{\mathrm{Pin}}}
\def\sF{{\sf F}}
\def\sT{{\sf T}}
\def\Det{{\mathrm{Det}}}
\def\1{{\bf 1}}
\def\2{{\bf 2}}
\def\3{{\bf 3}}
\def\4{{\bf 4}}

\def\y{{\sf y}}
\def\I{{\mathcal I}}

\def\\SPT{\sf{\SPT}}
\def\a{{\sf a}}
\def\upeta{{\eta_D}}

\def\CS{{\textit{CS}}}
\def\O{{\mathrm O}}
\def\SO{{\mathrm{SO}}}
\def\Spin{{\mathrm{Spin}}}
\def\SU{{\mathrm{SU}}}
\def\D{{\mathcal D}}

\def\Pf{{\mathrm{Pf}}}
\def\W{{\mathcal W}}

\def\tCS{{\textit{CS}}}

\def\tr{{\mathrm{tr}}}
\def\\SRT{{\sf{\SRT}}}

\def\h{\widehat}
\def\bar{\overline}

\def\ra{\rangle}

\def\bp{\begin{pmatrix}}
\def\ep{\end{pmatrix}}
\def\la{\langle}
\def\R{{\Bbb{R}}}\def\Z{{\Bbb{Z}}}

\font\teneurm=eurm10 \font\seveneurm=eurm7 \font\fiveeurm=eurm5
\newfam\eurmfam
\textfont\eurmfam=\teneurm \scriptfont\eurmfam=\seveneurm
\scriptscriptfont\eurmfam=\fiveeurm

\font\teneusm=eusm10 \font\seveneusm=eusm7 \font\fiveeusm=eusm5
\newfam\eusmfam
\textfont\eusmfam=\teneusm \scriptfont\eusmfam=\seveneusm
\scriptscriptfont\eusmfam=\fiveeusm

\font\tencmmib=cmmib10 \skewchar\tencmmib='177
\font\sevencmmib=cmmib7 \skewchar\sevencmmib='177
\font\fivecmmib=cmmib5 \skewchar\fivecmmib='177
\newfam\cmmibfam
\textfont\cmmibfam=\tencmmib \scriptfont\cmmibfam=\sevencmmib
\scriptscriptfont\cmmibfam=\fivecmmib

\numberwithin{equation}{section}

\def\d{\mathrm d}

\def\L{{\mathcal L}}
\def\\S{{\Bbb \S}}
\def\Z{{\Bbb Z}}
\def\bar{\overline}
\def\A{{\mathcal A}}

\def\bar{\overline}

\begin{document}
\begin{titlepage}
\begin{flushright}
\end{flushright}
\vskip 1.5in
\begin{center}
{\bf\Large{Anomaly Inflow and the $\eta$-Invariant}}
\vskip
0.5cm {Edward Witten$^1$ and Kazuya Yonekura$^2$} \vskip 0.05in 
{\small{ \textit{
$^1$School of Natural Sciences, Institute for Advanced Study}\vskip -.4cm
{\textit{Einstein Drive, Princeton, NJ 08540 USA}} 
\vskip 0 cm 
\textit{$^2$Department of Physics, Tohoku University, Sendai 980-8578, Japan }

}}
\end{center}
\vskip 0.5in
\baselineskip 16pt
\abstract{Perturbative fermion anomalies in spacetime dimension $d$ have a well-known relation to Chern-Simons functions in dimension $D=d+1$.
This relationship is manifested in a beautiful way in ``anomaly inflow'' from the bulk of a system to its boundary.    Along with perturbative
anomalies, fermions also have global or nonperturbative anomalies, which can be incorporated by using the $\eta$-invariant of Atiyah, Patodi, and Singer
instead of the Chern-Simons function.
   Here we give a nonperturbative description of anomaly inflow,
involving the $\eta$-invariant.  This formula has been expected in the past based on the Dai-Freed theorem,
but has not been fully justified.  
It leads to a general description of perturbative and nonperturbative fermion anomalies in $d$ dimensions in terms of an $\eta$-invariant in $D$ dimensions.
This $\eta$-invariant is  a cobordism invariant whenever perturbative anomalies cancel.}
\end{titlepage}
\def\Hom{\mathrm{Hom}}
\def\H{{\mathcal H}}
\def\d{{\mathrm d}}
\def\t{\widetilde}
\def\U{{\mathcal U}}
\def\UU{{\mathrm U}}
\def\V{{\mathcal V}}
\def\st{{\sf t}}
\def\O{{\mathcal O}}
\def\i{{\mathrm i}}
\def\A{{\mathcal A}}
\def\be{\begin{equation}}
\def\ee{\end{equation}}

\tableofcontents

\def\Sp{{\mathrm{Sp}}}
  
\def\aa{{\sf a}}
\def\bb{{\sf b}}
\def\a{{\bf a}}
\def\x{{\bf x}}
\def\y{{\bf y}}
\def\z{{\bf z}}
 \def\sym{{\mathrm{sym}}}
 \def\zotimes{{\otimes N}}
 \def\cl{{\mathrm{cl}}}
\def\h{\widehat}
\def\t{\widetilde}
\def\tt{\sf t}
\def\sign{{\mathrm{sign}}}
\def\tr{{\mathrm{tr}}}
\def\d{{\mathrm{d}}}
\def\H{{\mathcal H}}
\def\O{{\mathrm O}}
\def\cO{{\mathcal O}}
\def\Tr{{\mathrm{Tr}}}
\def\diag{{\mathrm{diag}}}
\def\la{\langle}
\def\ra{\rangle}
\def\sC{{\sf C}}
\def\pin{{\mathrm{pin}}}

\def\beq#1\eeq{\begin{align}#1\end{align}}
\def\alert#1{{\color{red}(#1)}}
\def\CD{ {\cal D}}
\def\CL{ {\cal L}}
\def\CH{ {\cal H}}
\def\SL{ {\sf L}}
\def\APS{ {\sf APS}}
\def\sPsi{{\sf \Psi}}
\def\U{{\mathrm U}}
\def\R{{\Bbb R}}
\def\slD{{\slashed{D}}}
\def\beq{\begin{equation}}
\def\eeq{\end{equation}}
\newcommand{\vev}[1]{ \left\langle {#1} \right\rangle }
\newcommand{\bra}[1]{ \langle {#1} | }
\newcommand{\ket}[1]{ | {#1} \rangle }
\newcommand{\et}[1]{  {#1} \rangle }

\def\d{{\mathrm d}}\def\t{\widetilde}
\def\be{\begin{equation}}
\def\bar{\overline}
\def\i{{\mathrm i}}
\def\ee{\end{equation}}

\section{Introduction}\label{intro}

Shoucheng Zhang made many outstanding contributions in condensed matter physics.   He started his career in particle physics
and was always very interested in the interplay between condensed matter physics and relativistic physics.  
He made numerous lasting contributions in areas that combine ideas of the two fields.

Fermion anomalies 
-- originally the one-loop triangle anomaly of Adler, Bell, and Jackiw  \cite{Adler,BJ}  -- have been important in particle physics since their discovery,
and  have proved to be important in condensed matter physics.   In particular, they played a role in some of Shoucheng Zhang's important contributions.
The lecture of one of us at the Shoucheng Zhang Memorial Workshop was devoted to a particular question about fermion anomalies \cite{WittenLecture}.   (The question
is briefly described at the end of section \ref{d1}.)  The conference organizers suggested that written contributions to the proceedings could be on broader themes
related to the topics of the lectures.   Here we will describe a nonperturbative approach to fermion anomaly inflow.   

 A rather well-known fact about perturbative fermion anomalies
is that the perturbative anomaly in spacetime dimension $d$ is related to a Chern-Simons function in dimension $D=d+1$ \cite{Jackiw,Zumino,Stora}.   This idea
has a beautiful manifestation in the idea of anomaly inflow \cite{CallanHarvey}, which is important in both particle physics and condensed matter physics.
The classic example in condensed matter physics is the integer quantum Hall effect.   The worldvolume of a quantum Hall sample has spacetime dimension $D=3$
and its boundary has dimension $d=2$.   The ``edge mode'' of a quantum Hall system is -- in relativistic language -- a massless electrically charged chiral fermion.
The coupling of this mode to electromagnetism has an anomaly -- analogous to the original triangle anomaly.    In the  bulk of a quantum Hall system,  the
effective action  for the gauge field $A$ of electromagnetism includes a multiple of the Chern-Simons coupling $\int A\wedge \d A$.   This coupling is not gauge-invariant on
a manifold with boundary, and that failure of  gauge invariance cancels the quantum anomaly of the edge mode.   

Along with perturbative anomalies, fermions can also have global or nonperturbative anomalies that are not visible in perturbation theory  \cite{WittenSUtwo}.
The relation between perturbative anomalies in $d$ dimensions and a Chern-Simons function in $d+1$ dimensions can be generalized
to include nonperturbative anomalies \cite{Witten:1985xe}.  One just has to replace the Chern-Simons function by the $\eta$-invariant of Atiyah, Patodi, and Singer (APS) \cite{Atiyah:1975jf}.
The $\eta$-invariant is equivalent in perturbation theory to a Chern-Simons function, but it contains additional nonperturbative information.
The relation between anomalies and the $\eta$-invariant motivated a mathematical result that we will call the Dai-Freed theorem \cite{Dai:1994kq}.
See \cite{Yonekura:2016wuc} for an explanation of the Dai-Freed theorem from a physical point of view.   The Dai-Freed theorem makes it possible
to get more precise results about  fermion anomalies than were available otherwise; see for example \cite{WittenDBrane}, section 2.2.

Since anomaly inflow is an important aspect of understanding perturbative fermion anomalies, and since fermions do have nonperturbative  anomalies, one would
like to formulate anomaly inflow in a way that incorporates nonperturbative anomalies as well as perturbative ones.   The Dai-Freed theorem strongly suggests how
to do  this. The basic idea is
to replace the Chern-Simons function with an $\eta$-invariant defined with APS boundary conditions.  
The resulting nonperturbative formula for anomaly inflow (eqn. (\ref{torro}) below) was described in \cite{Witten:2015aba}.    However,
this formula has  not been fully justified.    The main goal of the present article is to do this.

We present our nonperturbative approach to fermion anomaly inflow in section \ref{precise}.   We take a viewpoint that is natural in modern condensed
matter physics.   An arbitrary, possibly anomalous, fermion system in dimension $d$ can be regarded as the boundary state of an anomaly-free gapped
fermion system in dimension $D=d+1$.    Then the anomaly inflow formula is simply obtained, in principle, by integrating out the gapped fermions in the bulk.
Since it is already known that integrating out a massive fermion gives a Chern-Simons term perturbatively \cite{Redlich} and an $\eta$-invariant nonperturbatively
\cite{AlvarezGaume:1984nf}, it is fairly clear that this process will give an answer of a familiar kind but with Chern-Simons replaced with $\eta$.
Less obvious is why the $\eta$-invariant should be computed with APS boundary conditions,
as suggested by the Dai-Freed theorem.   Explaining this is one of the main tasks of section \ref{precise}.    To avoid extraneous  details, we begin section \ref{precise}
assuming that the bulk fermion is a Dirac fermion, but then we extend the discussion to apply to the anomalies of a completely general relativistic fermion system. (In a somewhat
similar way, though presented more abstractly, the Dai-Freed theorem was generalized  recently to an arbitrary relativistic fermion system in the appendix of
\cite{Freed:2019sco}.)

In section \ref{anomaly}, we explain that the nonperturbative formula for anomaly inflow implies a general  description of perturbative and nonperturbative
fermion anomalies in dimension $d$ in terms of an exponentiated $\eta$-invariant in dimension $d+1$.  This generalizes the global anomaly formula obtained
in  \cite{Witten:1985xe} (where $d$ was assumed to be even and only orientable manifolds were considered), 
and also gives more precise information about the fermion path integral in anomaly-free examples.
  In section \ref{examples}, we use the description via the $\eta$-invariant
to analyze potential global anomalies in various examples in dimensions $d=1,2,3,4$.  One of the examples we consider involves a celebrated contribution
by Shoucheng Zhang \cite{ZQH}, involving the electromagnetic $\theta$ angle in the bulk of a $3+1$-dimensional topological insulator.
  Another example we consider is the Standard Model of particle physics.

Though in the present article we consider only fermion anomalies, we should note that in modern condensed matter and particle physics, anomalies in bosonic
systems play a role as well.   In particular, in condensed matter physics, a symmetry protected topological (SPT) phase of matter is a gapped phase with
some global symmetry group $G$, such that the phase is in some sense topologically nontrivial but would become trivial if the symmetry is explicitly broken.  
A version of anomaly inflow is important in understanding SPT phases \cite{Chen,Chen2,Senthil}, with group cohomology playing the role that for fermion anomalies is played by the
$\eta$-invariant.  In all cases, the anomaly in dimension $d$ is related to an ``invertible topological field theory'' in dimension $d+1$.  For a general description of
this point of view about anomalies, see \cite{Freed:2014iua,Monnier}.

\section{A Precise Formula For Anomaly Inflow}\label{precise}

In this section, we study massive fermions on a manifold $Y$ with boundary $W$.  $W$ will have dimension $d$, while $Y$ has dimension $D=d+1$.
 The goal is to get a precise formula for anomaly inflow from $Y$ to $W$.

\subsection{Massive Dirac Fermion With Local Boundary Condition}\label{sec2}
We begin with  a massive Dirac fermion $\Psi$ on a $D $-dimensional manifold $Y$.     A Dirac fermion is just a fermion 
field $\Psi$ whose components
are all charged under a global or gauge $\U(1)$ symmetry.   The conjugate field $\bar\Psi$ carries opposite $\U(1)$ charges.
The assumption of a $\U(1)$ symmetry
 is certainly nongeneric, but we begin with this case in order to explain the main idea of the derivation without potentially distracting details.

The discussion will be quite general; $D$ may be even or odd,
$Y$ may be orientable or unorientable, and $\Psi$ may be coupled to gauge fields as well as to the geometry of $Y$.     
We assume only that $Y$ is endowed with the appropriate structure (such as a spin or pin structure, a spin$_c$ structure, etc.) so that a suitable action for $\Psi$
exists, of the usual form for a Dirac fermion:
\beq 
I = -  \int_Y \d^D x \sqrt{g}\, \overline{\Psi} (\slashed{D}_Y+m)\Psi. \label{eq:action}
\eeq   $\Psi$ is coupled to the Riemannian metric $g$ of $Y$ and possibly to a background gauge field $A$.
The Dirac operator is defined in the usual way, $\slashed{D}_Y=\sum_{\mu=1}^D \gamma^\mu D_\mu$, with $\{\gamma_\nu,\gamma_\nu\}=2g_{\mu\nu}$.   
On a manifold without boundary, the operator $\D_Y=\i \slashed{D}_Y$ is self-adjoint.

Now suppose that the manifold $Y$ has  boundary $W = \partial Y$, with a metric that looks like a product near the boundary.   Thus near the boundary
the metric of $Y$ is $\d s^2_Y=\d \tau^2+\d s^2_W$, where $\d s^2_W$ is a metric on $W$, and $\tau$ parametrizes the normal direction.   We can normalize
$\tau$  to vanish along $W$ and to be negative away from $W$.   We assume this product description is valid at least in a range $(-\epsilon,0]$.

We impose on $\Psi$ the following local boundary condition:
\beq 
\SL :  (1 - \gamma^\tau) \Psi|_{\tau=0} = 0 \label{eq:b.c.}
\eeq
where $\gamma^\tau$ is the gamma matrix in the direction $\tau$.
Before proceeding with the analysis, we will say a word about this boundary condition.

The boundary condition (\ref{eq:b.c.}) is elliptic but not self-adjoint.\footnote{Self-adjoint local boundary conditions are also possible, of course, at least in some cases.
For recent analysis of a self-adjoint local boundary condition that leads to no localized boundary mode (but may explicitly violate symmetries such as time-reversal or reflection
symmetry),  see e.g. \cite{Kurkov:2017cdz,Kurkov:2018pjw,Fialkovsky:2019rum}.}   Ellipticity of the boundary condition is necessary and sufficient to ensure that the Dirac operator
with this boundary condition has the properties needed to make Euclidean field theory well-defined.   For example, on a compact manifold, the Dirac operator with an elliptic
boundary condition has at most a finite-dimensional space of zero-modes, and, after removing those zero-modes, 
 it has a Green's function with the usual properties, such as short distance singularities of a standard form.  
 Technically, the condition for ellipticity is as follows.   Suppose that $W=\R^d$ and 
 $Y=\R^d\times \R_-=\R^{d+1}_-$, where $\R_-$ is a half-line parametrized by $\tau\leq 0$.   A boundary condition on the Dirac equation at $\tau=0$
 is elliptic if, after dropping from the equation lower order terms (notably mass terms and couplings to background fields),
 or equivalently after taking the momentum along $\R^d$ to be extremely large, the equation and its adjoint have
 no solutions that satisfy the boundary condition and vanish for $\tau\to-\infty$.   
 The local boundary
 condition  (\ref{eq:b.c.}) has this property, because any solution of the free Dirac equation on $\R^{d+1}_-$ that has nonzero momentum along the boundary and vanishes
 for  $\tau\to-\infty$ has components with $\gamma^\tau=-1$ as well as components with $\gamma^\tau=+1$.
 
 A boundary condition on the Euclidean Dirac operator  is self-adjoint if, with this boundary condition, $\i\slD_Y$  is self-adjoint on a manifold
 with boundary.  As usual, a self-adjoint operator has a well-defined eigenvalue problem with
  real eigenvalues.    The boundary condition  (\ref{eq:b.c.}) is
 not self-adjoint; an attempt to prove self-adjointness will fail because of a surface term at $\tau=0$.   
 A self-adjoint boundary condition could not lead to the chiral, potentially anomalous physics that we are interested in.
    In Euclidean signature, self-adjointness of a local boundary condition is not a physical requirement.
 The actual condition that a local boundary condition should satisfy in order to be physically sensible
 is the following.   Suppose that one analytically continues $W$ to Lorentz signature, with time coordinate $x^0$.    One wants the boundary
 condition to be such that the Hamiltonian that propagates a state in the $x^0$ direction is self-adjoint.   
 One can verify that this is true for the boundary condition $\SL$, using the fact that $\{\gamma^0,\gamma^\tau\}=0$.

The action (\ref{eq:action}) can be written near the boundary as
\beq
I = -  \int_Y  \d^D x\sqrt g\, \overline{\Psi} \gamma^\tau \left( \frac{\partial }{\partial \tau} + \CD_W +  \gamma^\tau m \right)\Psi. 
\eeq
where\footnote{\label{equivalence} $\D_W$ differs from  the usual Dirac operator $\i\sum_{\mu=1}^d \gamma^\mu D_\mu$ of $W$ only by
the use of a different set of gamma matrices ($-\i\gamma^\tau\gamma^\mu$ instead of $\gamma^\mu$) that obey the same algebra
and lead to the same rotation generators $\frac{1}{4}[\gamma_\mu,\gamma_\nu]$.   So the two operators are equivalent.}
\beq
\CD_W =\sum_{\mu \neq \tau}  \gamma^\tau  \gamma^\mu D_\mu
\eeq
is a self-adjoint Dirac operator on the boundary $W$. Notice that $\gamma^\tau$ and $\CD_W$ anticommute with each other.
If $d$ is even, the operator $\gamma^\tau$ measures what is usually called the ``chirality'' of a fermion that propagates along $W$.
We will call $\gamma^\tau$ a chirality operator in any case, though for odd $d$ this is not standard terminology.  

Our discussion so far makes sense for either sign of $m$, but in
what follows, given our choice of boundary condition, the interesting case is $m<0$.   
For  $m<0$, the Dirac equation $(\slashed{D}+m) \Psi =0$ has a  mode localized near the boundary and given by
\beq\label{edgemode}
\Psi =  \chi \exp ( |m| \tau), \qquad
(1-\gamma^\tau)\chi=0, \qquad
\CD_W \chi =0,
\eeq
where $\chi$ is a fermion field on $W$.
Since it vanishes exponentially for $\tau \ll 0$, this mode is localized along $W$.   Since  it satisfies $\CD_W\chi=0$, 
it propagates along $W$ as a massless fermion.   Finally, as it obeys $\gamma^\tau\chi=\chi$, it is a chiral fermion along $W$.  For $m>0$, there is no such
boundary-localized mode.

To quantize the fermion field $\Psi$, some sort of regulator is needed.   A useful choice for our purposes is a simple Pauli-Villars regulator, defined by adding a very
massive field of opposite statistics, satisfying the same Dirac equation (possibly with a different mass parameter) and the same boundary condition.
We take the Pauli-Villars regulator field to have a positive mass parameter, since we do not want the regulator field to have a low energy mode propagating along the
boundary, which would be quite unphysical.

Now we want to compute the partition function of the above massive fermion $\Psi$ on the manifold $Y$ with the boundary condition $\SL$. 
For this purpose, it is useful to think of  $\tau$ as a Euclidean time coordinate, and the boundary $W = \partial Y$ as a time-slice.
In this point of view, the path integral over $Y$ gives a physical state vector which we denote as $\ket{Y}$.
This is a state vector in the Hilbert space $\CH_W$ of $W$.
The boundary condition $\SL$ also corresponds to a state vector $\ket{\SL} \in \CH_W$, which we will later discuss explicitly. 
Then, the partition function on $Y$ with the boundary condition $\SL$ is just given by
\beq
Z(Y, \SL) = \bra{\SL} \et{ Y}. \label{eq:p.t.}
\eeq

The important point is now as follows. We take the mass $|m|$ to be very large so that its Compton wavelength $1/ |m|$ is much shorter than the typical scale of the manifolds $Y$ and $W$;
equivalently, we take the length scale of $Y$ and $W$ to be much larger than $1/|m|$.
In this limit, there is a large mass gap in the Hilbert space $\CH_W$. Moreover, the path integral on a cylindrical region $(-\epsilon, 0] \times W$ gives 
the Euclidean time evolution $e^{-\epsilon H}$ where $H$ is the Hamiltonian.
If the mass gap is large so that $\epsilon |m| \gg 1$,
the factor $e^{-\epsilon H}$ plays the role of the projection operator to the ground state $\ket{\Omega}$:
\beq
e^{-\epsilon H} \simeq \ket{\Omega} \bra{\Omega} \qquad ( \epsilon |m| \gg 1).
\eeq
This is valid up to errors which are exponentially suppressed by $e^{ - \epsilon |m|}$, which we neglect.
In this limit, $\ket{Y}$ is proportional to the ground state,
\beq
\ket{Y} \propto \ket{\Omega},
\eeq
where we may assume $  \la \Omega|  \Omega\ra=1$.
So we can rewrite the partition function as
\beq\label{factored}
Z(Y, \SL) = \bra{\SL} \et{\Omega} \bra{\Omega} \et{ Y}.
\eeq
In this way we can split the bulk contribution $\bra{\Omega} \et{ Y}$ and the boundary contribution $ \bra{\SL} \et{\Omega}$.

The ground state $\ket{\Omega}$ has a phase ambiguity, and hence the splitting into $\bra{\Omega} \et{ Y}$ and $ \bra{\SL} \et{\Omega}$ 
also has this ambiguity. In the context of the Dai-Freed theorem~\cite{Dai:1994kq}, a physical interpretation of this ambiguity is as follows~\cite{Yonekura:2016wuc}.
The space of ground states is a  one-dimensional subspace of the Hilbert space $\CH_W$. We denote this one-dimensional subspace as $\CL_W \subset \CH_W$.
Now suppose that we make an adiabatic change of  background fields such as, e.g., the metric $g$ of $W$ or an  external gauge field $A$. The ground state changes adiabatically,
and evolves with a Berry phase. Let $\W$ be the parameter space of the background fields.
The existence of such a Berry phase means that we may get a nontrivial holonomy when going around a loop in $\W$. The one-dimensional spaces $\CL_W$
at each point of $\W$ combine into a rank one complex vector bundle (a complex line bundle) over $\W$ with nontrivial holonomies determined by the Berry phases. 
Then instead of $\bra{\Omega} \et{ Y}$ and $ \bra{\SL} \et{\Omega}$, which do not have well-defined phases,  we can define the following:
\beq
\ket{\Omega}\bra{\Omega} \et{ Y} \in \CL_W, \qquad  \bra{\SL} \et{\Omega} \bra{\Omega} \in \CL_W^{-1}.
\eeq
These formulas are well-defined, since they do  not depend on the phase of the state $|\Omega\ra$.
But instead  of a number $\la\Omega|Y\ra$ we get a vector $\ket{\Omega}\bra{\Omega} \et{ Y} \in \CL_W$, and likewise instead of a number $\la\SL|\Omega\ra$
we get a vector $ \bra{\SL} \et{\Omega} \bra{\Omega} \in \CL_W^{-1}.$
All this is related to the following fact   \cite{Atiyah:1984tf}: the partition function of the chiral fermion 
$\chi$ on the boundary $W$ is naturally understood as a section of a line bundle $\CL_W^{-1}$ (called the determinant line bundle) rather than 
a complex number.   
Also, the factor $\bra{\Omega} \et{ Y} $ is  an exponentiated $\eta$-invariant that can be regarded as an element of $\CL_W$~\cite{Dai:1994kq}.
See \cite{Yonekura:2016wuc} for more details on these points, from a physical perspective.

Somewhat surprisingly, however, as long as we are in a situation in which the Dirac operator on $W$ generically has no zero-mode,\footnote{The case that $\D_W$
does generically have a zero-mode will be analyzed in section \ref{sec:zeromode}.   As explained there, this case is somewhat more exotic since, for example, it
does not arise if $W$ is connected.} it is possible
to avoid introducing the line bundle $\L_W$.   
One can avoid the line bundle $\CL_W$ by using instead of $\ket{\Omega}$  a state which has no phase ambiguity.
As long as the boundary Dirac operator $\CD_W$ has no zero modes, we can consider the global Atiyah-Patodi-Singer (APS) boundary 
condition~\cite{Atiyah:1975jf}, which we denote as $\APS$.   This boundary condition will be described explicitly in section \ref{aps}.  The path integral on $Y$ with boundary condition $\APS$ 
 corresponds to a state vector $\ket{\APS} \in \CH_W$, which we will also describe explicitly in section \ref{aps}.  Because $\ket{Y} $ is a multiple of $ \ket{\Omega}$, we can 
 rewrite (\ref{factored}) in the following longer but useful form:
\beq
Z(Y, \SL) =\frac{ \bra{\SL} \et{\Omega} \bra{\Omega} \et{\APS} }{ | \bra{\APS} \et{\Omega} |^2} \cdot  \bra{\APS} \et{ Y} . \label{eq:rewrite}
\eeq

In this formula, the inner product  $\bra{\APS} \et{Y}$ is the partition function of the massive fermion $\Psi$  on $Y$ with the APS boundary condition.  In section \ref{aps}, we show
that at long distances (that is, if $Y$ and $W$ are large compared to $1/|m|$), we have, modulo nonuniversal factors that do not affect the analysis of anomalies,\footnote{These
are factors that can be removed by adding  to the action the integral of a gauge invariant function of the background fields $g,A$.  For example, the ground  state
energy per unit volume of the $\Psi$ field is a nonuniversal effect that can be removed by adding to the action a multiple of the volume of $Y$.}
\beq\label{boldo} \bra{\APS}\et{Y} =\exp\left(-\i \pi \upeta\right), \eeq
where $\upeta$ is the APS $\eta$-invariant (defined with APS boundary conditions along $W=\partial Y$) for the Dirac fermion $\Psi$;  
 $\eta_D$  will be introduced in section \ref{aps}.   
The remaining factor $\bra{\SL} \et{\Omega} \bra{\Omega} \et{\APS} /| \bra{\APS} \et{\Omega} |^2$ in $Z(Y,\SL)$ will be analyzed in section \ref{sec3}, and we will see that
(modulo nonuniversal factors) it equals $|\Det \, \CD_W^+|$,
the absolute value of the path integral on $W$ for the massless chiral fermion $\chi$.    (The superscript $+$ in $\CD^+_W$ means that
here $\CD_W$ is taken to act only on the field $\chi$ of positive chirality.)     It is essential  that an absolute value comes in here.
In general, the path integral of the boundary mode $\chi$ might be anomalous, so the corresponding determinant, without taking an absolute value, may
not be well-defined.   However, an anomaly always affects only the phase of the path integral, so the absolute value $|\Det \CD^+_W|$ is always anomaly-free.

Combining these claims about the various factors in $Z(Y,\SL)$, we get the following formula for the partition function on $Y$ with boundary condition $\SL$:
\beq\label{refined} Z(Y,\SL)=|\Det \CD^+_W|\exp(-\i\pi\upeta). \ee
We claim that (for  fermions charged under a $\U(1)$ symmetry) this is the general statement of anomaly inflow, including all perturbative and nonperturbative anomalies.

If one is only interested in perturbative anomalies, then eqn. (\ref{refined}) can be replaced by a  slightly simpler and probably more familiar version:
\beq\label{pertthy} Z(Y,\SL)\overset{?}{=}\Det \CD^+_W\,\,\exp(-\i {\textit{CS}}(g,A)). \eeq
Here ${\textit{CS}}(g,A)$ is a Chern-Simons interaction that depends on the background fields $g,A$ on $Y$.
In perturbation theory, this formula is a satisfactory expression of the idea of anomaly inflow \cite{CallanHarvey}.  The idea behind this formula is that the
 path integral of the boundary fermion is
the determinant $\Det \CD^+_W$, and integrating out the massive fermion $\Psi$ on $Y$ induces the Chern-Simons coupling ${\textit{CS}}(g,A)$.  When one proves
the gauge-invariance of $\tCS(g,A)$, one encounters a surface term, as a result of which $\tCS(g,A)$ is not gauge-invariant on a manifold with boundary. 
This failure of gauge invariance of $\tCS(g,A)$  on a manifold with boundary precisely matches the perturbative anomaly of a chiral fermion on the boundary
 \cite{Jackiw, Zumino, Stora}, as a result of which the formula (\ref{pertthy}) is  satisfactory  as far as perturbative anomalies are concerned. 

In general, however, in a topologically nontrivial situation, the result of integrating out a massive Dirac fermion, even on a manifold without boundary,
is not $\exp(-\i\tCS(g,A))$ but $\exp(-\i\pi \upeta)$ \cite{AlvarezGaume:1984nf,Witten:2015aba}.  This fact will be reviewed in section \ref{aps}. 
The two expressions are equivalent in perturbation theory around Euclidean space, that is in perturbation theory around $Y=\R^D $, $g_{\mu\nu}=\delta_{\mu\nu}$, and $A=0$.
But in general, the description by $\exp(-\i\tCS(g,A))$ misses many topological subtleties that are important in an understanding of global or nonperturbative anomalies 
and anomaly inflow and that are properly described 
by $\exp(-\i\pi\upeta)$. This is true for all $D$, but perhaps it is most obvious if $D$ is even.   If $D$ is even, there are no conventional Chern-Simons
couplings, and in perturbation theory around Euclidean space, one sees nothing.   But $\upeta$ can still be nontrivial and describes anomaly inflow.

Bearing in mind that $\exp(-\i\tCS)$ should be replaced by $\exp(-\i\pi\upeta)$ if we want a nonperturbative description,
 let us ask what might be the correct general formula for anomaly inflow, replacing (\ref{pertthy}).    Part of the Dai-Freed theorem  \cite{Dai:1994kq}, which has been elucidated from a physical point of view in
 \cite{Yonekura:2016wuc}, is an abstract definition of $\exp(-\i\pi \upeta)$ (with APS boundary conditions along $W=\partial Y$) as a section of the line bundle $\L_W$.
 In that framework,  the obvious candidate for improving on (\ref{pertthy}) is $Z(Y,\SL)=\Det\CD^+_W\,\cdot \,\exp(-\i\pi\upeta)$.   Neither  factor is  gauge-invariant
 separately, but the Dai-Freed theorem shows that the product is well-defined and completely free of any perturbative or nonperturbative anomaly.
 
This formulation is correct but unnecessarily abstract.    As long as the operator $\CD_W$ generically has no zero-mode, it is possible to simply  define
 $\upeta$ as  a number; this is the route that  we will follow.\footnote{Fundamentally,  it is not necessary to introduce the line bundle $\L_W$ 
 because as long as  $\CD_W$  has no zero-mode,
  $\Det\,\CD_W^+$ is nonzero and  trivializes  $\L_W$.  Relative to this trivialization, $\Det\,\D_W^+$ is positive (so it is replaced by its absolute  value
   $|\Det\,\D_W^+|$) and $\exp(-\i\pi\eta_D)$
  becomes a number.    
  To understand what follows,  it is not necessary to think in such abstract terms.}  Then  the formula takes the form of eqn. (\ref{refined}).   In  this formulation, the two
 factors are separately gauge-invariant, but  they are not  separately physically sensible.   
  Indeed, because of the absolute value,
 $|\Det\,\CD_W^+|$ does not vary smoothly near a value
 of background fields $g,A$ at which $\CD_W$ has a zero-mode (and therefore $\Det\CD_W^+=0$), while $\exp(-\i\pi\upeta)$ does not vary smoothly near such a point because the
 definition of $\upeta$ breaks down.   The Dai-Freed theorem says that the product in (\ref{refined}) is always smoothly varying.  
 
 Thus eqn. (\ref{refined}) is the natural generalization of the more familiar but imprecise formula (\ref{pertthy}).   
 But to our knowledge a full justification of eqn. (\ref{refined})  has not previously been given.    
 We will address this issue in sections \ref{sec3} and \ref{sec4}, after first explaining the basic
 facts about the $\eta$-invariant.

\subsection{Integrating Out A Fermion}\label{aps}

Though our main interest in the present paper is in integrating out a massive fermion, it is also of interest to know what happens if one integrates
out a massless fermion.  It turns out to be useful to consider this case first.

We consider a Dirac fermion  $\Psi$ on a $D$-manifold $Y$, initially without boundary.  $\Psi$ has a self-adjoint Dirac operator $\D_Y=\i\slashed{D}_Y$.   It has
real eigenvalues that we denote $\lambda_k$.   

Formally, the path integral of $\Psi$ is $\Det\,\D_Y=\prod_k\lambda_k$.    This of course needs to be regulated.   We do so with the aid of a Pauli-Villars regulator of mass $M$.
A regulated version of the determinant is\footnote{Actually, to completely regularize the determinant, one needs several Pauli-Villars fields with different masses
and statistics.   The details do not really modify the following derivation.  The same remark applies whenever we consider expressions that require regularization.}
\be\label{moddo}\Det\,\D_Y= \prod_k \frac{\lambda_k}{\lambda_k+\i M}. \ee  

Now let us determine the phase or argument of $\Det\,\D_Y$.    This is ill-defined if any of the $\lambda_k$ vanish, so we assume they are nonvanishing.
Assuming $M>0$ and $M\gg |\lambda_k|$, we see that the argument of $\lambda_k/(\lambda_k+\i M)$ is approximately
$-\frac{\i\pi}{2}\sign(\lambda_k)$.     So the argument of $\Det\,\D_Y$
is a regularized version of $-\frac{\i\pi}{2}\sum_k \sign(\lambda_k)$.      

The Atiyah-Patodi-Singer $\eta$-invariant is a regularized version of $\sum_k \sign(\lambda_k)$.   The precise regularization does not matter.
We can take, for example,\footnote{The subscript $D$ in $\upeta$ is for Dirac; for a Dirac fermion $\Psi$, we define $\upeta$ by summing over modes
of $\Psi$, ignoring the modes of the conjugate field $\bar\Psi$.    We will shortly introduce a related invariant $\eta$ that is defined by summing over
all fermion modes regardless of charge.}
\be\label{etadef} \upeta=\lim_{\epsilon\to 0^+}\sum_k  \exp(-\epsilon|\lambda_k|) \sign(\lambda_k). \ee
The Pauli-Villars regularization in (\ref{moddo})  gives a different regularization of $\sum_k\sign(\lambda_k)$ (it gives a regularization since 
the argument of $\lambda_k/(\lambda_k+\i M)$ vanishes for $|\lambda_k|\to\infty$).   The two regularizations
are equivalent in the limit $M\to\infty $ or $\epsilon\to 0^+$.

The phase of the path integral for a massless fermion (with a positive regulator mass) is therefore $\exp(-\i \pi\upeta/2)$ \cite{AlvarezGaume:1984nf}.   
This is a refinement of a well-known perturbative
calculation, expanding around Euclidean space with a flat metric and $A=0$, in which one finds a Chern-Simons coupling \cite{Redlich}.  
The nonperturbative statement with $\upeta$ can be related to the perturbative statement based on a Chern-Simons coupling  by
using  the APS index theorem \cite{Atiyah:1975jf} for a manifold with boundary.
Recall first that if $X$ is a $D+1$-manifold without boundary, then the original Atiyah-Singer index theorem \cite{AtiyahSinger} gives a formula for the index $\I$ of a Dirac operator
on $X$ as the integral over $X$ of a certain differential form\footnote{\label{explicit}Concretely, for the case that $\Psi$ transforms in a representation $V$
of some gauge group, $\Phi_{d+2}=\h A(R)\, \Tr_V\exp(F/2\pi)$, where $R$ is the Riemann tensor,
$\h A(R) $ is a certain polynomial in $R$, and $F$ is the gauge field strength.}  $\Phi_{d+2}$ on $X$:    $\I =\int_X \Phi_{d+2}$.   If instead $X$ has a nonempty boundary $Y$, the APS index theorem gives a more general formula
\be\label{hurro} \I=\int_X\Phi_{d+2} -\frac{\upeta}{2}, \ee where $\upeta$ is the $\eta$-invariant of a suitable fermion field on $Y$.
(Roughly, a fermion field $\Psi$ on $X$ reduces along $Y=\partial X$ to two copies of a fermion field on $Y$, and $\upeta$ is here the $\eta$-invariant
of one such copy.  For more detail, see section \ref{cobordism}.)   In perturbation theory around Euclidean space, one can assume that a suitable $X$ exists with $\partial X=Y$ (if $Y=\R^D$, $X$ can be a half-space in
$\R^{D+1}$).   Moreover in perturbation theory,
one  does not see the integer $\I$.  Under these conditions, eqn. (\ref{hurro}) reduces to $\upeta/2=\int_X\Phi_{d+2}$, so one can 
think of $\upeta/2$ as a boundary term related to the characteristic class $\Phi_{d+2}$, or  in other words as a Chern-Simons interaction in a generalized sense.
In a topologically nontrivial situation, $X$ may not exist and if it does the integer $\I$ cannot necessarily be neglected.  So one has to describe the phase of the path
integral for a massless fermion as $\exp(-\i\pi\upeta/2)$, not as the exponential of a Chern-Simons coupling.   

Now let us consider integrating out a massive fermion, still on a $D$-manifold $Y$ without boundary.  In this case, the path  integral is  $\Det\,(\D_Y+\i m)$.   This
 formally equals $\prod_k(\lambda_k+\i m)$.   A regularized version is $\prod_k(\lambda_k+\i m)/(\lambda_k+\i M)$.   This can be factored as a ratio of two factors
 that we have already studied:
 \be\label{twofactors} \prod_k \frac{\lambda_k}{\lambda_k+\i M}\prod_{k'}\frac{\lambda_{k'}+\i m}{\lambda_{k'}}. \ee
 The resulting phase is trivial if $m$ and $M$ have the same sign.  Indeed, in the previous derivation, only the sign of $M$ was relevant, so if $m$ and $M$ have the
 same sign, we can set $m=M$.   But then the product in eqn. (\ref{twofactors}) is simply 1.   The interesting case is therefore the ``topologically nontrivial''
 case in which the physical fermion and the regulator have mass parameters with opposite signs.   For example  (as in section \ref{sec2}), we may take $m<0$ and $M>0$.
 Then eqn. (\ref{twofactors}) is the product of a factor we have already studied divided by the complex conjugate of the same factor.   The
 argument of the path integral is hence $\exp(-\i\pi\upeta)$.

This derivation has been so general that it also applies if $Y$ has a boundary with a self-adjoint boundary condition, the role of self-adjointness being
to ensure that the Dirac eigenvalues are real.    In particular, an example of a self-adjoint boundary condition is the global boundary condition of Atiyah, Patodi, and Singer
\cite{Atiyah:1975jf}.  This accounts for the claim that $\langle \APS|Y\rangle =\exp(-\i\pi\upeta)$ (eqn. (\ref{boldo})).  

But what is the APS boundary condition?  It is not a local boundary condition but a rather subtle global one.   
The original  APS definition (in the notation of the present paper) was  
that $\Psi$, restricted to $W=\partial Y$, is required to be a linear combination of eigenfunctions of $\D_W$ with positive eigenvalue.\footnote{This boundary condition does
not have a sensible continuation to the case that $W$ has Lorentz signature.
The properties of the APS boundary condition such as nonlocality and the absence of a continuation to Lorentz signature motivated some study of the APS theorem in a physical context~\cite{Fukaya:2017tsq}.  
See also \cite{Dab} for a new treatment of the APS index theorem.   For an early use of the APS boundary condition in the physics literature in a different context, see \cite{Hortacsu:1980kv}.
Notice that the local boundary condition $\SL$ is completely physically sensible, and we just introduce the APS boundary condition as a way to compute the partition function~\eqref{eq:p.t.}.}
 For our purposes, however, the most useful way to describe the APS boundary condition is to say that  $|\APS\rangle$ is the state $|\Omega_{m=0} \rangle$
that would  be the  fermion ground state if one sets $m=0$.    This can be regarded as a state in the $m\not=0$ Hilbert space (though it is certainly not the ground
state of the $m\not=0$ Hamiltonian);   this is made explicit in section \ref{sec3}.
  Because of  our  assumption  that  the operator $\CD_W$ has no zero-mode, the $m=0$ ground state $|\Omega_{m=0} \rangle$ is uniquely
determined up to phase.\footnote{If $\CD_W$ has zero-modes, then quantizing these modes gives a space of degenerate ground states of the $m=0$ theory and the simple
definition of the APS  state and boundary condition breaks down.   As a result, a more elaborate definition of APS boundary conditions is needed
in that situation. This case is treated in section \ref{sec:zeromode}.}
After Pauli-Villars regularization, there is  no phase ambiguity in the state $|\APS\rangle$, because in defining $|\APS\rangle$, we use  the
same massless ground state $|\Omega_{m=0} \rangle$, with the same phase, for the physical fermion and for the regulator.   The choice
of  phase cancels between the physical fermion field and the regulator.

We conclude with three remarks that are important in generalizations.

First, what happens if we replace the local boundary condition $\SL$ of equation (\ref{eq:b.c.})  with the opposite boundary condition $\SL'$ defined  by
$(1+\gamma^\tau)\Psi|_W=0$?    In this case, we get no boundary-localized mode if we set $m<0$, but for $m>0$, we get a boundary-localized mode with the
opposite chirality to that in eqn. (\ref{edgemode}).  In quantizing this theory, we have to take $M<0$, since we do not want the regulator field to have a massless mode
propagating on the boundary.   Reversing the sign of both $m$ and $M$ relative to the above derivation has the effect of complex-conjugating the product in eqn.  (\ref{twofactors}),
so now the phase is $\exp(\i\pi \upeta)$.   In other words, anomaly inflow for a boundary mode of one chirality requires a phase $\exp(-\i\pi\upeta)$, and anomaly
inflow for the opposite chirality requires a complex conjugate phase $\exp(\i\pi\upeta)$.    
The fact that the sign of the inflow depends on the chirality of the boundary mode is a standard result, usually deduced perturbatively in terms of Chern-Simons
couplings rather than an $\eta$-invariant.

Second, in this derivation, we have assumed that $\Psi$ carries a $\U(1)$ gauge or global symmetry. 
A 	``Dirac fermion'' is usually understood as a fermion field such that $\Psi$ has positive $\U(1)$ charge and its conjugate $\bar\Psi$ 
(which
appeared in the original action (\ref{eq:action}) though we have not had to discuss it subsequently)
has negative $\U(1)$ charge.
The Dirac operator $\D_Y$ could be taken to act on fermions of either positive or negative charge, that is, on either $\Psi$ or  $\bar\Psi$ .
  We defined 
$\upeta$ by summing over eigenvalues of $\D_Y$ acting on $\Psi$ only.    Obviously, we could have defined a similar invariant $\eta$ by summing over
all eigenvalues of $\D_Y$, acting on either $\Psi$ or $\bar\Psi$.   The relation between $\eta$ and $\upeta$ is simply $\eta=2\upeta$, since complex conjugation
can be used to show that the eigenvalues of $\D_Y$ acting on $\Psi$ or on $\bar\Psi$ are the 
same.\footnote{We could have made the whole derivation thinking of $\D_Y$ as an operator acting on $\bar\Psi$ rather than $\Psi$, leading to an $\eta$-invariant
defined in terms of the eigenvalues of $\D_Y$ acting on $\bar\Psi$.  So $\eta$-invariants defined using eigenvalues of $\D_Y$ acting on $\Psi$  or on $\bar\Psi$ must
be equal.}  
In the absence of a $\U(1)$ symmetry, there is no distinction between $\Psi$ and $\bar\Psi$, so the only natural definition  involves a sum over
all eigenvalues of $\D_Y$.    In other words, to express our results in a way that generalizes naturally to an arbitrary fermion system, we should use $\eta$
rather than $\upeta$.   In terms of $\eta$, the effect of integrating out a massive fermion (with $m<0$, $M>0$) is a factor $\exp(-\i\pi\eta/2)$.   This is the formula
that we will use when we drop the assumption of $\U(1)$ symmetry, starting in section \ref{sec4}.    

Finally, in the above derivation, we assumed that $\D_Y$ has no zero-modes, so that all $\lambda_k$ are nonzero.  We did this because we started
with a massless fermion; the path integral of a massless fermion vanishes when there is a zero-mode, and does not have a well-defined phase.
However, a zero-mode of $\D_Y$ causes no problem in defining the phase that is induced by integrating out a {\it massive} fermion.  Taking $m=-M$, each mode
contributes to the regularized fermion path integral a factor $(\lambda-\i M)/(\lambda+\i M)$, which is $-1$ if $\lambda=0$.   So in including zero-modes,
we want each such mode to give a factor of $-1$.
  For this, we define
\be\label{signdef}\sign(\lambda)=\begin{cases} 1 & ~\mathrm{if}~\lambda\geq 0\cr -1 & ~{\mathrm {if}}~\lambda<0,\end{cases}\ee
and then we define $\upeta$ (and $\eta=2\upeta$) precisely as in (\ref{etadef}).   Then the phase induced by integrating out a fermion with $m<0$, $M>0$ is
$\exp(-\i\pi\upeta)$ or $\exp(-\i\pi\eta/2)$, whether zero-modes are present or not.
We should warn the reader, however, that this definition of $\eta$ or $\upeta$ including the zero-modes
is not completely standard.  In the original APS paper  \cite{Atiyah:1975jf} (and many subsequent mathematical references), such a sum
with the zero-modes included is called $2\xi$ rather than $\eta$; $\eta$ is defined precisely as in eqn. (\ref{etadef}), but summing over nonzero modes only.

\subsection{The Chiral Fermion Partition Function as a State Overlap}\label{sec3}

To complete the proof of the formula (\ref{refined}) for the partition function, we must calculate 
the state overlaps $\bra{\SL} \et{\Omega} $ and $ \bra{\APS} \et{\Omega}$
that appear in eqn. (\ref{eq:rewrite}).
We will see that $\bra{\SL} \et{\Omega} $ is essentially the partition function of the boundary chiral fermion $\chi$.\footnote{
A similar phenomenon that a state vector in $d+1$-dimensions is related to a partition function in $d$-dimensions was also observed
in a different system~\cite{Belov:2004ht}. There, the $d$-dimensional theory is a free $\U(1)$ Maxwell theory with $d=4$ and 
the $D=d+1$-dimensional theory is a gauge theory which flows to a topological field theory in the low energy limit.
Their result might be interpreted as anomaly inflow of the global anomaly of the Maxwell theory under S-duality~\cite{Witten:1995gf,Seiberg:2018ntt,Hsieh:2019iba}.}

They are computed by a straightforward quantization of the fermion $\Psi$ on the space $W$.
We make a Wick rotation $\tau \to \i t$, and the Hamiltonian is easily read off from the action \eqref{eq:action} as
\beq
H = \int_W  \d^d x\sqrt g\, \Psi^\dagger ( \CD_W + \gamma^\tau m ) \Psi,
\eeq
where we have used the usual relation $\overline{\Psi} = \Psi^\dagger \gamma^\tau = \Psi^\dagger ( \i \gamma^t)$  (for Dirac fermions).   

The operators $\gamma^\tau$ and $\CD_W$ anticommute, so we can make a  mode expansion of the following kind.
All modes of the $\Psi$ field on $W$  can be expressed as linear combinations of  pairs of modes $(\psi_{+, a},\psi_{-,a})$ which satisfy
\beq
\gamma^\tau \psi_{ \pm, a} = \pm \psi_{ \pm, a}, \qquad \CD_W \psi_{ +, a} = \lambda_a \psi_{ -, a}, \qquad \CD_W \psi_{ -, a} = \lambda_a \psi_{ +, a}.
\eeq
Here $\lambda_a$ are the positive square roots of the eigenvalues of $(\CD_W)^2$.  Since we assume that $\D_W$ has no zero-mode, all modes
have $\lambda_a>0$ and are paired in the above fashion.  We expand
\beq
\Psi = \sum_a ( A_{+, a} \psi_{+, a} +A_{-, a} \psi_{-, a} ).
\eeq
The Hamiltonian is given by
\beq
H = \sum_a (A_{+, a}^\dagger, A_{-, a}^\dagger) 
\left( \begin{array}{cc} 
m & \lambda_a \\
\lambda_a & -m 
\end{array}
\right)
\left( \begin{array}{c} 
A_{+, a} \\
A_{-, a}
\end{array}
\right).
\eeq
The operators $A_{\pm,a}$ are annihilation operators, and $A^\dagger_{\pm, a}$ are the corresponding creation operators.

We can consider each value of $a$ separately, so we omit $\sum_a$ (or $\prod_a$, depending on the context) until the end of this section.
After defining
\beq
( \cos (2 \theta_a), \sin (2\theta_a)) =  \frac{(m, \lambda_a) }{\sqrt{m^2 +\lambda_a^2} },
\eeq
the Hamiltonian is easily diagonalized as
\beq
H = \sqrt{\lambda_a^2+m^2} (  B_{1,a}^\dagger  B_{1,a} - B_{2,a}^\dagger  B_{2,a} ),
\eeq
where
\beq
 B_{1,a} = \cos \theta_a A_{+,a} + \sin \theta_a A_{-,a}, \qquad  B_{2,a} = - \sin \theta_a A_{+,a} + \cos \theta_a A_{-,a}.
\eeq
The ground state $\ket{\Omega}$ is specified by the conditions $B_{1,a}\ket{\Omega}=0$ and $B_{2,a}^\dagger \ket{\Omega}=0$, or
$0=\bra{\Omega} B_{1,a}^\dagger =\bra{\Omega} B_{2,a}$.

The local boundary condition $\SL$ given in \eqref{eq:b.c.} requires the condition that $A_{-,a} =0$ at the boundary, while $A_{+,a}$ is free.
Then the corresponding state $\ket{\SL}$ is specified by the condition that $\bra{ \SL} A_{-,a}=0$ and $ \bra{ \SL}A_{+,a}^\dagger  = 0$. 
(See \cite{Yonekura:2016wuc} for more details about the relation between boundary conditions and state vectors.)
The APS boundary state $\ket{\APS}$ is the ground state with $m$ set to 0, which corresponds to $\theta_a=\pi/4$ for all $a$ and hence $\cos\theta_a=\sin\theta_a$.
So  $\ket{ \APS}$ is characterized by $\bra{\APS}( A_{-,a} - A_{+,a})=0$ and $\bra{\APS} (A_{-,a} + A_{+,a})^\dagger =0$.

These conditions leave us free to multiply the states $\bra{\SL} $ and $\bra{\APS}$  by arbitrary nonzero complex numbers.    We could constrain them to  have unit norm,
but this leaves an undetermined  phase.
However, the ambiguities do not matter. This is because we will eventually take the ratio between the negative mass theory $m <0$ (for the physical fermion) and the positive mass theory $m>0$ (for the Pauli-Villars regulator).   The ambiguities cancel out in the ratio, because
the boundary conditions $\bra{\SL} $ and $\bra{\APS}$ do not depend on the mass parameter. The ground state $\ket{\Omega}$ has a phase ambiguity, which is related to the appearance of the line bundle $\CL_W$, as discussed in section \ref{sec2}. 
This ambiguity is physically unavoidable because of Berry phases when background fields are varied. But this ambiguity cancels out 
in the product $\bra{\SL} \et{\Omega} \bra{\Omega} \et{\APS}$. 
Thus overall normalization factors will always cancel out and we can use any explicit state vectors satisfying the above conditions.

Let $\ket{E}$ be a state vector with 
\beq
A_{+, a}\ket{E} = A_{-, a}\ket{E}=0.
\eeq
Then we can realize the relevant state vectors as
\beq
\ket{\Omega}= B_{2,a}^\dagger \ket{E}, \qquad \bra{\SL} = \bra{E}A_{-,a}, \qquad \bra{\APS} = \bra{E}(A_{-,a} - A_{+,a} ).
\eeq
From these expressions, we easily get
\beq
\bra{ \APS } \et{\Omega} = \cos \theta_a + \sin \theta_a   \to 1  \qquad  (  \lambda_a \ll |m|),
\eeq
and 
\beq
\bra{ \SL } \et{\Omega}  = \cos \theta_a  
 \to \left\{ \begin{array}{ll}
1 &  (m > 0,  \quad  \lambda_a \ll |m|) \\
\lambda_a/ (2|m|) &  (m<0,  \quad  \lambda_a \ll |m|).
\end{array}
\right.  
\eeq
Essentially, $\cos \theta_a$ for $m<0$ is the eigenvalue $\lambda_a$ normalized by $2|m|$ as long as $\lambda_a \ll |m|$. But  $|m|$ plays the role of
a regulator. In the limit $\lambda_a/|m| \to \infty$, we have $\cos \theta_a \to 1/\sqrt{2}$,  independent of $a$ or $m$.
Upon taking the ratio between the theories with $m<0$ and $m>0$, the factors of $\cos \theta_a$ associated to eigenvalues with $|\lambda_a|\gg m$ cancel out,
 and hence the ultraviolet is regularized.
Therefore, after taking the ratio, we finally get 
\beq\label{chdet}
\frac{ \bra{\SL} \et{\Omega} \bra{\Omega} \et{\APS} }{ | \bra{\APS} \et{\Omega} |^2}  = \prod_a \left( \frac{\lambda_a}{2|m|} \right)_{\rm reg}  = | {\rm Det}( \CD_W^{+} ) |
\eeq
where the subscript ``${\rm reg}$'' means that the product is regularized by $\cos \theta_a$ in the way just mentioned.
Note that the determinant of the nonchiral Dirac operator $\D_W$ would have a factor of $\lambda_a^2$ for each pair of modes, while in  eqn. (\ref{chdet}), there is just one
factor of $\lambda_a$ for each pair.  That is why the right hand side of eqn. (\ref{chdet}) is  $|\Det\,\D_W^+|$ (where as before $\D_W^+$ is the chiral Dirac operator on $W$), not $\Det\,\D_W$. Note that the $\lambda_a$ are all positive, and that $|\Det\,\D_W^+|$ is the same as $|\Det\, \D_W|^{1/2}$.

Combining this result with eqn.  (\ref{eq:rewrite}) and with what we learned in section \ref{aps}, it follows that the
 total partition function of the bulk massive fermion $\Psi$ with the boundary condition $\SL$ is  given by
\beq
Z(Y, \SL) =  | {\rm Det}( \CD_W^{+} ) |\exp( -\pi \i \upeta).
\eeq
as we claimed in eqn. (\ref{refined}), and as 
suggested by the Dai-Freed theorem~\cite{Witten:2015aba,Witten:2016cio}.

\subsection{The General Case}\label{sec4}

We have so far described anomaly inflow for a fermion all of whose components are charged under a $\U(1)$ symmetry.  
Here we want to generalize the construction to describe anomaly inflow for an arbitrary set of relativistic boundary fermions.
For this, we will start with an arbitrary relativistic fermion field $\chi$ on the $d$-manifold $W$, and then find a massive fermion system on $Y$
that is related to $\chi$ by anomaly inflow.    This problem has also been studied recently in the appendix of \cite{Freed:2019sco}, in the context of generalizing the
Dai-Freed theorem to an arbitrary system of relativistic fermions.

We will ultimately work in Euclidean signature, but the necessary conditions to define a physically sensible theory are most naturally stated in Lorentz signature.
The field $\chi$ at each point in $W$ will take values in a vector space $S$ that will be a representation of a group that includes Lorentz symmetries
possibly together with some gauge or global symmetries.  In the Lorentz  group, we  possibly  include disconnected components, 
related to time-reversal and reflection symmetry; also the symmetry group that acts on the fermions is really a $\Z_2$ central extension of a product
of the Lorentz group with a group of gauge or global symmetries.   The central $\Z_2$ subgroup is generated by the operator $(-1)^\sF$ that distinguishes bosons
and fermions.
This $\Z_2$ central extension may  be a spin group, a pin group, or a refinement such as ${\mathrm {spin}}_c$. We treat all cases uniformly.\footnote{\label{exception} The symmetry group of a relativistic fermion system in any dimension $d\geq 2$
is as just described.   For $d=1$, relativity loses its force and there are more general possibilities.  We will run into this in section \ref{d1}.}

 We will denote the components of $\chi$ generically as $\chi^a$, where $a$ runs over a basis of $S$. The only assumption that we make about $\chi$ is that it has a first order action 
of the general form
\be\label{first}\frac{\i}{2} \int  \d^dx \sqrt g\,\sum_{a,b} \chi^a \sigma^\mu_{ab}D_\mu \chi^b, \ee
with some matrices $\sigma^\mu_{ab}$, which obey constraints that will be described in a moment.   We do not make any assumption about $S$
beyond whatever is needed to ensure the existence of a physically sensible action of this form. 
We have not written  a  mass term in eqn. (\ref{first})   and the more
interesting case (since anomalies are possible) is that the symmetries do not allow a mass term.  
Possible mass terms and possible additional couplings, such as couplings to scalar fields, do not modify the following analysis in any essential way.

Because  $\chi$ satisfies fermi statistics, and in view of the possibility of integration by parts, only the
symmetric part of  $\sigma^\mu$ really contributes to the action, so we can assume that 
$ \sigma^\mu_{ab} = \sigma^\mu_{ba}$. In Lorentz signature, the $\chi^a$ are real (after quantization, they become
hermitian operators) and the matrices $\sigma^\mu$ are real, so as to make the action and the Hamiltonian real.\footnote{The real and imaginary parts
of a complex field are real, so there is no loss of generality in assuming the $\chi^a$ to be real.   If there are gauge and global
symmetries, then their generators acting on $\chi$ are real matrices.  Although $\sigma^\mu$ and the gauge and global symmetry generators are all real, in general one
cannot write $S=S'\otimes S''$ with $S'$ and $S''$ being real vector spaces such that the $\sigma^\mu$ are  bilinear forms on $S'$ and the gauge and global symmetry
generators act only on $S''$.   The existence or not of such a decomposition will play no role in the present paper.}
 Also, in Lorentz signature, one wants $\sigma^0$ to be a positive-definite matrix for unitarity
 (positivity of $\sigma^0$ ensures  that if $\beta=\sum_a r_a\chi^a$ with real coefficients $r_a$, then after quantization,
 $\{\beta,\beta\}>0$, consistent with  positivity of the Hilbert space inner product).  Here $\sigma^0$ is the time component of $\sigma^\mu$ in any local Lorentz frame.
 (In some particular local Lorentz frame, one can make a linear transformation of the $\chi^a$ to set $\sigma^0=1$, but this statement would then only hold in that
 particular frame.)

The Dirac equation is
\be\label{second} \sigma_{ab}^\mu D_\mu \chi^b=0.  \ee
The left hand side of this equation is not valued in the same vector space $S$ in which $\chi$ takes values,
but rather in the dual space $\t S$.    The duality is clear from the fact that the action (which is a pairing of $\chi^a$ with the left hand side of eqn. (\ref{second})) exists.
   The $\sigma^\mu$ are not gamma matrices, since they map the vector space  $S$ not to itself, but rather to a dual
space $\t S$.  Fields valued in
$S$ and $\t S$ will be represented by spinor fields $\chi^a$ or $\t\chi_a$ with ``index up'' or ``index down.''  
 
To get a dispersion relation of the appropriate form for relativistic fermions, the  matrices $\sigma^\mu$  should satisfy the following.  
Let $g_{\mu\nu}$ be the spacetime metric (in a local Lorentz frame), and define $\bar\sigma_\mu=g_{\mu\mu}\sigma_\mu^{-1}$, with no summation over $\mu$.  
(Thus in Euclidean signature with $g_{\mu\nu}=\delta_{\mu\nu}$,
$\bar\sigma_\mu$ is just the inverse of $\sigma_\mu$.)
Since $\sigma_\mu$ maps $S$ to $\t S$,
its inverse matrix $\sigma_\mu^{-1}$ maps $\t S$ to $S$, and therefore $\bar\sigma_\mu$ does the same.   Then we want
\be\label{again}\bar\sigma_\mu\sigma_\nu +\bar\sigma_\nu \sigma_\mu=2g_{\mu\nu}. \ee
This equation makes sense: the   $\sigma_\mu$ map $S $ to $\t S$ 
and  the $\bar\sigma_\mu$ map in the opposite direction, so the product maps $S$ to itself and can be a $c$-number.
From eqn. (\ref{again}) we can deduce that $\chi$ obeys the expected dispersion relation, since eqn. (\ref{second}) implies (in flat space) that
\be\label{impl}0 =(\bar\sigma^\nu\partial_\nu)(\sigma^\mu\partial_\mu) \chi =g^{\mu\nu}\partial_\mu\partial_\nu \chi.\ee  
Therefore we require \eqref{again} as a condition on $\sigma^\mu$.
Note that the Clifford algebra of  eqn. (\ref{again}) implies Lorentz invariance with generators $\sigma_{\mu\nu}=\frac{1}{4}( \bar\sigma_\mu\sigma_\nu -\bar\sigma_\nu \sigma_\mu)$. 
These generators map $\chi$ to $\chi$, so the original theory of $\chi$ only was in fact Lorentz-invariant.

As an example of this, in two dimensions with Euclidean signature, $\sigma_1$ and $\sigma_2$ can be the $1\times 1$ matrices 1 and $\i$.   Then
$\bar\sigma_1$ and $\bar\sigma_2$ are $1$ and $-\i$.  The fermion $\chi$ has a single component, with definite chirality.  In Lorentz signature,
$\chi$  is called a Majorana-Weyl fermion. In this example, the conjugate fermion $\t \chi$ that is introduced in a moment has opposite chirality.

We now introduce a second spinor field $\t \chi_a$ that transforms in the dual representation $\t S$.   (In a particular example, the original representation
might be self-dual so it may happen that $\chi$ and $\t\chi$ actually transform the same way.   We proceed the same way whether this is so or not.  Note that even
if $\chi$ and $\t\chi$ transform the same way, a mass term for $\chi$ might be forbidden by fermi statistics, so anomalies may be possible.)
For $\t \chi$ we can always write an action
\be\label{firsto} -\frac{\i}{2} \int \d^dx \sqrt{g} \,\sum_{a,b} \t\chi_a \bar\sigma^{\mu\,ab}D_\mu \t\chi_b, \ee
since the matrices $\bar\sigma^{\mu}$  map $\t S$ back to $S$.   
Here we have put a minus sign in the action relative to eqn.~\eqref{first}, because $\bar\sigma^0 = - (\sigma^0)^{-1}$ is negative-definite in Lorentz signature.
We write $\D_W^+= -\sigma^\mu D_\mu$ and $\D_W^-=\bar\sigma^\mu D_\mu$.

Now we can combine $\chi$ and $\t \chi$ to a fermi field 
$\Psi=\begin{pmatrix} \chi \cr \t \chi\end{pmatrix}$.    Acting on $\Psi$, we define the matrices
\be\label{third}\gamma_\mu =\begin{pmatrix} 0 & \bar\sigma_\mu\cr \sigma_\mu & 0 \end{pmatrix}.  \ee 
These are gamma matrices $\{ \gamma_\mu, \gamma_\nu \} = 2 g_{\mu\nu}$ by virtue of eqn. (\ref{again}).   

The action for $\Psi$ is the sum of the actions for $\chi$ and for $\t\chi$ plus a possible mass term:
\be\label{fourth}  -\i  m\int \d^dx \sqrt{g} \,(\t\chi,\chi) 
\ee  
where $( \t\chi,\chi) = \t\chi_a\chi^a$.   
We can  introduce an antisymmetric bilinear form $\langle ~,~\rangle$ on the space $S \oplus \t S$, invariant under all symmetries,
 by
\beq\label{biform}
\langle \Psi_1, \Psi_2 \rangle = (\t\chi_1)_a (\chi_2)^a - (\chi_1)^a(\t\chi_2)_a.
\eeq
One can verify that
\be\label{verify}
\la \gamma^\mu\Psi_1,\Psi_2\ra =-\la\Psi_1,\gamma^\mu \Psi_2\ra. \ee
The action can be written in the manifestly Lorentz-invariant form.
\beq
- \frac{ \i}{2} \int \d^dx \sqrt{g} \,   \langle \Psi, (\gamma^\mu D_\mu +m) \Psi \rangle .\label{action1}
\eeq
The identify (\ref{verify}), together with integration by parts and the antisymmetry of $\la~,~\ra$, can be used to verify that the equation of motion derived from this action
is the expected $(\slashed{D}+m)\Psi=0$.

Crucially, it  is also possible to add another dimension.    One adds a new coordinate $\tau$ and a new gamma matrix
\be\label{north} \gamma^\tau =\begin{pmatrix} 1 & 0 \cr 0&-1\end{pmatrix}\ee
that obeys $(\gamma^\tau)^2=1$ and anticommutes with the previous gamma matrices. 
One can verify the analog of (\ref{verify}):
\be\label{verity} \la\gamma^\tau\Psi_1,\Psi_2\ra =-\la\Psi_1,\gamma^\tau\Psi_2\ra. \ee 
With a new gamma matrix, one can extend $d$-dimensional Lorentz symmetry to $D=(d+1)$-dimensional Lorentz symmetry,
the new generators being $\frac{1}{4}[\gamma^\tau,\gamma^\mu]$.    The form $\la ~,~\ra$ possesses $D$-dimensional Lorentz symmetry.\footnote{To 
check invariance under the additional
generators, the identity we need is
$\langle \gamma^\tau \gamma^\mu \Psi_1, \Psi_2 \rangle + \langle \Psi_1, \gamma^\tau \gamma^\mu \Psi_2 \rangle =0$,
which follows  immediately from (\ref{verify}) and (\ref{verity}).    Note also that if the original $d$-dimensional theory has a time-reversal or reflection symmetry, the 
$D$-dimensional theory  has the same symmetry.   This follows from Lorentz invariance in $D$ dimensions together with the discrete symmetry in $d$ dimensions.}
Then there can be a new term in the action 
\be\label{orth} -\i \int  \d^dx \d \tau \sqrt{g} (\t\chi, D_\tau \chi) = -\frac{\i}{2}\int  \d^dx \d \tau \sqrt{g} \langle \Psi, \gamma^\tau D_\tau\Psi\rangle.  \ee
Combining \eqref{action1} (or rather its integral over $\tau$) and \eqref{orth}, we construct the action of the massive fermion $\Psi$ in $D=d+1$ dimensions 
\beq
- \frac{ \i}{2} \int \d^{d+1}x \sqrt{g} \,   \langle \Psi, (\slashed{D} +m) \Psi \rangle .\label{action2}
\eeq 
From this action, one can derive  the expected
$D$-dimensional Dirac equation $(\slashed{D}+m)\Psi=0$ (all facts used to prove this statement in dimension $d$ also hold in dimension $D=d+1$).
The action and the Dirac equation have  manifest $D$-dimensional Lorentz invariance.
Therefore, this  action can be defined on any $D$-manifold $Y$ that is endowed with all the structures\footnote{\label{spinisom}By a spin
structure or gauge bundle on $Y$, we always mean a spin structure or gauge bundle on $Y$ that extend the spin structure and gauge bundle of $W$.
Along $\partial Y$, a specific isomorphism is chosen between the spin structure and gauge bundle of $Y$ and those of $W$.} 
(such as orientations, spin structures, gauge bundles) that were needed to define the original fermion field $\chi$ on $W$.

With the $d$-manifold $W$ still understood to  have Lorentz  signature, the original $d$  gamma matrices (\ref{third}) are all real, and of course $\gamma^\tau$  as
defined  in eqn. (\ref{north})  is also real.    So all gamma  matrices are real, and therefore  the Lorentz generators are also real. $\chi$ was real to begin with, and
$\t\chi$, transforming as the dual to $\chi$, is also real.    So there is a physically
sensible theory of a real $\Psi$ field in $D$ dimensions  with Lorentz signature.   This is important in order to make the formula we will get for anomaly inflow physically meaningful.

Thus, on a completely general $D$-manifold with Lorentz signature, $\Psi$ can be considered real.   What happens when we go to Euclidean signature?   Then
$\Psi$ is generally no longer real, but in fact it becomes pseudoreal.\footnote{In some cases, $\Psi$ can also be given a real structure in Euclidean signature.
This happens if $\Psi$ has $2k$ components, and the full symmetry group including rotations and gauge and global symmetries is a subgroup of $\U(k)$.  This
group is a subgroup of $\Sp(2k)$ (corresponding to
the pseudoreal structure of $\Psi$ described in the text) and of $\O(2k)$ (corresponding to an additional  real structure).   When $\Psi$ does have a real structure, this can be used to
simplify the analysis of anomalies.    However,  we will focus on the pseudoreal structure that $\Psi$ carries universally.}   We rotate to Euclidean signature by a Wick rotation of $W$, say $x^0\to -\i x^0_E$, where $x^0$ parametrizes
a time direction in $W$.   The corresponding
transformation of gamma matrices is $\gamma^0=-\i\gamma^0_E$.   In Lorentz signature, all gamma matrices were real, so in Euclidean signature, they are all
real except for one, namely  $\gamma^0_E$.
   Let $*$ be complex conjugation.   Evidently, $*$ anticommutes with $\gamma^0_E$ and commutes with other
gamma matrices.  Accordingly,
\be\label{antiunitary} \sC = * (-\i\gamma^0_E) =*\gamma^0 \ee
anticommutes with all gamma matrices.   $\sC$ therefore commutes with the  $D$-dimensional rotation generators $\frac{1}{4}[\gamma_\mu,\gamma_\nu]$, and of course it commutes with all gauge and global symmetry
generators (which are real and commute with the gamma matrices).   So $\sC$ commutes with all symmetries.
$\sC$ is antilinear and satisfies
\be\label{square}\sC^2=-1. \ee
The existence of an antilinear operator $\sC$  that commutes with all symmetries and obeys $\sC^2=-1$ means that $\Psi$ is in a pseudoreal representation of the symmetry
group.   (This statement follows more abstractly from the existence of the antisymmetric bilinear form $\la~,~\ra$ that preserves all symmetries.  But that reasoning would
not give an explicit construction of $\sC$.)  Since $\sC$ is antilinear and anticommutes with gamma matrices, it commutes with the self-adjoint
Dirac operator $\D_Y=\i\sum_{\mu=1}^D\gamma^\mu D_\mu$.   Moreover, the fact that $\sC$ is invariant under all symmetries means that the construction can be made on an arbitrary
$D$-manifold
$Y$ that admits all the relevant structures, though we started by singling out a particular ``time'' direction.\footnote{Concretely, to make this construction in
curved spacetime, one uses the vierbein formalism that is anyway necessary to define fermions in curved spacetime, and one uses the above formulas in a locally
Euclidean frame.  Because $\sC$ commutes with the rotation generators in any locally Euclidean frame, the definitions of $\sC$ in  different locally Euclidean
frames are compatible.}

Using the antilinear operator $\sC$ that was just introduced, we can define a hermitian form on the space of $\Psi$ fields,
\be\label{hermform}   (\Psi_1, \Psi_2) =  \la \sC\Psi_1,\Psi_2\ra  , \qquad (\Psi_1,\Psi_2)_Y =\int_Y \d^Dx \sqrt g\, (\Psi_1, \Psi_2).   \ee
Calling $(~,~)$ a hermitian form means that it is linear in the second variable and antilinear in the first.   This hermitian form is clearly invariant under
all symmetries, since it was constructed from the invariant ingredients $\la ~,~\ra$ and $\sC$.    Explicitly if $\Psi=\begin{pmatrix}\chi\cr \t\chi\end{pmatrix}$, 
then $\sC\Psi= \begin{pmatrix} \bar\sigma^0\t\chi^* \cr \sigma^0\chi^*\end{pmatrix}$, where $^*$ is complex conjugation.   
So
\be\label{posmeas}(\Psi,\Psi) = \sigma^0_{ab}\chi^{a*}\chi^b -\bar\sigma^{0ab}\t\chi_a^*\t\chi_b  . \ee
Since $\sigma^0$ is positive-definite and $\bar\sigma^0$ is negative-definite, we see that $(\Psi,\Psi)$ is positive-definite.  Moreover, the identities that have been
described imply that, assuming $Y$ has no boundary,  $\D_Y=\i \gamma^\mu D_\mu$ is self-adjoint with respect to the hermitian form $(~,~)_Y$:
\be\label{hermop} (\D_Y\Psi_1,\Psi_2)_Y=(\Psi_1,\D_Y\Psi_2)_Y. \ee

Though $\Psi$ is not real on a general $D$-manifold with Euclidean signature, it can be considered real on a $D$-manifold that is presented 
with a given factorization as $W\times \R$, where $W$ has Euclidean signature  (and we only care about the behavior
of $\Psi$  under symmetries that preserve this factorization).   One way to explain this statement is to observe that whether
the $\R$ direction is Lorentzian or Euclidean does not affect the behavior under symmetries of $W$, and if the $\R$ direction is Lorentzian, we are on a Lorentz
signature manifold, so $\Psi$ can be considered real.     Since the point is important, we will give another explanation.
The vector spaces $S$ and  $\t S$ in which $\chi$ and $\t\chi$  take values were dual to each other when $W$ has Lorentz signature, and this duality persists after
continuation to Euclidean signature.   But in Euclidean signature, the symmetry group -- a double cover of the group of rotations
and possible  gauge or  global symmetries -- is compact.   The dual of a representation  of a compact Lie group  is isomorphic to the complex
conjugate representation.   This means that when $W$ has Euclidean signature,  as far as symmetries of $W$ are concerned, 
$S$ and $\t S$  can be regarded as complex  conjugate vector  spaces, while in Lorentz
signature they were each real and were not related to each other by complex conjugation. 
Hence $\chi$ and $\t\chi$ transform as complex conjugates and $\Psi$ can be considered real and takes values in a real vector
space $\sf S$.   Informally ${\sf S} =\S\oplus \t S$ (the precise statement is that $\sf S$ is a real vector space whose complexification  has such a decomposition).

We stress that there are two senses in which $\Psi$ might be considered real.  In the starting point, in Lorentz signature, $\Psi$ is naturally real.
   This is important for showing that our study of anomaly inflow applies to a physically sensible quantum field theory.   In Lorentz signature, for $\Psi$ to be real
   just means that $\chi$ and $\t\chi$ are both real.
But in our calculations, we will be in Euclidean signature, and then it is useful that a different real structure can be defined in the case of a $D$-manifold that
locally is a product $\R\times W$.   In this second real structure, $\t\chi$ is proportional to the complex conjugate of $\chi$.   The relationship between them is
described more precisely presently.

The Dirac operator $ \D_W=\sum_{\mu=1}^d \gamma^\tau\gamma^\mu D_\mu$  (see footnote \ref{equivalence} for its relation to the usual Dirac operator $\i\sum_{\mu=1}^d\gamma^\mu D_\mu$) 
can be written $\D_W=\begin{pmatrix} 0& \D_W^-\cr \D_W^+ & 0\end{pmatrix}$, where $\D_W^+=-\sigma^\mu D_\mu$, $\D_W^-=\bar\sigma^\mu D_\mu$
are the Dirac operators for $\chi$, $\t\chi$ that we had in the beginning before going to $D$ dimensions.

Now consider the theory of the $\Psi$ field on a $D$-manifold $Y$ with boundary $W$, with the local boundary condition  $\SL$ defined by $\t\chi|_W=0$.   For $m<0$, and treating
$Y$ as a product near the boundary, $\Psi$ has a boundary-localized mode given by the ansatz of eqn. (\ref{edgemode}).  In particular, $\t\chi$ vanishes identically (in the approximation that $Y=W\times \R_-$),
 and $\chi$ vanishes exponentially fast away from $W$.   
The effective theory for this boundary-localized mode is the original purely $d$-dimensional theory (\ref{first}) for $\chi$ only.    
This means that anomaly inflow from $Y$, generated by integrating out the massive
field $\Psi$, will cancel any anomaly of the original $d$-dimensional theory of $\chi$.    

To describe this anomaly inflow precisely, we would like to generalize to this situation the formula (\ref{refined}) for $Z(Y,\SL)$.   It is not difficult to guess the generalization.   First of all, integrating out the massive field $\Psi$ will generate a factor of $\exp(-\i\pi\eta/2)$.   (When there is no $\U(1)$ symmetry, we have to use $\eta$,
defined by summing over all eigenvalues of $\D_Y$, rather than $\upeta=\eta/2$, defined by summing only over eigenvalues of positive charge.   This was explained
at the end of section \ref{aps}.)     Also, by fermi statistics, the kinetic operator $\D_W^+ = - \sigma^\mu D_\mu$ of $\chi$ can be viewed as an antisymmetric matrix.  The path integral 
of $\chi$ is its Pfaffian, $\Pf(\D_W^+)$.    When there is a $\U(1)$ symmetry carried by all the fermions, this Pfaffian can be viewed as the 
determinant of a kinetic operator that acts on a smaller
set of fields (those of positive $\U(1)$ charge), but in the absence of a $\U(1)$ symmetry, the path integral is best understood as a Pfaffian.
So the natural analog of eqn. (\ref{refined}) is
\be\label{torro} Z(Y,\SL)= |\Pf(\D_W^+)|\exp(-\i\pi\eta/2), \ee
and we will aim to justify this formula by adapting the derivation of section \ref{sec3}.

The first step is straightforward.  $Z(Y,\SL)$ can be expressed in terms of inner products 
 by the same logic as before, leading to eqn. (\ref{eq:rewrite}). 
Here  $\langle \APS|Y\rangle $ is known from the arguments of section \ref{aps}, and we will slightly modify the arguments of section \ref{sec3} to compute the inner products $\bra{\SL} \et{\Omega}$ and $\bra{\APS} \et{\Omega}$.

In general, when one quantizes fermions, to get a Hilbert space with a positive-definite inner product, the fermions that are being quantized carry a real structure.
(If complex fermion fields are present, one can take their real and imaginary parts.)   Real fermion fields become hermitian operators after quantization.
And on the space of real fermion fields, there is a positive-definite inner product that appears in the canonical anticommutation relations.  
So in order to quantize the $\Psi$ field on $W$, we want to describe explicitly the real structure that is appropriate if $Y=\R\times W$, and the natural
positive-definite inner product in this real structure.

From our definitions, it follows immediately that if $W$ has Lorentz signature with time coordinate $x^0$, and $\gamma^0$ is the corresponding
gamma matrix (defined in eqn. (\ref{third})), then  $-\langle \Psi_1,\gamma^0\Psi_2\rangle = \sigma^0_{ab}\chi_1^a\chi_2^b-\bar\sigma^{0 \,ab}(\t\chi_1)_a (\t\chi_2)_b$.
Since $\sigma^0$ is positive-definite and $\bar\sigma^0$ is negative-definite, it follows that the inner product $(\Psi_1,\Psi_2)_0=-\langle \Psi_1,\gamma^0\Psi_2\rangle$
is positive-definite, as long as $\chi$ and $\t\chi$ are real.   We actually want to take $W$ to be Euclidean and make a Wick rotation $\tau=\i t$ of the coordinate
orthogonal to $W$.   Lorentz invariance
of the pairing $\la\Psi_1,\Psi_2\rangle$ means, of course, that  any statement  with $x^0$ viewed as a time coordinate and $\tau$ as a space coordinate
has an analog if the roles are reversed.    Thus, setting $\gamma^t=-\i\gamma^\tau$, and now assuming $W$ to be Euclidean, we define the inner product 
\beq
(\Psi_1, \Psi_2)_t = - \langle \Psi_1, \gamma^t \Psi_2 \rangle =   \langle \Psi_1, \i \gamma^\tau \Psi_2 \rangle,  \label{eq:inner}
\eeq
where $\gamma^\tau$ was defined in eqn. (\ref{north}).
This will be positive-definite if we place on $\Psi$ the appropriate Wick-rotated reality condition.   What reality condition will do the job?
The positivity of $(~,~)_t$  is ensured if we impose on $\Psi$ a reality condition such that
\beq
(\Psi_1, \Psi_2)_t  =  (\Psi_1, \Psi_2), 
\eeq
where the right-hand-side is the positive definite hermitian inner product introduced in eqn.~\eqref{hermform}.
Notice that the left-hand-side is linear in $\Psi_1$, while the right-hand-side is antilinear in $\Psi_1$, so this equation imposes a certain reality condition on $\Psi_1$.
Explicitly, since $(\Psi_1, \Psi_2)_t =   \langle -\i \gamma^\tau  \Psi_1, \Psi_2 \rangle$ and $(\Psi_1, \Psi_2) =  \la \sC\Psi_1,\Psi_2\ra$,
the reality condition is given by 
\beq
\i \gamma^\tau  \sC\Psi = \Psi.\label{eq:reality}
\eeq
This is a consistent reality condition,  since $\sC$ and $\gamma^\tau$ anticommute
and hence $(\i \gamma^\tau  \sC)^2 = - \sC^2=1$. More explicitly, in  terms of $\chi$ and $\t\chi$, eqn. (\ref{eq:reality}) becomes
$\t\chi_a = -\i \sigma^0_{ab}(\chi^b)^*$ and $ \chi^a = \i  \bar\sigma^{0\, ab} (\t\chi_b)^*$, where $^*$ is complex conjugation.\footnote{This way of writing the
reality condition looks noncovariant, but it actually is covariant, since
 eqn. (\ref{eq:reality}) is a manifestly covariant version.  
  As always, formulas such as $\t\chi_a = -\i \sigma^0_{ab}(\chi^b)^*$ are written in a locally Euclidean frame.
  With the representation of the gamma matrices that led to the explicit definition (\ref{antiunitary}) of $\sC$,
 some rotation generators of the tangent space of $W$ are real and some are imaginary; when this is taken into account, the formula $\t\chi_a = -\i \sigma^0_{ab}(\chi^b)^*$ is
covariant under a change of the locally Euclidean frame, as is the definition of $\sC$.   
A different representation of the gamma matrices (in which $\gamma_\tau$ is imaginary and the gamma matrices of $W$ are real) would make the relation
between $\t\chi$ and $\chi^*$ look more natural while obscuring the relationship to the starting point, which was a physically sensible theory of the $\chi$ field in Lorentz
signature. }
This is consistent with the claim that  when $W$ has Euclidean signature,  $\t\chi$ and $\chi$ transform as complex conjugates.

We define $\bar\chi$ as
\beq
\bar\chi_a = \sigma^0_{ab}(\chi^b)^*. \label{eq:reality2-1}
\eeq
Then the reality condition is
\beq
\t\chi = - \i \bar \chi. \label{eq:reality2-2}
\eeq
We impose this condition in canonical quantization of the system when $W$ has Euclidean signature and $t = - \i \tau$ is the time coordinate.
In particular, the product $\bar\chi \chi = \bar\chi_a \chi^a = (\chi^b)^* \sigma^0_{ab} \chi^b$ is positive-definite and invariant under all symmetries on $W$.
The matrix $\sigma^0$ plays the role of a hermitian metric on $S$, invariant under all symmetries.

Using $\t\chi = - \i \bar{\chi}$,
the action after the Wick rotation $\tau \to \i t$ is 
\beq
I =  \int  \d^dx \d t\sqrt g \,   \left[ \i \bar \chi \partial_t \chi - m \bar \chi \chi +  \frac{\i}{2}  \chi\sigma^\mu D_\mu \chi  + \frac{\i}{2} \bar \chi \,\bar\sigma^\mu D_\mu \bar \chi  \right],\label{eq:action3}
\eeq
where $\bar\chi \partial_t \chi = \bar \chi_a \partial_t \chi^a,~ \chi\sigma^\mu D_\mu \chi =\chi^a \sigma^\mu_{ab} D_\mu\chi^b$ and so on.
The Hamiltonian is
\beq
H =  \int  \d^dx \sqrt g \,   \left[  m\bar \chi \chi + \frac{\i}{2}  \chi \CD_W^+ \chi  + ({\rm h.c.})  \right]  
\eeq
where $\CD_W^+ = - \sigma^\mu D_\mu$ and $({\rm h.c.} )$ is the hermitian conjugate of $ \frac{\i}{2}  \chi \CD_W^+ \chi  $.

We work in the subspace in which $(\CD_W^+)^\dagger (\CD_W^+) \chi = \lambda_a^2 \chi$. 
We can assume that $\lambda_a>0$, since until section \ref{sec:zeromode}, we assume  there are no zero-modes.
The  existence of the action $\int \chi\D_W^+\chi$, consistent with fermi statistics, means that the
  differential operator $\CD_W^+ $ can be regarded as an antisymmetric matrix.
Thus we can put it in a block diagonal form with $2 \times 2$ blocks of the form $ \lambda_a \epsilon_{ij} $,
where $\epsilon_{ij}$ is the totally antisymmetric $2 \times 2$ matrix with $\epsilon_{12} = 1$.
We can normalize the corresponding orthonormal eigenmodes $\psi_{1}$ and $\psi_{2}$ such that
\beq
 \int  \d^dx \sqrt{g}\, {\bar \psi}^{\, i} \psi_j = \delta^i_j, \quad
 \int  \d^dx \sqrt{g}\,   \psi_{  i} \CD_W^+ \psi_{ j} = \i \lambda_a \epsilon_{ij}, \label{eq:pf}
\eeq
where ${\bar \psi}^{\, i}_a = \sigma^0_{ab} (\psi_i^b)^*$. The phase of $ \int  \d^dx \sqrt{g}\,   \psi_{  1} \CD_W \psi_{ 2}$ can be freely chosen by rotating 
the phases of $\psi_1$ and $\psi_2$, and we have chosen it to be $\i$ to slightly simplify the later equations.

Using these modes, we expand $\chi$ and $\bar \chi$ as
\beq
\chi = A_+ \psi_1 + A_-^\dagger \psi_2, \qquad \bar \chi =  A_+^\dagger {\bar \psi}^{ 1} + A_- {\bar \psi}^{ 2}.
\eeq
The reason for the notation $A_+$ and $A_-^\dagger$ for the coefficients of the expansion of $\chi$ is to make the Hamiltonian below to be of the same form as in the case of a Dirac fermion studied in 
section~\ref{sec3}.
By canonical quantization, the above action gives the anticommutation relations $\{ A_+, A_+^\dagger \} = 1$, $\{ A_-, A_-^\dagger\} =1$, with others  zero.
The Hamiltonian is 
 \beq
 H = m (A_+^\dagger A_+ -   A_-^\dagger A_-) + \lambda_a (  A_-^\dagger A_+  + A_+^\dagger A_-  )
 = 
 (A_{+}^\dagger, A_{-}^\dagger) 
\left( \begin{array}{cc} 
m & \lambda_a \\
\lambda_a & -m 
\end{array}
\right)
\left( \begin{array}{c} 
A_{+} \\
A_{-}
\end{array}
\right).
 \eeq
This is the same as we had for a Dirac fermion.

The boundary conditions have basis-independent characterizations as follows.
We denote the Hamiltonian with the parameters $m, \lambda_a$ as $H(m,\lambda_a)$.
The ground state $\ket{\Omega}$ is the lowest energy state of the Hamiltonian $H(m,\lambda_a)$.
The APS boundary condition $\ket{\APS}$ is the lowest energy state of the Hamiltonian $H(0, \lambda_a)$ (i.e. the Hamiltonian of the massless theory $m=0$).
The state $\ket{\SL}$ defined by the local boundary condition $\SL$
is the lowest energy state of the Hamiltonian $H( |m|, 0)$ (i.e. the Hamiltonian of the positive mass theory in the limit $\lambda_a/|m| \to 0$).

 Now the problem reduces to the case of the Dirac fermion studied in the previous section. The Hamiltonian as well as the boundary conditions are completely the same.
 The only change is that the infinite product over $a$ now gives a Pfaffian:
 \beq\label{helpful}
\frac{ \bra{\SL} \et{\Omega} \bra{\Omega} \et{\APS} }{ | \bra{\APS} \et{\Omega} |^2}  = \prod_a \left( \frac{\lambda_a}{2|m|} \right)_{\rm reg}  = | {\rm Pf}( \CD_W^{+} ) |.
 \eeq
Indeed, eqn.  (\ref{eq:pf}) shows that $|\Pf(\D_W^+)| $ is a regularized version of $\prod_a\lambda_a$.

Together with the fact that $\langle\APS|Y\rangle$ is the path integral of the massive fermion field $\Psi$ on $Y$ with APS boundary conditions, and so
is equal to $\exp(-\i\pi\eta_Y/2)$, eqn. (\ref{helpful}) is what we need to justify eqn. (\ref{torro}).

\subsection{Treatment of Zero-Modes}\label{sec:zeromode}
So far we have assumed that the boundary Dirac operator $\D_W$ has no zero-modes.     We have justified the anomaly inflow formula (\ref{torro})
under this assumption.   As long as there are no zero-modes for generic background fields $g,A$, this formula gives a good characterization of
anomaly inflow.  (The formula  remains
 valid if the background fields are varied so that zero-modes appear, since the left  and right hand sides of eqn. (\ref{torro}) both
vanish in that case.)

We need a new derivation for the case that for arbitrary background fields, $\D_W$  has zero-modes.    This might happen
because of a nonzero index or mod 2 index that implies the existence of zero-modes.\footnote{The mod
2 index $\zeta_W$ of the Dirac operator $\D_W^+$  is the number of zero-modes of $\chi$ mod 2.   As we will explain in section \ref{d1}, fermi statistics imply that it is a topological invariant.
 When $\zeta_W\not=0$, $\D_W^+$ must have a zero-mode.   (By complex conjugation, $\D_W^-$ has the same number of zero-modes as $\D_W^+$.)  An ordinary Dirac
index can likewise imply the existence of zero-modes.   For example, in $d=4$, suppose that $\chi$ is a Majorana fermion coupled
to gravity only.    Then the number of positive chirality zero-modes of $\chi$ minus the number of negative-chirality zero-modes of $\chi$  is the Dirac index $\mathcal I$; when
it is nonzero, again $\D_W^+$ must have zero-modes. }
    Actually, in the anomaly inflow problem, we consider a $d$-manifold $W$
that by definition is the boundary of some $Y$.    The index and the mod 2 index of the Dirac operator on $W$  are cobordism invariants, so they vanish.\footnote{\label{cob}Cobordism
invariance of the mod 2 index $\zeta_W$ of $\D_W^+$  is an easy consequence of the anomaly inflow construction. Because the equation $(\slashed{D}_Y+m)\Psi =0$ on $Y$ with the local boundary condition $\SL$ that leads to anomaly inflow
can be derived from an action consistent with fermi statistics, the number $\zeta_Y$ of zero-modes of this
equation mod 2 is a deformation invariant and in particular independent of $m$ (see section \ref{d1} for this
argument).   Taking $m\gg 0$, the equation $(\slashed{D}_Y+m) \Psi=0$ has no approximate solutions on $Y$
that satisfy the boundary condition, so $\zeta_Y=0$.   Taking $m\ll 0$, the approximate solutions of the equation are the same as the zero-modes of $\chi$ on $W$,
so $\zeta_W=\zeta_Y$ and hence $\zeta_W=0$.   Note that zero-modes of $\D_W^+$ may not correspond to {\it exact} zero-modes of $\slashed{D}_Y+m$, but small corrections
to the spectrum do not affect the number of zero-modes of an antisymmetric matrix mod 2, and so do not affect the statement that $\zeta_W=\zeta_Y$.  For the ordinary index $\I$, the Atiyah-Singer
theorem gives a formula $\mathcal I=\int_W \Phi$ for some characteristic class $\Phi$.   If $W=\partial Y$ and the structures needed to define $\Phi$ extend over $Y$ then
$\mathcal I=\int_W\Phi=\int_Y \d \Phi=0$, so $\mathcal I$ is likewise invariant under cobordism.   This statement can also be proved without knowing the
index formula by adapting the proof we explained  for the mod 2 index.}
 But it may
happen that $W$ is the union of disconnected components $W_i$.    In that case, although the overall Dirac index and mod 2 index on $W$ will vanish,
on individual components there may be a nonvanishing index or mod 2 index.   That is the situation in which generically (and in fact always) $\D_W$  has zero-modes.  Though this
situation may at first sight seem rather esoteric, it actually plays a role   in understanding relatively simple examples, as we will see in section \ref{examples}.
Therefore, it is important to generalize the anomaly inflow formula (\ref{torro}) to cover this situation.

In doing  so, we will use the fact that there is always  a generic number of zero-modes, the minimum number allowed by any index or mod 2 index theorem.
Let $b_i$ be the generic number of zero-modes of the chiral Dirac operator $\D_W^+$ on  $W_i$.   Then summing over all components, the generic number of zero-modes of the chiral
Dirac operator $\CD_W^+$  on $W$ is
$b=\sum_i b_i$. Since (when $W$ has Euclidean signature), complex conjugation exchanges $\chi$ and $\t\chi$, the number of zero-modes of the opposite
chirality Dirac operator $\CD_W^-$ is also $b$.  In the anomaly inflow problem, the
 number $b$ is always even, because the mod 2 index is a cobordism invariant and $W$ is assumed to be the boundary of some $Y$.   So we set $b=2\nu$.

Generalizing eqn. (\ref{torro}), we will describe a formula that characterizes anomaly inflow as long as  $\CD_W^+$ has the generic number $2\nu$ of zero-modes.
When the background fields are varied so that additional zero-modes appear, the formula will remain valid but will become a trivial identity $0=0$.   

We denote orthonormal bases of the spaces of
 zero-modes of $\CD_W^+$ as $\psi_i$, where $i = 1, \ldots, 2\nu$.
 By definition, the $\psi_i$ are modes of the positive chirality field $\chi$. The zero modes for the negative chirality field $\t\chi$ are $\bar \psi^{\, i} = \sigma^0 (\psi_i)^*$.  
 We impose the orthonormality condition 
 \beq
 \int \d^d x\sqrt{g}\, \bar\psi^{\, i} \psi_j  = \delta^i_j. \label{eq:orthonormal}
\eeq

We neglect nonzero modes because their treatment is completely the same as before.
We expand $\Psi=\begin{pmatrix}\chi\cr\t\chi\end{pmatrix}$ as  
\beq
\chi =\sum_{i=1}^{2\nu}  A^{i}\psi_i , \qquad \t\chi = -\i \sum_{i=1}^{2\nu}  \bar A_{i} \bar\psi^{\, i} .
\eeq
When the theory is formulated on the Euclidean signature manifold $W$ or on $W\times \R$ (where $\R$ parametrizes the ``time'' in the Hamiltonian framework), 
the reality condition on $\Psi$ is $\t\chi = - \i \bar\chi$ and hence $\bar A_i$ is the complex conjugate of $A^i$.
Upon quantization, they become hermitian conjugate operators, 
the nonzero canonical anticommutators  being \be\label{canonical} \{A^i, \bar A_j \} =  \delta^i_j .\ee The Hamiltonian is
\begin{align}
H = \ \int \d^d x \sqrt{g}\, m\bar\chi \chi   
 = m \sum_{i=1}^{2\nu} \bar A_i A^i.
\end{align}
If the mass parameter is positive $m > 0$, the ground state is specified by $A^i \ket{\Omega} = 0$, while if it is negative $m<0$, 
the condition is $\bar A_i \ket{\Omega} = 0$. The local boundary condition $\SL$ is $\bra{\SL} \bar A_i =0$, which is the same as the 
one for the ground state of the positive mass theory.

When we formulate the theory on a general Euclidean signature manifold $Y$, $\Psi$ is no longer real\footnote{In fact, the reality condition \eqref{eq:reality} cannot be extended into the bulk of Euclidean $Y$,
because this condition involves $(\Psi_1,\Psi_2)_t = \langle \gamma^t \Psi_1, \Psi_2 \rangle$ which is not invariant under $D$-dimensional Lorentz transformations in $Y$.
It is invariant only under $d$-dimensional Lorentz transformations in $W$.}  and we consider the Dirac operator $\D_Y$ acting
on a complex-valued field $\Psi$.  
If $Y$ has  boundary $W$, it would be meaningful to say that $\Psi$ is real along the boundary, 
but neither the local boundary condition $\SL$ nor the APS boundary condition, which we describe shortly, imposes such a constraint.   
We have already noted that  $\SL$  constrains $\bar A_i$ but not $A^i$ along the boundary, 
and similarly the APS boundary condition 
will put a constraint on  $\bar A_i$ and $A^i$ that is not consistent with $\Psi$ being real along the boundary.  
Quantum mechanically, the constraint will mean that certain non-hermitian linear combinations
of $A^i$ and $\bar A_i$ annihilate the state $|\APS\ra$ (similarly to the fact that the non-hermitian operators $\bar A_i$ annihilate the state $\la\SL|$).   Note
that because of the canonical anticommutation relations (\ref{canonical}), there is no state that is annihilated by a hermitian linear combination of the $A^i$ and $\bar A_j$,
since the square of such an operator is strictly positive.   So no reasonable boundary condition can make $\Psi$ real along the boundary at the classical level.  

Let us proceed to the discussion of the APS boundary condition.
When zero-modes are present, the APS boundary condition is subtle.    Remember that in the absence of zero-modes, the state $\ket{\APS}$ associated with the APS
boundary condition is the ground state of the $\Psi$ field for $m=0$.    When the operator $\D_W$ has zero-modes, the $m=0$ theory does not have a unique ground state
  (even up to an overall scalar multiple) but a nontrivial space of ground states, obtained by quantizing the zero-modes.    Hence, to define an APS boundary
condition, it is necessary to pick a particular state (or more precisely, a particular one-dimensional subspace) 
 in the space of ground states.   A completely arbitrary choice will not do, as we want the APS boundary condition
to be such that the path  integral of the $\Psi$ field on $Y$ with APS boundary conditions will generate the usual phase $\exp(-\i\pi\eta_Y/2)$.    In order for this
to happen,  the equation $(\D_Y+\i m)\Psi=0$ should be the equation of motion derived from the fermion action, and in addition $\D_Y$ must be self-adjoint.   These  were inputs
to the derivation in section \ref{aps} (the first condition is needed so that the fermion path integral is the Pfaffian of $\D_Y+\i m$, and the second makes the eigenvalues
of $\D_Y$ real, as assumed in the derivation).   
Each of these requirements is nontrivial.

First, we ask whether the equation of motion derived from the usual Dirac action $\int_Y \d^Dx \sqrt g \langle \Psi, ( \slashed{D}+m)\Psi\rangle$ is the Dirac
equation $(\CD_Y+\i m)\Psi=0$, or whether the equation of motion contains additional delta function terms supported on the boundary.
To avoid such terms, we need
\beq\label{desired}
\langle \Psi_1, \CD_Y \Psi_2  \rangle_Y = \langle \CD_Y \Psi_1,  \Psi_2 \rangle_Y 
\eeq
 for $c$-number (commuting) fermion fields $\Psi_1,\Psi_2$,
where
\beq
\langle \Psi_1, \Psi_2 \rangle_Y = \int \d^D x \sqrt{g}\, \langle \Psi_1, \Psi_2 \rangle.
\eeq 
To prove eqn. (\ref{desired}), we have to integrate by parts, and we encounter a surface term proportional to 
\beq
 -\int  \d^d x \sqrt{g}\, \langle \Psi_1 ,  \gamma^t \Psi_2 \rangle =   \int  \d^d x \sqrt{g}\, ( \Psi_1 ,  \Psi_2 )_t,
\eeq  
where $(~,~)_t$ was introduced in eqn.~\eqref{eq:inner}.
We want this expression to vanish whenever $\Psi_1$ and $\Psi_2$ satisfy the APS boundary condition.    
Expanding 
\beq
\Psi_1=\begin{pmatrix}\chi_1\cr\t\chi_1\end{pmatrix}=\sum_{i=1}^{2\nu} \begin{pmatrix} A^{i}\psi_i \cr -\i \bar A_i \bar \psi^{\, i} \end{pmatrix}, 
\qquad \Psi_2=\begin{pmatrix}\chi_2\cr\t\chi_2\end{pmatrix}=\sum_{i=1}^{2\nu} \begin{pmatrix} B^{i}\psi_i \cr -\i \bar B_i \bar \psi^{\, i}  \end{pmatrix},
\eeq
we get
\beq
\int  \d^d x \sqrt{g}\, ( \Psi_1 ,  \Psi_2 )_t = \sum_{i=1}^{2\nu} (A^i  \bar B_{i} + \bar A_i B^i ), \label{surface1}
\eeq
where we have used the orthonormality condition \eqref{eq:orthonormal}.
As noted earlier, along $W$, $\Psi$ is not constrained to be real, so in this formula, $A^i$ and $\bar A_j$ (and similarly $B^i$ and $\bar B_j$) should
be understood as independent complex variables.

Now let us look at the condition for self-adjointness of $\D_Y$.  When $Y$ has no boundary, $\D_Y$ is hermitian with respect to the hermitian form $(~,~)_Y$
that was introduced in eqn. (\ref{hermform}):
 $(\Psi_1, \CD_Y \Psi_2)_Y = (\CD_Y \Psi_1, \Psi_2)_Y$. When $Y$ has a boundary, integration by parts generates a surface term 
  $\int  \d^d x \sqrt{g}\,  \langle \sC \Psi_1, \gamma^\tau \Psi_2 \rangle$.
 From the definition of $\sC$ given in eqn.~\eqref{antiunitary} along with $\sigma^0 (\psi_i)^* = \bar \psi^{\, i} , ~ \bar\sigma^0 (\bar\psi^{\, i})^* = - \psi_i$, we get
 \beq
 \Psi_1=\sum_{i=1}^{2\nu} \begin{pmatrix} A^{i}\psi_i \cr -\i \bar A_i \bar \psi^{\, i} \end{pmatrix} \quad \Longrightarrow \quad \sC \Psi_1 
 =\sum_{i=1}^{2\nu}   \begin{pmatrix} -\i  (\bar A_{i})^*\psi_i \cr  ( A^i)^* \bar \psi^{\, i} \end{pmatrix}.
 \eeq
Therefore, the surface term is given by
\beq
\int  \d^d x \sqrt{g}\,  \langle \sC \Psi_1, \gamma^\tau \Psi_2 \rangle = \sum_{i=1}^{2\nu} \left( (A^i)^*   B^{i} - (\bar A_i)^* \bar{B}_i \right). \label{surface2}
\eeq

For the  surface terms (\ref{surface1}) and (\ref{surface2})  to vanish, we impose a boundary condition of the following sort. 
Let $J_{ij} = - J_{ji}$ be a matrix that is antisymmetric and also unitary. Then we impose
\beq\label{constraints}
\bar A_i = \sum_j J_{ij}A^j.
\eeq
(with the same condition on $B^i$, $\bar B_i$, since  $\Psi_1$ and $\Psi_2$ take values in the same space).
The antisymmetry of the matrix $J$ guarantees that the term \eqref{surface1} vanishes and hence the Pfaffian of $\CD_Y$ is well-defined.
The unitarity of $J$ guarantees that the term \eqref{surface2} vanishes and hence the eigenvalues are real.
For the existence of an antisymmetric unitary $J$, it is crucial that the total mod 2 index is zero and hence the index $i$ takes the values $i = 1,\ldots, 2\nu$.
If the number of zero modes were odd, an antisymmetric matrix necessarily would have a zero eigenvalue and could not be unitary.

There is no unique way to choose the matrix $J_{ij}$.   That leads to the non-uniqueness of APS boundary conditions in the presence of zero modes.
We allow any choice of $J$.   But we note that it is always possible to choose a basis of zero modes $\psi_i$ to put $J$ in the standard form
\beq
J_{2i-1, 2i} = -J_{2i, 2i-1} = -1, \qquad \text{other }J_{ij}=0.\label{standardJ}
\eeq

Quantum mechanically, $A^i$ and $\bar A_j$ become operators that satisfy the canonical anticommutation relations (\ref{canonical}).
The condition (\ref{constraints}) means that the state $|\APS\rangle$ satisfies
\beq
\bra{\APS}(\bar A_i - \sum_j J_{ij} A^j ) =0.
\eeq
The operators $\bar A_i - \sum_j J_{ij} A^j $ are a maximal set of anticommuting operators constructed from the zero-modes, so these constraints
are consistent and uniquely determine 
 a linear combination of the ground states (up to a scalar multiple).   On the other hand, not every linear combination
of the ground states satisfies a condition of this form.   The above detailed analysis has singled out a preferred class of ground states.    There is, however,
no way to avoid making a choice of the  matrix $J$.     For nonzero modes, there was no need to make such a choice; one simply says that the state $|\APS\rangle$
is annihilated by the negative energy modes of $\Psi$.

Let us go back to the Hamiltonian framework on $W$. 
Let $\ket{E}$ be a state vector with $A_i \ket{E}=0$, $i=1,\cdots,2\nu$.
For notational simplicity we focus on a block of four operators $(A^{2k-1}, A^{2k}, \bar A_{2k-1}, \bar A_{2k})$ (with some fixed value of $k$) and omit $\prod_k$ in the following equations.
Then we have
\beq
\bra{\SL} =  \bra{E}, \qquad \bra{\APS} = \bra{E}(-\bar A_{2k} + A^{2k-1} )(\bar A_{2k-1} + A^{2k} ),
\eeq
and
\beq
\ket{\Omega}=\left\{
\begin{array}{ll}
  \ket{E}, & m>0 \\
\bar A_{2k} \bar A_{2k-1}  \ket{E} & m<0
  \end{array}
  \right.
\eeq
From this, we get
\beq\label{usefulone}
\bra{\APS}\et{\Omega}=1
\eeq
and
\beq
\bra{\SL}\et{\Omega}=\left\{
\begin{array}{ll}
1 & m>0 \\
0 & m<0
  \end{array}
  \right.
\eeq
This result could have been anticipated. For $m>0$, there are no localized chiral fermions on the boundary and hence nothing happens.
On the other hand, when $m < 0$, we have a localized chiral fermion whose partition function ${\rm Pf}(\CD_W^+)$ vanishes because of the zero-modes.

To get something nonzero for $m<0$, we have to insert an operator.   We choose some operator $O$ that is a function of the field $\chi$ on $W=\partial Y$.
Then instead of a vacuum path integral, we compute a path integral with an insertion of $O$.   We write $\langle O\rangle_{Y,\SL}$ for an unnormalized path integral on $Y$
with this insertion, and with boundary condition $\SL$.     Before going on, let us note that $\langle O\rangle_{Y,\SL}$ is completely well-defined and anomaly-free.
It is an observable in the $D$-dimensional theory of the massive field $\Psi$; because this field is massive, it admits Pauli-Villars regularization and is free of any anomaly.

We can get  an illuminating formula for $\langle O\rangle_{Y,\SL}$ as follows.
 Reasoning as in the derivation of eqn. (\ref{factored}), in the long distance limit we have $\langle  O\rangle_{Y,\SL} =\langle \SL  |O|\Omega\rangle\langle\Omega|Y\rangle.$
 Choosing an APS boundary condition and repeating the derivation that led to eqn. (\ref{eq:rewrite}) gives
\beq
\langle O\rangle_{Y,\SL} =\frac{ \langle\SL|O| \Omega\rangle  \bra{\Omega} \et{\APS} }{ | \bra{\APS} \et{\Omega} |^2} \cdot  \bra{\APS} \et{ Y} . \label{more}
\eeq
In other words, the only change from the previous derivation is that $\langle \SL|\Omega\rangle$ is replaced by $\langle \SL|O|\Omega\rangle$.

It does not matter very much which $O$ we pick, as long as $\langle\SL|O|\Omega\rangle $ is generically 
nonzero.   A minimal choice is to take $O=\chi(x_1)\chi(x_2) \cdots \chi(x_{2\nu})$ for
points $x_1,x_2,\ldots, x_{2\nu}\in W$.   The zero-mode part of $\chi$ is
$\chi(x)  = \sum_i A^i \psi_i(x)$.
The matrix element $\bra{\SL} \chi(x_1) \ldots \chi(x_{2\nu}) \ket{\Omega} $ is a product of a zero-mode factor and a factor coming from nonzero  modes.
The factor that comes from nonzero modes can be analyzed precisely as in sections \ref{sec3} and \ref{sec4} and is $|\Pf'(\D_W^+)|$, where $\Pf'$ is the Pfaffian
in the space orthogonal to the zero modes.   The zero-mode factor is 
\beq\label{zm}
\bra{\SL} \chi(x_1) \ldots \chi(x_{2\nu}) \ket{\Omega}_0=  \sum_{\sigma } \sign (\sigma) \prod_{i=1}^{2\nu}  \psi_i (x_{\sigma (i)}) 
\eeq
where the sum is over all permutations $\sigma$ of $2\nu$ numbers, and we have restored $\prod_i$ in the notation.  This follows from the formulas
given earlier for the states $|\SL\rangle$ and $|\Omega\rangle$.
Combining the two factors,
\be\label{combined}
\bra{\SL} \chi(x_1) \ldots \chi(x_{2\nu}) \ket{\Omega}=  |\Pf'(\D_W^+)| \sum_{\sigma } \sign (\sigma) \prod_{i=1}^{2\nu}  \psi_i (x_{\sigma (i)}) \, .
\ee

The bulk contribution is exactly the same as before, namely
\beq\label{general}
\bra{\APS} \et{Y} = \exp\left( -{\i \pi} \eta_Y /2\right),
\eeq   where $\eta_Y$ is computed using the chosen APS boundary conditions.
In fact, we have defined this boundary condition so that $\D_Y$ has all the properties that were needed in section \ref{aps} for the  derivation of eqn. (\ref{general}).  

Combining these results, we get
\begin{align}
\langle \chi(x_1) \ldots \chi(x_{2\nu}) \rangle_{Y,\SL} &= \frac{ \bra{\SL}  \chi(x_1) \ldots \chi(x_{2\nu}) \ket{\Omega} \bra{\Omega} \et{\APS} }{ | \bra{\APS} \et{\Omega} |^2} \bra{\APS} \et{Y} \nonumber \\
&= | {\rm Pf}'( \CD_W^{+} ) | \exp\left(  -{\i \pi} \eta_Y /2 \right)  \sum_{\sigma } \sign (\sigma) \prod_{i=1}^{2\nu}   \psi_i (x_{\sigma (i)}) .
\end{align}

This is the result with a minimal operator insertion that gives a nonzero result.
We could also  consider some other choice of $O$.   
The only consequence is to modify the factor $\langle \SL|O|\Omega\rangle$.   The anomaly inflow factor $\exp(-\i\pi\eta_Y/2)$  is unchanged.   
This is important because -- in the context of the derivation explained in section \ref{anomaly} -- it will  imply that the anomaly of the path integral
of the original $\chi$ field on $W$   does not depend on what operator insertion is made. 
This is equally true whether $\nu=0$ -- the case assumed in sections \ref{sec3} and \ref{sec4} --  or $\nu>0$, as analyzed here.  The only difference is  that for $\nu=0$,
the minimal choice of $O$ is $O=1$, so in the previous analysis, we did not introduce $O$ explicitly.

We conclude with one last remark.
One might worry that the above result depends on the choice of a basis of zero modes $\psi_i$,
because if we transform the basis as $\psi_i \to \sum_j \psi_j U^j_i$ with  a unitary matrix $U^i_j$, the factor $ \sum_{\sigma } \sign (\sigma) \prod_{i=1}^{2\nu}  \psi_i (x_{\sigma (i)}) $ is multiplied  by 
$\det U$. However, if we change the basis in this way, we are also changing the APS boundary condition, since we have fixed $J$ to the standard form \eqref{standardJ}.
The Dai-Freed theorem~\cite{Dai:1994kq} states that if we change the APS boundary condition in this way, the exponentiated $\eta$-invariant $\exp (  -{\i \pi} \eta_Y /2  ) $ changes by $\det U^{-1}$
so that the above correlation function is independent of $U$. In this way the final result is independent of any choice of basis vectors or APS boundary condition. Alternatively, since in this derivation we knew at the beginning that  $\langle O\rangle_{Y,\SL}$ is well-defined, and the APS boundary condition was only
introduced as a way to calculate it, one may see the above result as a physical proof of this part of the Dai-Freed theorem~\cite{Yonekura:2016wuc}.

Of course, the result for  $\langle O\rangle_{Y,\SL}$ will in general depend on the choice of $Y$  and of its spin structure.   Here is a special case -- an important one, though
unfortunately somewhat technical to describe.
    In general, if $W$ has $s$ connected components $W_1,W_2,\cdots, W_s$,  and we are given a spin structure 
    on $Y$ whose restriction to the boundary is isomorphic to some given spin structures on $W_1,W_2,\cdots, W_s$, then there are $2^s$ such
    isomorphisms,\footnote{We assume that a specific isomorphism between the gauge bundle of $\partial Y$ and that of $W$ has been fixed.   In general, if the gauge
    group has a nontrivial center, then $\langle O\rangle_{Y,\SL}$ may depend on the choice of this isomorphism.}  as a given isomorphism on any given component can be multiplied by $(-1)^\sF$.   According to  the definition of footnote \ref{spinisom} in section
    \ref{sec4}, the choice
    of such an isomorphism is part of the definition of a spin structure on $Y$.   However, an overall $(-1)^\sF$ gauge transformation on $Y$, changing the sign of all fermions,
    would ``flip'' the isomorphism on each of the $W_i$.    So spin structures on $Y$ come in groups of $2^{s-1}$ that differ only by the chosen isomorphisms
    with the given spin structures on $W_1,W_2,\cdots, W_s$.   Suppose that we ``flip'' the isomorphism on one of the boundary components, say $W_1$.    This is gauge-equivalent
    to changing the APS boundary condition on $Y$ without changing the bases of zero-modes on $W$, 
    so it multiplies $\exp(-\i\pi \eta_Y/2)$ by $\det\, U^{-1}$, with no compensating factor
    of $\det\,U$. Here $U$ is the matrix that acts as $-1$ on zero-modes supported on $W_1$, and $+1$ on zero-modes supported on other components.
      As a result, $\langle O\rangle_{Y,\SL}$ is multiplied by $(-1)^{\zeta_1}$, where $\zeta_1$ is mod 2 index on $W_1$ (the number of $\chi$ zero-modes
    on $W_1$, mod 2).   If $(-1)^{\zeta_1}$ can be nontrivial, then the product of an odd number of $\chi$ fields on $W_1$ can have an expectation value in the original
    $d$-dimensional theory on $W_1$.   Since $\chi$ is odd under $(-1)^\sF$, this represents an anomaly in $(-1)^\sF$ symmetry.    This particular anomaly
    is reproduced in the $D$-dimensional formalism by the spin structure dependence that was just described.

 \section{The Anomaly}\label{anomaly}
 
 In section \ref{precise}, we have obtained a general formula describing anomaly inflow for an arbitrary fermion field $\chi$ on a manifold $W$.   Implicit in this formula,
 as we will now explain,
 is a description of the anomaly of $\chi$.   The anomaly inflow involved an $\eta$-invariant on a manifold $Y$ with boundary $W$,
 and this will enable us to express the anomaly in $d$ dimensions in terms of an $\eta$-invariant  in dimension $D=d+1$.

 The fact that an anomaly in $d$ dimensions is naturally related to some quantity in $d+1$ dimensions is familiar for perturbative anomalies.   The most familiar version of the statement is
 that the perturbative anomaly in $d$ dimensions is related to a Chern-Simons function in $d+1$ dimensions \cite{Jackiw, Zumino, Stora}.  Our point here is to explain that nonperturbative or global fermion
 anomalies can be incorporated in this statement by just replacing the Chern-Simons function with $\eta$.   
 
 The relation between the global fermion anomaly in $d$ dimensions and an $\eta$-invariant in $d+1$ dimensions was originally found in \cite{Witten:1985xe}.  The original derivation involved
 computing what one might call the holonomy of a Berry connection on the determinant (or Pfaffian) line bundle and expressing this in terms of $\eta$.   Our derivation here, inspired by the Dai-Freed theorem 
 \cite{Dai:1994kq,Yonekura:2016wuc} (which itself was partly inspired by the computation in \cite{Witten:1985xe}), is more general as it applies in any dimension, even or odd, and on any manifold, orientable or not, on which a fermion system can be defined.  
 (The Dai-Freed theorem has  been extended to this more general context in  the appendix to \cite{Freed:2019sco}.)
  Also, our derivation
 gives a more precise answer than was sought in the original work.  When there is no anomaly, we determine the actual phase of the fermion path integral on $W$.
 In early work on anomalies, one aimed to show that the overall phase of the path integral was not affected by any inconsistency or anomaly, but
 one did not aim to get a formula for that phase.   To determine the phase is much more precise than just showing that the phase can be defined.
   For  the use of the Dai-Freed theorem and related
 ideas  to define the absolute phase of a path integral in  particular cases see \cite{WittenDBrane} section 2.2, \cite{DMW} section 2, and \cite{FM}.

 \subsection{Defining the Phase of the Path Integral}\label{defphase}
 
 Let us start with a fermion field $\chi$ on a $d$-manifold $W$, and try to define the corresponding partition function. 
 If $\chi$ can have a bare mass, the partition function can always be defined
  using Pauli-Villars regularization.  Even if a bare mass is not possible, Pauli-Villars regularization can always be used to define the absolute value of the fermion
 partition function.  But in general, if a bare mass is not possible, there is no simple direct way to define the phase of the fermion partition function, and there may be an anomaly that makes it impossible to get
 a satisfactory definition of this phase.   
 
 However, from section \ref{precise}, we know how to define the partition function of a modified system
  if $W$ is the boundary of some manifold $Y$ over which all the  structures (such as a spin or pin structure and possibly a gauge bundle)
 needed to define the original fermion field $\chi$ on $W$ have been extended.    We regard $\chi$ as a boundary mode of a massive fermion field $\Psi$ on $Y$, whose partition function is 
 \beq\label{candidate}
  Z(Y, \SL) =  | {\rm Pf}( \CD_W^{+} ) | \exp\left(  -{\i \pi} \eta_Y /2 \right),
 \eeq
 where we put the subscript $Y$ on $\eta$ to indicate that it is computed on the manifold $Y$.  This is the partition function of a combined system consisting of a massless fermion on $W$ with anomaly inflow
 from a massive fermion in bulk.   
 
 If the expression (\ref{candidate})  actually does not depend on the choice of $Y$, we can regard it as a definition of the path integral for the $\chi$ field on $W$.   So let us investigate the dependence on $Y$.
 (If a suitable $Y$ does not exist at all, then it is necessary to generalize the procedure.  We postpone this issue for the moment.)  
 To investigate the dependence on $Y$, 
let $Y'$ be another manifold with the same boundary $\partial Y' = W$.
 The ratio between the partition functions $Z(Y, \SL)$ and $Z(Y' , \SL)$ is given by $\exp( - \i\pi (\eta_Y - \eta_{Y'} )/2 )$.
 Let $\bY$ be the closed manifold\footnote{A closed manifold is a compact manifold without boundary. In what follows, $\bY$ is always a closed $D$-manifold,
 and $Y$ is a $D$-manifold that might have a boundary.}
  which is constructed by gluing $Y$ and the orientation reversal of $Y'$ along their common boundary $W$.
 The gluing theorem for $\eta$~\cite{Dai:1994kq} says that we have 
 \beq
 \exp \left( - \i\pi (\eta_Y - \eta_{Y'} )/2 \right) = \exp\left( - \i\pi\eta_{\bY}/2 \right) . \label{eq:glue}
 \eeq
 The physical interpretation of this gluing theorem is as follows. The partition function of the massive fermion on the closed manifold $\bY$ is represented in the path integral formulation as
 \beq
 Z(\bY) = \bra{Y'}\et{Y},
 \eeq
 where $\ket{Y}$, $\ket{Y'}$ are physical states on the Hilbert space $\CH_W$ as introduced in section~\ref{sec2}.
 As explained in that section, these states are proportional to the ground state $\ket{Y} \propto \ket{\Omega}$, $\ket{Y'} \propto \ket{\Omega}$, and hence
 we can write
 \beq
 \bra{Y'}\et{Y} = \frac{ \bra{Y'}\et{\APS} \bra{\APS}\et{Y} }{|\bra{\APS} \et{\Omega} |^2  }.
 \eeq
 By computing  as in the  section \ref{aps}, we get the gluing formula \eqref{eq:glue}.  The universal phase in the numerator on the right hand side
 is $ \exp \left( - \i\pi  (\eta_Y - \eta_{Y'} )/2 \right)$ (note that reversing the orientation
 of $Y'$ reverses the sign of $\eta_{Y'}$) and the denominator is  positive.   The universal phase on the left hand side is $\exp(-\i\pi\eta_{\bY}/2)$.
 
 Therefore, the dependence of the partition function $Z(Y,\SL)$ on $Y$ is characterized by
 \beq
 \frac{Z(Y, \SL) }{Z(Y', \SL)} = \exp\left( -  \i\pi \eta_{\bY}/2 \right).
 \eeq
 If 
 $\Upsilon_{\bY}=\exp( - \i\pi  \eta_{\bY}/2 )$ is always equal to 1 for any closed manifold $\bY$,  
 $Z(Y,\SL)$  does not depend on the choice of $Y$.
 In that case, $Z(Y,\SL)$ can serve as a definition of the path integral $Z(W)$ for the original fermion system on $W$, which therefore is completely  anomaly-free.
   If instead $\Upsilon_{\bY}$ is nontrivial, then $Z(Y,\SL)$ does depend on $Y$ and cannot serve as a satisfactory definition of $Z(W)$.
However, we should ask if there is some other definition
 that we should use instead of $Z(Y,\SL)$.   This will be discussed in section \ref{converse}.     The conclusion will be to show that the anomaly can be eliminated
 by using a better definition if and only if $\eta_{\bY}$ can be written as a local integral over $\bY$.

 In the above derivation, we have implicitly assumed that the operator $\D_W$ has no zero-modes, so that the anomaly can be probed by studying the partition function $Z(Y,\SL)$.
When that is not the case, to get something nonzero, one has to consider a path integral $\langle O\rangle_{Y,\SL}$ with some operator insertion, as analyzed in section \ref{sec:zeromode}, and the APS
 boundary condition is no longer unique.   However, the
 above derivation is still valid in this more general situation.   The only properties of the APS boundary condition that are needed in the derivation are that
 the universal phase of $\langle \APS|Y\rangle$ is $\exp(-\i\pi\eta_Y/2)$, and that the matrix element $\langle\APS|\Omega\rangle$ is nonzero.   The APS boundary
 condition was chosen in section \ref{sec:zeromode}  to ensure the first property, and the second property also holds in general (eqn. (\ref{usefulone})).

 When $\exp\left(-\i\pi\eta_{\bY}/2\right)=1$ for any closed manifold $\bY$ that carries the appropriate structures, this tells us in a very strong sense that the original fermion system on $W$
 was anomaly-free.  Not only is there no anomaly that would prevent us from determining the phase of the path integral of the original fermion system, but we have actually determined this phase in eqn. (\ref{candidate}).
 In sections \ref{cobordism} and \ref{converse}, we will explore this result more fully.   But first we fill a gap in the explanation that we have given so far.

 In this derivation, we assumed that $W$ is the boundary of some manifold $Y$ over which all relevant structures are extended.  What happens if a suitable $Y$ does not exist?
 This actually does not mean that the original theory on $W$ cannot be defined.   It means that if it is possible to define this theory, then the definition is not unique.  
The condition under which it is possible to define the theory is the same as before: $\exp\left(-\i\pi\eta_{\bY}/2\right)$ should equal 1 
for any closed manifold $\bY$ carrying the appropriate structures (or a slight generalization of this described in section \ref{converse}).

 The nonuniqueness arises for the following reason.   The very fact that $W$
 is not the boundary of any $Y$ over which the appropriate structures extend means that there is a nontrivial  cobordism group $\Gamma$ whose elements are $d$-manifolds carrying such structures modulo those that are boundaries.
 The group operation in $\Gamma$ is disjoint union of manifolds.   Any homomorphism $\varphi$  from $\Gamma$ to $\U(1)$ gives the partition function of an ``invertible'' topological field theory 
\cite{FM,Kapustin:2014dxa,Freed:2016rqq,Yonekura:2018ufj}.  This is a purely $d$-dimensional theory whose partition function on a manifold $W$ is $\varphi(W)$.    When $\Gamma$ is nontrivial, we cannot expect to uniquely determine
the partition function of the original fermion theory on $W$ by any general arguments, because any definition that is consistent with all general principles of quantum field theory could always be modified
by multiplying it by $\varphi(W)$.    In general, different regularizations of the same theory will give results that differ by such a factor.

We can proceed as follows, as in section 2.6 of \cite{Witten:2016cio}.  Rather than being abstract, we will assume that  $\Gamma=\Z_k$ for some integer $k$.  This means that there is some manifold $W_0$
such that $W_0$ is not the boundary of any suitable $Y$, but the disjoint union of $k$ copies of $W_0$ is such a boundary.   Writing $W'$ for this disjoint union, we determine 
the partition function $Z(W')$ from the formula (\ref{candidate}).   Since $W'$ is the disjoint union of $k$ copies of $W_0$, we interpret $Z(W')$ as $Z(W_0)^k$.   Now we define $Z(W_0)$ as $(Z(W'))^{1/k}$, with some
choice of the $k^{th}$ root.  
It is in this choice of a $k^{th}$ root that the nonuniqueness enters.   Now given any $W$, since $W_0$ generates the cobordism group, there is some integer $r$ such that $W''$, defined as the disjoint union
of $W$ with $r$ copies of $W_0$, is the boundary of some $Y$.   Thus we can use eqn. (\ref{candidate}) to define $Z(W'')$.  We interpret this as $Z(W)Z(W_0)^r$, since $W''$ is a disjoint union of $W$ with $r$ 
copies of $W_0$, 
and we define $Z(W)=Z(W'')/Z(W_0)^r$.   

In this construction, changing  the $k^{th}$ root in the definition of $Z(W_0)$ will multiply $Z(W)$ for any $d$-manifold $W$ by $\varphi(W)$, for some $\varphi:\Gamma\to \U(1)$.   Since $W_0$ was assumed
to generate $\Gamma$, $\varphi$ is completely determined by $\varphi(W_0)$.

 \subsection{Topological Field Theory And Cobordism}\label{cobordism}
 
  To understand the global anomaly more deeply, we have to return to the APS index theorem \cite{Atiyah:1975jf}, which was briefly introduced in section \ref{aps}.  
 
Suppose that the $d+1$-manifold $\bY$ is the boundary of a $d+2$-manifold $X$. 
Suppose also that we are given a fermion field $\Psi$ on $\bY$ with a self-adjoint Dirac operator $\D_\bY=\i \sum_{\mu=1}^{d+1}\gamma^\mu D_\mu$,  and that all structures
(such as  spin structures and gauge bundles) needed to define $\D_\bY$ have been extended over $X$.    The APS index
theorem relates $\eta_\bY$, the $\eta$-invariant of $\D_\bY$,  to the index $\I$ of a certain Dirac operator $\D_X$ on $X$.   $\D_X$ is defined using the same doubling
procedure that we used in section \ref{sec4} to go from $W$ to $\bY$.   In brief, we  introduce a second copy $\t\Psi$ of $\Psi$ and a combined field $\h\Psi=\begin{pmatrix}\Psi\cr\t\Psi\end{pmatrix}$.
Acting on $\h\Psi$, we define $d+2$ gamma matrices rather as in  eqns. (\ref{third}) and (\ref{north}):
\begin{equation}\Gamma^\mu =\begin{pmatrix} 0 & \gamma^\mu \cr \gamma^\mu & 0\end{pmatrix}, ~~~~~~\Gamma^\tau=\begin{pmatrix} 0 & -\i \cr \i & 0 \end{pmatrix}. \ee
This enables us to define a Dirac operator $\D_X = \i (\Gamma^\mu D_\mu+\Gamma^\tau D_\tau)$ on $\bY \times \R_-$.   By Lorentz invariance of the construction, this operator can be defined on any $X$
with boundary $\bY$ such that the relevant structures on $\bY$ have been extended over $X$.  

There is one very important difference from the previous case.  In section \ref{sec4}, we started with a $d$-dimensional fermion field $\chi$ that did not necessarily have a self-adjoint
Dirac operator.  Doubling involved introducing a dual field $\t\chi$ that did not necessarily transform the same way as $\chi$ under the symmetry group.    As a result, in general the nonzero blocks
in eqn. (\ref{third}) are not equal; they consist of distinct matrices $\sigma_\mu$ and $\bar\sigma_\mu$.    But in the present discussion, we started with a $d+1$-dimensional fermion field that by hypothesis
does have a self-adjoint Dirac operator, and the additional field $\t\Psi$ that we introduced in the doubling was just a second copy of $\Psi$.   Accordingly the nonzero blocks in the definition of $\Gamma^\mu$ are equal.
This gave more freedom in the definition of $\Gamma^\tau$ than we had in the previous case, and we have taken advantage of that freedom in making a convenient choice of $\Gamma^\tau$.
The extra freedom also means that in addition to $\Gamma^\tau$, we can define a ``chirality'' operator:
\be\label{chir}\bar\Gamma =\begin{pmatrix} 1&0 \cr 0 & -1\end{pmatrix}. \ee
Because $\bar\Gamma$ anticommutes with all gamma matrices $\Gamma^\mu$ and $\Gamma^\tau$, it anticommutes with the Dirac operator $\D_X=\i(\Gamma^\mu D_\mu +\Gamma^\tau D_\tau)$.   
Hence we can define the index $\I$ of
$\D_X$:  the number of zero-modes of $\D_X$ with $\bar\Gamma=1$ minus the number with $\bar\Gamma=-1$.  This did not have an analog when we went from $d$ dimensions to $d+1$ dimensions.

We stress that this index can in general be nonzero in any dimension, even or odd.  The Atiyah-Singer index theorem implies that the index of an elliptic operator on an odd-dimensional manifold without boundary
vanishes, but this is not so on an odd dimensional manifold $X$ with boundary $\bY$.   In general, the index of $\D_X$ can be nonzero in any dimension.

So far, everything makes sense for any $d+1$-dimensional fermion field $\Psi$ with a self-adjoint Dirac operator $\D_\bY$.   But we are actually interested in the case that $\Psi$ was itself defined by the doubling
procedure starting from some fermion field $\chi$ in $d$ dimensions.   In that case, there is an antilinear operator $\sC$, defined in eqn. (\ref{antiunitary}), that
anticommutes with all gamma matrices while commuting with all symmetries and with the Dirac operator $\D_\bY$.  
After another round of doubling, the Dirac operator $\D_X$ on $X$ has an antilinear symmetry $\bar\sC$ that also obeys $\bar\sC^2=-1$.   The definition of $\Gamma^\tau$ was chosen to make the definition
simple:
\be\label{antilin}\bar\sC=\begin{pmatrix}\sC & 0 \cr 0 &\sC\end{pmatrix}. \ee
Given that $\sC$ anticommutes with the $\gamma^\mu$,  $\bar\sC$ anticommutes with $\Gamma^\mu$ and $\Gamma^\tau$, and hence commutes 
with $\D_X=\i(\Gamma^\mu D_\mu +\Gamma^\tau D_\tau)$.

The existence of an antilinear transformation $\bar\sC$ that commutes with $\D_X$ and squares to $-1$ implies that all eigenvalues of $\D_X$ have even multiplicity, by
a kind of Kramers doubling of eigenvalues.    In particular, the index $\I$ of $\D_X$ is even.
We will see shortly why this is important.

Now let us look at the APS index formula for $\I$:
 \beq\label{apsindex}
 \I =  \int_X \Phi_{d+2}-\frac{\eta_\bY}{2}. 
 \eeq
If $\bY$ is empty, this reduces to the usual Atiyah-Singer index formula $\I=\int_X\Phi_{d+2}$.  $\Phi_{d+2}$ can be expressed in terms of the gauge field strength and the Riemann tensor, as already
remarked in footnote \ref{explicit} of section \ref{aps}.   However, for us, the most useful characterization of $\Phi_{d+2}$ is that it is the anomaly $d+2$-form of the $d$-dimensional fermion $\chi$.  
This follows directly from the APS index formula.   The anomaly $d+2$-form $\Phi_{d+2}$ is by definition the polynomial in the Riemann tensor and the gauge field strength whose  associated Chern-Simons
$d+1$-form is  related by anomaly inflow to the perturbative anomaly of  $\chi$.   But  the formula (\ref{apsindex}) shows that modulo the integer  $\I$ (which plays no role in perturbation theory) and provided $X$ exists (which can be assumed in perturbation theory)
 $\eta_\bY/2$ is the Chern-Simons form associated to $\Phi_{d+2}$.   Moreover, we have learned in section \ref{precise}
that $\eta_\bY/2$ is related by anomaly inflow to the perturbative (and nonperturbative) anomaly of $\chi$.  So $\Phi_{d+2}$ is the anomaly $d+2$-form.

Therefore, $\Phi_{d+2}$ vanishes if and only if the original theory of the $\chi$ field in $d$ dimensions is free of perturbative anomalies.   For example, this is always the case for odd $d$; in odd dimensions,
there are no perturbative anomalies, and the Atiyah-Singer index theorem shows that $\Phi_{d+2}=0$ for odd $d$.  For even $d$, perturbative anomalies are possible, of course, but many interesting  theories -- such as the Standard Model of particle physics -- 
 are free of them.   These are again theories with $\Phi_{d+2}=0$.   

When there is no perturbative anomaly, there may still be a global anomaly.  As we discussed in section \ref{defphase}, the anomaly is governed by $\exp\left( -\i\pi \eta_\bY/2\right)$ for closed $d+1$-manifolds $\bY$.
This can definitely be nontrivial even when there is no perturbative anomaly.    However, from the APS index formula, we can reach an important conclusion about the function $\exp\left(-\i\pi \eta_\bY/2\right)$
when $\Phi_{d+2}=0$.  Suppose that $\bY$ is the boundary of a $d+2$-manifold $X$ over which the relevant structures extend.
The index formula then reduces to $\eta_\bY=-2\I$.  Since $\I$ is an even integer, $\eta_\bY$ is an integer multiple of 4, and therefore in this situation $\exp\left(-\i\pi\eta_\bY/2\right)=1$.
In other words, we have learned that when there is no perturbative anomaly, the function $\exp\left(-\i\pi\eta_\bY/2\right)$ that governs the global anomaly is a cobordism invariant: it is trivial on any $\bY$ that
is the boundary of some $X$.  
It may be a nontrivial cobordism invariant; if  $\bY$ is not the boundary of any  $X$, it may happen that  $\exp\left(-\i\pi\eta_\bY/2\right)\not=1$.
This is precisely the case that the original theory of the $\chi$ field in dimension $d$ has a nontrivial global anomaly. 

 As a special case of what we have said, $\exp\left(-\i\pi\eta_\bY/2\right)$ is a topological
invariant, unchanged in any continuous deformations of metrics and gauge fields.   Cobordism invariance is much stronger than topological invariance.

The APS index formula and the associated cobordism invariance of the global anomaly when perturbative anomalies vanish was used in \cite{GlobalString} (without using the term ``cobordism'')
for certain applications to string theory.  The framework was more restrictive than that of the present paper: in relating anomalies to $\eta$-invariants,
$d$ was assumed to be even and all manifolds were assumed orientable. 
Nowadays, there is a more general framework for understanding the role of cobordism invariance.   Any gapped theory with no topological order reduces at long distances (modulo nonuniversal
 terms that can be removed by local counterterms)  to an ``invertible'' topological
quantum field theory whose partition function is a cobordism invariant.    
This was conjectured in \cite{Kapustin:2014dxa}, and proved under some axioms of locality and unitarity in \cite{Freed:2016rqq,Yonekura:2018ufj}.
A system of free massive fermions coupled to background fields is an example of a gapped system with no topological order, and our discussion establishes directly the claim about cobordism invariance at long distances for such a system.

 \subsection{Deformation Classes And Anomalies}\label{converse}

In section \ref{defphase}, assuming that $\Upsilon_{\bY} = \exp(-\i\pi\eta_{\bY}/2)$ is trivial for every closed $D$-manifold $\bY$, 
we have found a satisfactory partition function for the  purely  $d$-dimensional  theory of the  $\chi$ field,\footnote{Refinements discussed
in section \ref{defphase}, where $\D_W$  has  zero-modes (so that the path integral measure cannot be characterized by a partition function)
or $W$ is not a boundary (leading to a slightly more elaborate discussion), can be straightforwardly included in the following.  We omit them for brevity.}
  namely  $Z_W^0
=|\Pf\,\D_W^+|\exp(-\i\pi\eta_Y/2)$, where $Y$ is any appropriate manifold
 with boundary $W$.   
We call this $Z_W^0$ (and not just $Z_W$) because we will consider some more general possibilities in a moment. 

What happens if $\Upsilon_{\bY}$ is not always trivial?  Then the formula $Z_W^0$ is not
a satisfactory definition  for
the partition function of the $\chi$ field as a purely $d$-dimensional theory,  but we should ask if this  formula can be improved.
Let us try some other definition $Z_W=|\Pf\,\D_W^+|\exp(-\i\pi \eta_Y/2)\exp(\i Q_Y)$, where to begin with
$Q_Y$ is some unknown function of the background fields  $g,A$ on $Y$ (we can assume that $Q$ is real as the anomaly only affects the phase
of the partition  function).

In  order for $Z_W$ to be independent of the background fields $g,A$ on $Y$  once $Y$  has been chosen, $Q_Y$ must be
unaffected by any variation of the background fields  on $Y$ that leaves them fixed along $W$.  (Therefore, when presently we discuss the
variation of $Q_Y$ under a change of $g,A$ along $W$, it will not matter how $g,A$ are being changed away from $W$.) 
For $Z_W$ to be independent of the choice of $Y$,   
we need $\exp(\i Q_Y)$ to satisfy conditions similar to those that were used in the previous analysis of $Z_W^0$:
$\exp(\i Q_Y)$ must satisfy the same gluing law (\ref{eq:glue}) as $\exp(-\i\pi\eta_Y/2)$, and for a closed $D$-manifold $\bY$,
we need
\be\label{wobble} \exp(-\i\pi\eta_{\bY}/2)\exp(\i Q_{\bY})=1. \ee

These conditions could be trivially satisfied with $\exp(\i Q_{Y})=\exp(\i\pi \eta_{Y}/2)$.   However, we need more.
Since $Z_W^0$ is  the partition function  of a physically sensible system  consisting of the $\chi$ field
plus massive ($D$-dimensional) regulator degrees of freedom, it is manifestly physically  sensible (at least in a $D$-dimensional sense) and we did not
have to discuss what properties make it physically  sensible.   An important aspect is the following.   Consider making a small variation of
the background fields $g,A$  along $W$.   The variation of the logarithm of the partition function of $\chi$ should be given by the one-point function of the
stress tensor $T$ or the current operator $J$ of  $\chi$.   In any theory free of perturbative anomaly, these  one-point functions $\la T\ra$ and $\la J\ra$
can be regularized in a way consistent with all physical principles, including conservation of $T$ and $J$.   (If there is a global anomaly,
it may appear when one tries to integrate  $\delta \log Z/\delta g$ and $\delta \log Z/\delta A$ to determine $Z$.)    The regularization is unique up to the
possibility of adding to $T$ and $J$ a function of the background fields that is local, gauge-invariant, and conserved.   
The derivation  that led to the formula $Z_W^0$ for the partition of the $\chi$ field plus massive degrees of freedom  makes it manifest that 
$\delta\log Z_W^0/\delta g$ and $\delta  \log Z_W^0/\delta A$ are related in the expected way to $\la T\ra$ and $\la J\ra$.

What  happens if we replace $Z_W^0$ with $Z_W=Z_W^0 \exp(\i  Q_Y)$?  Since $\log Z_W=\log Z_W^0+\i Q_Y$, this shifts
$\delta  \log Z/\delta g$ and $\delta \log Z /\delta A$ by $\i \delta Q_Y/\delta g$  and $\i \delta  Q_Y/\delta A$.
For $Z_W$ to be a physically sensible candidate formula for a partition function of the $\chi$ field, it must be possible
to interpret $\i\delta Q_Y/\delta g$ and $\i\delta  Q_Y/\delta A $ as contributions to $T$ and $J$.  So $\delta Q_Y/\delta g$ and $\delta Q_Y/\delta A$  must be gauge-invariant, local
(and conserved) functions of the background fields along $W$.

For this to be the case, $Q_Y$ must be the integral over $Y$ of some local operator $\Phi$:   $Q_Y=\int_Y \Phi$. 
$\Phi$ must be such that $Q_Y$ respects all symmetries of the theory, including possible time-reversal or reflection symmetries.
   Eqn. (\ref{wobble})  tells us that in the case of a closed manifold
$\bY$, 
\be\label{goodformula}\exp(-\i\pi \eta_{\bY}/2)=\exp\left(-\i\int_{\bY}\Phi\right),\ee
so $\exp\left(-\i \int_{\bY}\Phi\right)$ is a topological invariant  and in fact a cobordism invariant.    For a local operator $\Phi$ to have that property, $\Phi$ must be a  $D$-form
constructed as a gauge-invariant
polynomial in the Riemann tensor $R$ and the gauge field strength $F$, plus a possible exact form $\d\Lambda$ (here $\Lambda$ is a gauge-invariant $(D-1)$-form,
locally constructed from $g,A$).   Adding $\d\Lambda$ to $\Phi$ will modify $Z_W$ by $Z_W\to Z_W\exp(\i \int_Y \d \Lambda)=Z_W\exp(\i\int_W \Lambda)$.
This is equivalent to adding to the action of the original theory on $W$ a $c$-number term $-\i\int_W \Lambda$ constructed from the background fields.
That does not affect the consistency of the theory, so an exact term in $\Phi$ is not important.     So in short, the case that $\Phi$ can help in eliminating an anomaly is
that $\Phi$ is a polynomial in $R$ and $F$.   $\Phi$ is supposed to be a $D$-form, so
this is only possible if $D$ is even.   For such a $\Phi$, $\int_{\bY}\Phi$ is a characteristic class.

Whenever $\Upsilon_{\bY}=\exp(-\i\pi\eta_{\bY}/2)$ can be expressed  as in eqn. (\ref{goodformula})  in terms of a characteristic class $\int_{\bY} \Phi$, we can define
a purely $d$-dimensional partition function for the field $\chi$.   Under these assumptions,
\be\label{goodone} Z_W=|\Pf\,\D_W^+|\exp(-\i\pi\eta_Y/2)\exp\left(\i\int_Y \Phi\right)\ee
depends only on $W$ and not on $Y$.\footnote{\label{reflectionpos}Actually, there is one last requirement to make the definition (\ref{goodone}) physically sensible: $\Phi$ should be odd under reflections, to ensure reflection positivity  of the theory.   Accordingly, the case that
$\Phi$ is the polynomial in $R$ related to the Euler characteristic is not satisfactory. Reflection positivity implies that the Euler characteristic should appear in the Euclidean
signature effective action with a real coefficient, not an imaginary one.}

When $Y$  has  boundary $W$, $\CS(g,A)=\int_Y \Phi$ is a generalized Chern-Simons function of the fields $g,A$ on $W$, and is independent of how those
fields are extended over $Y$ (and of the choice of $Y$) modulo the values of $\int_{\bY} \Phi$ for closed manifolds $\bY$.   (This is proved by using the analog of eqn.
(\ref{eq:glue}) with $\pi\eta_Y/2$ replaced by  $\int_{Y}\Phi$.)
  If  $\int_{\bY}\Phi$ takes values in  $2\pi\Z$, then $\exp(\i\CS(g,A))$ depends only on the background fields on $W$, and not on anything about  $Y$.
  In that case, we say that $\CS(g,A)$ is a properly normalized Chern-Simons action; it makes sense as a purely $d$-dimensional coupling.   
However, eqn. (\ref{goodformula}) tells us that whenever $\Upsilon_{\bY}$ is nontrivial to begin with, $\int_{\bY}\Phi$ is not valued in $2\pi\Z$ and 
$\CS(g,A)$ is not a properly  normalized Chern-Simons action.   It does not make sense by itself as  a $d$-dimensional action; rather, it is being
used to cancel an anomaly.   The classic example of this situation is the ``parity anomaly'' for odd $d$ \cite{Redlich}, in which $\CS(g,A)$ is $\frac{1}{2}$ 
of a properly normalized Chern-Simons coupling.\footnote{Somewhat similar are fractional Chern-Simons counterterms that arise in certain situations \cite{CDFKS}.}  We will discuss that case from the present point of view in section \ref{d3}.   

Whenever there is a $\Phi$ that satisfies (\ref{goodformula}),  it is not unique, because we can always shift $\Phi\to \Phi+\Phi'$, where $\int_{\bY}\Phi'$  is a characteristic
class normalized to take values in $2\pi\Z$.   This will have the effect of shifting the effective action on $W$ by a properly quantized, physically sensible
Chern-Simons function of the background fields.   This shift can be interpreted as the result of using a different regularization of the underlying theory.

Now let us discuss the same subject from the point of view of the massive theory in the $D$-dimensional bulk.    The massive $\Psi$ field has a single
ground state when quantized on any $(D-1)$-manifold, so at long distances the theory of the $\Psi$ field becomes an ``invertible topological field theory.'' The partition function
of this invertible topological field theory on a closed manifold $\bar Y$ 
is the cobordism invariant $\exp(-\i\pi\eta_{\bY}/2)$.    Let us ask what are the moduli of this theory as an invertible
topological field theory.   In general, a first order deformation with small parameter $\varepsilon$ of any quantum field theory 
 multiplies the partition function by $\exp\left(\i\varepsilon \int_{\bY}\la \cO \ra\right)$ for some local operator $\cO$.  In an invertible topological field theory,
 $\cO$ is just a function of background fields $g,A$ (since there are no other local operators), and so $\la\cO\ra$ reduces to the classical value of $\cO$
 in the given background fields.   For $\exp\left(\i\varepsilon\int_{\bY}\cO\right)$ to be a topological invariant (and in fact a cobordism invariant), the exponent
 must be a characteristic class.  In other words, the moduli of an invertible topological field theory precisely
 correspond to the possibility of multiplying the partition function by $\exp\left(\i\int_{\bY}\Phi\right)$, where as before $\Phi$ is a gauge-invariant polynomial in the curvature
 and field strength.
 
 We are free here to take $\Phi=\sum_i t_i \Phi_i$, where $\Phi_i$ are a basis of possible gauge-invariant polynomials and $t_i$ are arbitrary real numbers.
 In particular, we can interpolate continuously between any given $\Phi$ and $\Phi=0$.     Therefore, eqn. (\ref{goodformula}) is precisely the condition under which
 the invertible topological field theory with partition function $\exp(-\i\pi \eta_{\bY}/2)$ can be deformed, through a family of invertible topological field theories,
 to a trivial theory.   In other words, the anomaly of the original theory in $d$ dimensions is really associated not with the invertible topological field theory
 associated to $\exp(-\i\pi\eta_{\bY}/2)$ but with the deformation class of this theory.  There is no  anomaly if and only if
 this  invertible  topological field theory is  deformable to  a trivial one.

This result is natural from the point of view of condensed matter physics.  
In the discussion of symmetry protected topological (SPT) phases~\cite{Chen,Chen2,Senthil}, 
we consider two phases to be equivalent if they can be continuously deformed to each other without any phase transition while preserving the relevant symmetries.
In particular, since the coefficients with which the polynomials $\Phi_i$ appear in the effective action can be varied continuously, these coefficients are not
universal and are not part of the classification of SPT phases.   The SPT theory of the massive field $\Psi$ is considered nontrivial if the invertible topological
field theory associated to it, with partition function $\exp(-\i\pi\eta_{\bY}/2)$, is not deformable to a trivial theory.      As we have seen, this is the case in which the theory of the $\chi$ field cannot be defined
as a purely $d$-dimensional theory.

\section{Examples In Dimensions $d=1,2,3,4$}\label{examples}

In this section, we consider some examples of the use of the $\eta$-invariant to analyze global anomalies in dimensions $d=1,2,3,4$.

The cases of odd $d$ have a different flavor, because for odd $d$ there are no perturbative anomalies, and all anomalies are global from the beginning.
We will consider the odd $d$ cases first.   

To keep our examples simple, we will 
primarily work on orientable manifolds only; in other words, we will generally not incorporate time-reversal or reflection symmetry.
When $d$ is odd, the global anomaly is related to an $\eta$-invariant in an even dimension $D=d+1$.    On an even-dimensional orientable manifold,
the $\eta$-invariant reduces to a more familiar invariant -- an index or a mod 2 index.   Therefore our odd $d$ examples could be described more directly
in terms of the index or the mod 2 index rather than the $\eta$-invariant.  (See \cite{GlobalString} and
\cite{Witten:2015aba} for  such a treatment of the $d=1$ and $d=3$ examples, respectively.)   However, it seems useful to explain that all examples can be deduced from the
$\eta$-invariant.   

In even $d$, we consider examples in which perturbative anomalies cancel, and analyze the global anomalies using the $\eta$-invariant.

\subsection{$d=1$}\label{d1}

In $d=1$, we consider a system of $n$ real fermions with a classical $\O(n)$ symmetry and a simple action
\be\label{simple}I=\int \d t \frac{\i }{2}\sum_{j=1}^n \chi_j \frac{\d}{\d t} \chi_j.  \ee
For simplicity, to begin with we take $n$ even, do not incorporate time-reversal symmetry,
and consider background $\SO(n)$ gauge fields only.  We briefly explain at the end what happens if one relaxes one of those conditions.

A compact 1-manifold $W$ will have to be a circle.   The circle has two possible spin structures, as $\chi$ could be either periodic or antiperiodic in going
around the circle.   We will refer to these spin structures as Ramond (R) or Neveu-Schwarz (NS).

The Hamiltonian derived from the action $I$ simply vanishes.   Though we could add additional terms to the action to get a nonzero Hamiltonian,
instead we will turn on a background $\SO(n)$ gauge field on the circle.   We write $U$ for the holonomy of this gauge field.
In its canonical form, $U$ is block diagonal with $2\times 2$ blocks of the form
\be\label{blocks} \begin{pmatrix} \cos\theta_k & \sin\theta_k \cr -\sin\theta_k & \cos\theta_k \end{pmatrix}, ~~~k=1,\cdots, n/2. \ee
As this formula makes clear, shifting any of the $\theta_k$ by an integer multiple of $2\pi$ does not change $U$ and so is a gauge transformation of the background gauge field.   

Upon quantization, the $\chi_k$ satisfy $\{\chi_k,\chi_{k'}\}=\delta_{kk'}$ so (up to a factor of $\sqrt 2$) they are gamma matrices.
Hence the group that acts on the quantum Hilbert space $\H$ is not the classical symmetry group $\SO(n)$ but its double cover $\Spin(n)$. 
The fact that the group that acts quantum mechanically is a double cover of the classical
symmetry group is an anomaly of sorts.     To see that this can be understood as an anomaly in the conventional sense -- an ill-definedness of the path integral -- 
we consider the path integral on a circle in the presence of a background gauge field with holonomy $U$.

The path integral with antiperiodic boundary conditions for the fermions (NS spin structure) computes $\Tr_\H \,U$.   Using the explicit description of $\H$
as a spinor representation of $\SO(n)$,  we can write a formula for this trace:
\be\label{omigo} \Tr_\H \,U=\prod_{k=1}^{n/2} 2\cos(\theta_k/2). \ee
Here we see the anomaly: shifting any one of the $\theta_k$ by $2\pi$ is a gauge transformation of the background gauge field, but it changes the sign
of the path integral.

The path integral with periodic boundary conditions for the fermions (R spin structure) computes $\Tr_\H\,(-1)^{\sf F} U$.  Again we can write an explicit
formula:
\be\label{migo} \Tr_\H\,(-1)^{\sf F} U =\pm \prod_{k=1}^{n/2} 2\i \sin(\theta_k/2).    \ee
Again we see the anomaly, but now there is a new ingredient: the overall sign of the path integral depends on an arbitrary choice.   One way to explain this
fact is the following.   The operator $(-1)^{\sf F}$ is characterized by the fact that it anticommutes with the elementary fermions, and its square is 1.
But these conditions do not determine the overall sign of the operator $(-1)^{\sf F}$, and without more input there is no natural way to fix this sign.   
We can define $(-1)^{\sf F}$ up to sign by the product $\chi_1\chi_2\cdots \chi_n$, but as the $\chi_k$ anticommute with each other, there is no way to determine
the overall sign of this expression without knowing something about how the $\chi_k$ should be ordered.
We run into the same issue from the point of view of path integrals.   Consider the system of $n$ real fermions $\chi_k$ with periodic spin structure in the special case $U=1$.
Each of the $\chi_k$ has a zero-mode; let us call these modes $\chi_k^0$.    The measure for the fermion zero-modes is, up to sign,
$\d \chi_1^0\d \chi_2^0\cdots \d\chi_n^0$,   But again, the sign of this measure depends on how we order the $\d\chi_k^0$.   

Now let us look at this anomaly from the perspective of the $\eta$-invariant.    Applied to a 1-component real fermion in dimension $d=1$,
the doubling procedure  of  section \ref{sec4} produces a 2-component real fermion field $\Psi$ in dimension $D=d+1=2$.    So starting with the $n$
fields $\chi_1,\cdots,\chi_n$ in dimension 1 transforming in the fundamental representation of $\O(n)$,   we get in 2 dimensions an $n$-component Majorana fermion field $\Psi$ in the fundamental representation of $\O(n)$.

To study the anomaly by our general procedure, 	we regard the circle $W$ (at least in the NS case; see below for the R case) as the boundary of a two-manifold $Y$, over
which the spin structure of $W$ is extended.     
Since we do not incorporate time-reversal symmetry as part of the discussion, we can consider the original circle $W$ to be oriented, and then in the procedure
of section \ref{sec4}, we consider only oriented $Y$.  
The anomaly is given by $\exp\left(-\i\pi\eta_\bY/2\right)$, where $\eta_\bY$ is the $\eta$-invariant of the self-adjoint Dirac operator $\D_\bY$ of 
the $\Psi$-field on a closed manifold $\bY$.

On an even-dimensional orientable manifold $Y$, there is always a chirality operator $\bar\gamma$ (usually called $\gamma_5$ in four-dimensional particle physics)
that anticommutes with the self-adjoint Dirac operator 
$\D_Y$.   Accordingly, the nonzero eigenvalues of $\D_Y$ are equal and opposite in pairs:    if $\D_Y \Psi=\lambda\Psi$,
then $\D_Y (\bar\gamma\Psi)=-\lambda \bar\gamma\Psi$.  From the definition of the $\eta$-invariant
\be\label{etad} \eta_Y=\lim_{\epsilon\to 0}\sum_k \exp(-\epsilon|\lambda_k|)\sign(\lambda_k)\ee
(where the function $\sign(x)$ was defined in eqn. (\ref{signdef})), we see at once that a pair of eigenvalues $\lambda,-\lambda$ do not contribute.
Therefore, in this situation, $\eta_Y$ simply equals the number of linearly independent zero-modes of $\D_Y$.

A zero-mode of $\D_Y$ can have either positive or negative chirality.   In $d=2$, complex conjugation exchanges the two types of mode.   So the number of zero-modes
of $\D_Y$ is 2 times the number of zero-modes of positive chirality.

It is physically sensible in two dimensions to consider a fermion field $\psi$ of positive chirality only, transforming in a real representation of some symmetry group.
For our case, the relevant real representation is the vector representation of $\SO(n)$ or $\O(n)$.   Such a field can have an action
\be\label{actchiral}\int\d^2x\sqrt g \,(\psi, \D_Y^+\psi), \ee
where $\D_Y^+$ is the Dirac operator acting on a fermion  field of positive chirality.  By fermi statistics, we can here think of $\D_Y^+$ as an antisymmetric
matrix.   The canonical form of such a matrix is
\be\label{conform}\begin{pmatrix} 0&a_1& && && \cr
                                   -a_1 & 0 &&&& \cr
                                    && 0 & a_2 &&&\cr
                                     && -a_2&0 &&&\cr
                                     &&&& \ddots && \cr
                                     &&&&&0 & \cr
                                      &&&&&&0 \end{pmatrix},\ee
with skew ``eigenvalues'' $a_i$ that appear in $2\times 2$ blocks, and unpaired zero-modes.   The only way that the number of zero-modes can change
is that one of the $a_i$ can become zero or nonzero.   When that happens, the number of zero-modes jumps by $\pm 2$.   So if $\zeta$ is the number of
zero-modes mod 2, $\zeta$ is invariant in any continuous deformation.  $\zeta$ is called the mod 2 index of the chiral Dirac operator on $Y$.  The derivation shows
that  any fermion system with an action consistent with fermi statistics has a mod 2 index (this invariant is not always interesting as for many fermion systems it identically vanishes
or is the mod 2 reduction of a more familiar invariant).  In our problem of the chiral fermion field $\psi$ in the vector
representation of $\SO(n)$ or $\O(n)$, this means that the number $\zeta$ of zero-modes of $\psi$ mod 2 is a topological invariant: it is unchanged if one
varies the metric of $Y$ or the background gauge field.

Since $\eta_Y$ is twice the number of zero-modes  of $\psi$, we have $\eta_Y=2\zeta$ mod 4.   Hence $\exp\left(-\i\pi\eta_Y/2\right)=(-1)^\zeta$.
So the anomaly reduces to a sign $\pm 1$, as we saw more directly in eqn. (\ref{omigo}).   To evaluate  the  anomaly more explicitly, first note  that
an $\SO(n)$  bundle $E\to \bY$, for any  closed two-manifold $\bY$, is classified topologically  by its second Stieffel-Whitney class $w_2(E)$.
Hence the anomaly can only  depend on $w_2(E)$.     Furthermore, the structure group  of $E$ can be reduced to an $\SO(2)$ subgroup. (See the end of
section \ref{d4} for this argument.)  We have  $\SO(2)\cong \U(1)$.
From the point of view of $\U(1)$, the vector representation of $\SO(n)$ consists of $n-2$ neutral fermions and  
 two components of charge $\pm 1$.    In the field of a $\U(1)$ gauge field  with
 first Chern class $k$, generically one of the charged components has $|k|$ zero-modes, and the other has none.    Since we assume $n$ even, 
 the $n-2$ neutral fermions  do  not contribute to the  mod  2 index.   So $\zeta$  is the mod 2 reduction of $k$.  But the mod 2 reduction of $k$ coincides with $w_2(E)$.
 So $\zeta=w_2(E)$   and the    anomaly is $(-1)^{w_2(E)}$.
In particular, the anomaly does not depend on the spin structure and could have arisen in a purely  bosonic theory. 
 Since $w_2(E)$ is the obstruction  to lifting the structure group of $E$  from $\SO(n)$ to its double cover $\Spin(n)$,
 the anomaly would disappear if we view  the original system (\ref{simple}) as a system with $\Spin(n)$  rather than  $\SO(n)$ symmetry.    This is consistent with the  observation
 that we made at the outset: the Hilbert space $\H$ of this  theory furnishes a representation of  $\Spin(n)$, not  $\SO(n)$.

Now let us consider the case of Ramond spin structure.   A single circle $W$ with Ramond spin structure is not the boundary of any spin manifold $Y$.   But two such
circles are the boundary of such a $Y$ (we can take $Y$ to be a cylinder).  The cobordism group in this
problem is $\Z_2$; for a generator, we can pick a circle $W_0$ with Ramond spin structure and with any chosen holonomy $U_0$ for the background gauge field.
Since $W_0$ is not a boundary, our formalism gives no natural way to compute the sign $\Tr\, (-1)^{\sf F} U_0$.   But once we fix this sign, any other Ramond
sector path integral is determined by the procedure explained at the end of section \ref{defphase}.   Thus the Ramond sector path integral is uniquely determined
up to an arbitrary overall sign, as we saw more directly in eqn. (\ref{migo}). 

Finally, we briefly consider three generalizations that were mentioned at the outset. 

(1)  For odd $n$, consider the path integral of the theory (\ref{simple}) on $W=S^1$ with a spin structure of R type.   With such a spin structure, each of $\chi_1,\cdots,\chi_n$
has a zero-mode, so to get a nonzero path integral, we need to insert the product of all these fields.   With this insertion, we get $\la  \Tr\,(-1)^\sF\chi_1\chi_2\cdots\chi_n\ra
\not=0$.
But for odd $n$, the operator $\chi_1\chi_2\cdots\chi_n$ is odd under $(-1)^\sF$.   So we have found an anomaly  in $(-1)^\sF$.

Let us try  to recover this anomaly from a  $D=2$  point of view.   For odd $n$, 
the derivation showing that the anomaly is governed by $(-1)^\zeta$ is still valid.  The only difference is that $(-1)^\zeta$ can be nontrivial
even if we ignore the $\O(n)$  symmetry and do not turn on 
 any background $\O(n)$ gauge field.  For an example, take $Y$ to be a two-torus with fermions periodic in both directions.    Then the
 Dirac operator on $Y$ for a one-component chiral fermion $\psi$   has a single zero-mode (the  ``constant''  mode of $\psi$), and so  $(-1)^\zeta=-1$. 
 Taking $n$ identical chiral fermions does not change $(-1)^\zeta$ if $n$ is odd.    So there is an anomaly  even if the only symmetry we consider is
  $(-1)^\sF$.   This symmetry is implicit whenever we discuss fermions and  spin structures.
 
This anomaly in $(-1)^\sF$ 
has a variety of applications in different areas of physics; for example, see \cite{Kitaev,KS,DW,StaW}.   An application to Type II superstring theory,
closely related to the recent paper \cite{Tachikawa}, was actually the subject of the lecture by one of us  at the Shoucheng Zhang Memorial Workshop \cite{WittenLecture}.  

(2) Another  generalization is to allow background gauge fields of $\O(n)$ rather than $\SO(n)$.     In discussing  this, for simplicity
we  take $n$ even.   Let  $W$ be a circle with  NS spin structure and with a background $\O(n)$  gauge
field with monodromy ${\sf R}_1=\mathrm{diag}(-1,1,1,\cdots,1)$.     In this background, $\chi_1$ has a zero-mode and  $\chi_r$, $r>1$, does  not.   So again  we see
an anomaly in $(-1)^\sF$, this time a ``mixed anomaly'' between $(-1)^\sF$ and the $\O(n)$ symmetry, since to detect it we had to turn on a background  $\O(n)$ gauge field.

Replacing $\O(n)$ by a   double cover $\Pin^+(n)$ or $\Pin^-(n)$ will not eliminate this anomaly.    $\sf R_1$ can be lifted to $\Pin^+(n)$ or $\Pin^-(n)$, though
not uniquely, and the same analysis leads to the same anomaly in  $(-1)^\sF$ in the presence of a background field.

From a two-dimensional point of  view, the anomaly is still governed by$(-1)^\zeta$.    But  in contrast  to what we found  in discussing $\SO(n)$ for even $n$,  $(-1)^\zeta$ now  depends on the spin structure.   
To see this, take $Y$ to be a two-torus $S^1_A\times S^1_B$, where the first circle $S^1_A$ has trivial $\O(n)$ gauge field  and the second circle $S^1_B$ has
an $\O(n)$ gauge field with monodromy ${\sf R}_1$.    Then $(-1)^\zeta$ can  again be computed just by counting ``constant'' fermion modes; it equals 1 if $S^1_A$
has NS spin structure, and equals $-1$  if $S^1_A$ has spin structure of R type (regardless of the spin structure of $S^1_B$).    Thus there is an anomaly,
and, since $(-1)^\zeta$  depends on the spin structure, the anomaly is not given by a cohomological formula and cannot be reproduced in a  purely bosonic theory.   That is
consistent with the fact that in the boundary description, the anomaly involves $(-1)^\sF$.

One can relate the two ways of seeing the anomaly by trying to compute $\la \Tr\, (-1)^\sF \sR_1 \chi_1\ra^2$ via a path integral on the cylinder $Y=S^1\times I$,
where $S^1$ is a circle with Ramond spin structure and $I$ is an interval.   We assume that the gauge field is a pullback from $S^1$ and has monodromy $\sR_1$.
Once we specify the spin structure on $S^1$, there are two possible spin structures on $Y$, for a reason explained at the end of section \ref{sec:zeromode}.
  (After picking an isomorphism between the two boundary circles and their spin bundles, one can say that the two
spin bundles on $Y$ differ by the sign of parallel transport of a fermion from one end of the cylinder to the other.)   
As  explained at the end of section \ref{sec:zeromode},
 the sign one gets for $\la \Tr\, (-1)^\sF \sR_1\chi_1\ra^2$ depends on the spin structure of $Y$,   and this dependence reproduces the $(-1)^\sF$ 
anomaly  of the original $d=1$ theory on a single boundary component.

 The theory (\ref{simple}) can certainly be quantized, with the Hilbert space providing an irreducible representation of the Clifford algebra
 $\{\chi_i,\chi_j\}=\delta_{ij}$.   But  after quantization, the ``internal symmetry''
$\sf  R_1$ anticommutes  with the ``spacetime symmetry'' $(-1)^\sF$, rather than commuting.   As a  result, the symmetry  group after quantization does not
fit the general framework introduced at the start of section \ref{sec4}, where $(-1)^\sF$ is supposed to be central, commuting with the full symmetry group
 (as noted in footnote \ref{exception}, there are more general possibilities in $d=1$).
  The symmetry  group  of the quantized $d=1$ theory does not extend naturally to $D=2$
(in dimension $D\geq 2$, the element $(-1)^\sF\in \Spin(D)$ is always central in any relativistic fermion theory).   Hence
it is not immediately obvious how to deduce the symmetry
group of the quantized $d=1$ theory from  the $D=2$ anomaly.

(3)  To include time-reversal symmetry, we should consider $W$ to be unoriented and allow unorientable $Y$.  
The obvious time-reversal symmetry of eqn. (\ref{simple}) that acts on the fermions by $\sT \chi_j(t)\sT^{-1}=\chi_j(-t)$ 
satisfies $\sT^2=1$ at the classical level and
 corresponds to a $\pin^-$ structure in 1 dimension.
To incorporate this symmetry  in the boundary theory, one should allow $Y$ to be an unorientable manifold  endowed with a $\pin^-$ structure.
 In this case,  as there is no longer  a chirality operator $\bar\gamma$ that anticommutes with $\D_Y$, it is no longer true 
that  $\exp(-\i\pi\eta_\bY/2)=\pm 1$  for a closed manifold $\bY$, and instead $\exp(-\i\pi\eta_\bY/2)$, for a single Majorana fermion,
 can be an arbitrary eighth  root of 1. (This can be proved
along lines explained in Appendix C of \cite{Witten:2015aba} for the analogous problem in four dimensions.)   Indeed, in the time-reversal invariant
case, the theory (\ref{simple}) that we started with does  have a mod 8 anomaly from a  Hamiltonian  point of view \cite{FidKit}; that is, it is anomalous unless
$n$ is a multiple of 8.    The interpretation of
this in terms of cobordism was originally suggested in   \cite{Kapustin:2014dxa}.
If $n$ is even, it is possible to define a time-reversal symmetry such that $\sT^2=(-1)^{\sf F}$ at the classical
level.  This leads to a slightly different analysis, in which  the description by the $\eta$-invariant again reduces to a mod 2 index.

\subsection{$d=3$}\label{d3}

Our example in $d=3$ will actually involve one of the celebrated contributions of Shoucheng Zhang \cite{ZQH}. 

First consider in $d=3$   a massless Dirac fermion $\chi$ coupled 
with  charge 1  to a $\U(1)$ gauge field.    
Such a field could have a gauge-invariant bare mass, so there will be no anomaly spoiling the gauge symmetry.  However, a bare mass of the $\chi$
field would explicitly violate the time-reversal and reflection symmetries of the massless theory.   So there is a possibility of an anomaly that would spoil those
symmetries.  Indeed, this model is 
 the original context for the ``parity'' anomaly \cite{Redlich}: the field $\chi$ cannot be quantized, purely in three-dimensional terms, in a way
that preserves time-reversal and reflection symmetry.  We essentially computed the anomaly in section \ref{aps}, when we explained that
integrating out the $\chi$ field on a manifold $W$, with a particular regulator mass, generates a phase $\exp(-\i\pi\eta_{D,W}/2)$ (as we are dealing with a Dirac fermion,
we use $\eta_{D,W}$, the $\eta$-invariant on $W$ for charge 1 modes only, rather than $\eta_W=2\eta_{D,W}$, the $\eta$-invariant on $W$ for all modes of charge $\pm 1$).  
The partition function of $\chi$, including this phase, is
\be\label{pthree}|\Det\,\D_\chi|\exp(-\i\pi \eta_{D,W}/2), \ee
where $\D_\chi$ is the Dirac operator of the $\chi$ field.
This phase is odd under time-reversal or reflection symmetry and cannot be removed
by any counterterm. Reversing the sign of the regulator mass would give the opposite phase $\exp(\i\pi \eta_{D,W}/2)$, still violating
time-reversal and reflection symmetry.   The anomaly is a mod 2 anomaly, because a pair of $\chi$ fields could be quantized with regulator masses of opposite signs, in which case the phases cancel out
and all symmetries are preserved.      

We are interested here in the anomalous case with a single $\chi$ field.
In the theory of topological insulators, instead of trying to quantize  $\chi$  by itself in purely three-dimensional terms, one views it as a field that propagates
on the boundary of a four-manifold $Y$, the worldvolume of a topological insulator.   The combined system can then
be quantized in a way that preserves time-reversal and reflection symmetry.  In this interpretation, the $\U(1)$ gauge field is just the 
usual gauge field of electromagnetism.      The result of Shoucheng Zhang, together with
T. Hughes and X. Qi \cite{ZQH}, was to show that  the electromagnetic $\theta$-angle in the bulk of a topological insulator is equal to $\pi$.  (A striking
consequence of this can be explicitly demonstrated in a lattice model of a topological insulator \cite{RF}: a magnetic monopole
of unit charge immersed in a $3+1$-dimensional topological insulator will acquire a half-integral electric charge.)   Let us see how the
result that $\theta=\pi$ may be understood from the point of view of the present paper.

The doubling procedure applied to the $\chi$ field produces a massive charge 1 Dirac fermion $\Psi$ in dimension $d+1=4$.   
With the couplings, boundary conditions, and regulator described in sections \ref{sec2} and \ref{aps}, $\Psi$ provides a model of a topological insulator; in particular,
$\chi$ can be interpreted as a boundary localized mode of $\Psi$, and the path integral of $\Psi$ gives a consistent framework that describes $\chi$ together
with massive degrees of freedom in the bulk.   
According to the general formalism, the universal part of the path integral of $\Psi$ on a manifold $Y$ with boundary $W$ is 
\be\label{unipart}|\Det \D_W^+|\exp(-\i\pi \eta_{D,Y}). \ee

We want to show that this formula is consistent with time-reversal  and reflection symmetry, and moreover we would like to recover the result of Shoucheng Zhang
and his colleagues showing that the electromagnetic $\theta$-angle of the bulk theory is $\theta=\pi$.  For simplicity, we work on oriented manifolds
only.  The self-adjoint Dirac operator $\D_Y$
anticommutes with a chirality operator $\bar\gamma$, so just as in section \ref{d1}, its nonzero modes do not contribute to $\eta_{D,Y}$; $\eta_{D,Y}$
is simply equal to the number of linearly independent (charge 1) zero-modes of $\D_Y$.    However, unlike the two-dimensional case, 
in four dimensions, complex conjugation does
not exchange zero-modes of positive and negative chirality.   On the contrary, if $n_+$  and $n_-$ are the numbers of linearly independent zero-modes of
$\D_Y$ (acting on fermions of charge 1) with positive or negative chirality, then the index of $\D_Y$  is $\I=n_+-n_-$.  On the other hand, $\eta_{D,Y}$ is the
total number of zero-modes: $\eta_{D,Y}=n_++n_-$.   We see that $\eta_{D,Y}\cong \I$ mod 2, and hence $\exp(-\i\pi\eta_{D,Y})=(-1)^\I$.
Thus the combined path integral of the bulk and boundary modes of the topological insulator -- or more precisely its universal part -- is
\be\label{compath} |\Det\D_W|(-1)^\I.    \ee
This is real, and thus manifestly consistent with time-reversal and reflection symmetry.

To understand the result of  Hughes, Qi, and Zhang concerning the $\theta$-angle, we write $(-1)^\I =\exp(\i\pi \I)$
and think of $-\i \pi \I$ as a contribution to the effective action.    For discussing the bulk effective action, we can temporarily work on a closed manifold $\bY$.
In that case the  Atiyah-Singer index formula gives $\I=\int_{\bY}\Phi$ with
\be\label{elcon}\Phi=\h A(R)+ \frac{1}{2}\frac{F\wedge F}{(2\pi)^2},\ee   where $\h A(R)$ is a certain quadratic polynomial in the Riemann tensor.  
Because of the electromagnetic contribution to $\Phi$, the term $-\i\pi\I=-\i\pi\int \Phi$ in the action corresponds to an electromagnetic $\theta$-angle $\theta=\pi$.

A key fact in this derivation was that on a closed manifold, $\exp(-\i\pi \eta_{\bY,D})=\exp(\i\pi\I)= \exp\left(\i\pi \int_{\bY}\Phi\right)$ is the exponential of the integral of a characteristic class.   
Therefore we are in the situation that was analyzed in detail in section \ref{converse}.    As in eqn. (\ref{goodone}), it is possible to give a purely three-dimensional formula
for the partition function of $\chi$ by replacing $\exp(-\i\pi\eta_{Y,D})$ with $\exp(-\i\pi \eta_{Y,D})\exp(-\i\pi \int_Y\Phi)$:
\be\label{welgo}Z_W= |\Det \, \D_W^+| \exp(-\i\pi \eta_{D,Y}) \exp\left(-\i\pi\int_Y\Phi\right)= |\Det\,\D_W^+|\exp(\i\pi\I) \exp\left(-i\pi\int_Y\Phi\right). \ee
   The general formalism tells us that the right hand side of eqn. (\ref{welgo}) will make
sense in purely three-dimensional terms.   We can confirm this by using the APS index theorem:
\be\label{elgo}\I =\int_Y\Phi -\frac{\eta_{D,W}}{2}.\ee
So in fact eqn. (\ref{welgo}) is equivalent to
\be\label{lgo} Z_W=|\Det\,\D_W^+|\exp(-\i\pi \eta_{D,W}/2). \ee
This is the purely three-dimensional, but time-reversal violating, formula that was explained more directly in  eqn. (\ref{pthree}), except that after doubling, we
write $\D_W^+$ for $\D_\chi$.   Time-reversal
violation entered this derivation when we canceled the $Y$ dependence of $(-1)^\I=\exp(\pm \i \pi \I)$ with a factor of $\exp(-\i\pi \int_Y\Phi)$.   Time-reversal
would map this to a conjugate construction using $\exp(+\i\pi \int_Y\Phi)$.  

So we recover the familiar fact that this system can be quantized as a purely three-dimensional
theory if we are willing to give up time-reversal and reflection symmetry.   Alternatively, we can consider the $\chi$ field to propagate on the boundary
of a four-manifold, and use the time-reversal invariant partition function (\ref{compath}).    
If we wish to define the $\chi$ field  on unorientable three-manifolds, then time-reversal and reflection symmetry are essential and the purely three-dimensional
quantization is not available.   In this case, we have to consider the $\chi$ field as living on the surface of a topological insulator.   The appropriate formula for the partition
function is eqn. (\ref{unipart}), in which the  $\eta$-invariant no longer reduces to $\I$.

\subsection{$d=2$}\label{d2}

Our remaining examples will be cases with nontrivial cancellation of perturbative anomalies.

In $d=2$, we consider first a $\U(1)$ gauge theory with positive chirality Dirac fermions of charges $n_1,n_2,\cdots,n_s$
and negative chirality Dirac fermions of charges $m_1,m_2,\cdots, m_s$.   The perturbative gauge theory anomaly cancels if and only if
\be\label{wordo}\sum_{i=1}^s n_i^2=\sum_{i=1}^s m_i^2. \ee
By taking equal numbers of positive and negative chirality fermions, we have ensured cancellation of perturbative gravitational anomalies.
Therefore, when eqn. (\ref{wordo}) holds, this theory is free of perturbative anomalies.  

A possible global anomaly would be controlled as usual by an $\eta$-invariant in dimension $d+1=3$.   In the present case, let us write $\eta_{D,r}$
for the $\eta$-invariant of a charge $r$ Dirac fermion on a three-manifold $Y$. (We use $\upeta$ as we are dealing with Dirac fermions.)
The global anomaly of our given theory is controlled by $\Upsilon=\exp\left(-\i\pi\left(\sum_{i=1}^s\eta_{D,n_i}-\sum_{j=1}^s\eta_{D,m_i}\right)\right).$  
(As explained at the end of section \ref{aps}, reversing the sign of the fermion chirality in two dimensions reverses the sign with which the $\eta$-invariant
appears in the exponent.)    However, as is always the case when perturbative anomalies are absent, the factor $\Upsilon$ that controls the global anomaly
is a cobordism invariant.   The cobordism of a three-dimensional closed spin manifold $\bY$ with a $\U(1)$ gauge field is trivial (any such $\bY$ is the boundary
of some $X$ over which the spin structure and $\U(1)$ gauge field extend).   So $\Upsilon $ always equals 1 on a closed manifold, and a theory of this kind
that is free of perturbative anomalies is also always free of global or nonperturbative anomalies.\footnote{One could avoid relying on a knowledge of
the cobordism group by making an argument similar to what we will make for the Standard Model in section \ref{d4}.    A $\U(1)$ gauge field is classified topologically
by its first Chern class.   In three dimensions, this is dual to an embedded circle $C\subset \bY$.   Using this, and cobordism invariance of $\Upsilon$, one can
reduce as in section \ref{d4} to two special cases: (a)  the gauge field is trivial, or (b) $\bY=S^2\times S^1$ with a gauge field is a pullback from $S^2$.   In
case (a), trivially $\Upsilon=1$ since the $\U(1)$ charges do not matter, 
and in case (b), arguing as in section \ref{d4} one shows that $\Upsilon=1$.   This argument will also
work for a theory with gauge group $\U(1)^n$ for any $n$: if such a theory in two dimensions has no perturbative anomaly, it also has no global anomaly.}

We can get an example in $d=2$ that does have a nontrivial global anomaly if we replace $\U(1)$ by a $\Z_k$ subgroup, for some $k$.  Then there is no perturbative
anomaly as long as there are equally many positive and negative chirality fermions (to avoid a gravitational anomaly).    In particular, there is no condition
analogous to (\ref{wordo}).  For generic choices of the $\Z_k$ quantum numbers of the fermions, such a theory will have a global anomaly.

We will consider a special case in a moment, but first we explain the motivation to consider this special case.   Spacetime supersymmetry in string theory
was originally discovered by  Gliozzi, Olive, and Scherk (GOS) \cite{GOS}.   Their original insight was that the partition function of 8 chiral fermions in two
dimensions (in genus 1, where they computed explicitly) vanishes if it is summed over spin structures; this vanishing was  a reflection of spacetime supersymmetry.
(8 is the number of light cone oscillator modes in the Ramond-Neveu-Schwarz model.)
Though this point was not made explicitly until later, it only makes sense to add together the partition functions with different spin structures if the anomaly
does not depend on the spin structure.  Later discoveries involved elaborations of the original GOS analysis.  One always finds that
spacetime supersymmetry in string theory depends on the fact that the anomaly of 8 chiral fermions in two dimensions is
independent of the spin structure.   It turns out that this is true for a system of $k$ chiral fermions if and only if $k$ is divisible by 8.   

A closely related problem has  been much studied in the context of condensed matter physics.   In that context, 
one studies ``symmetry protected topological'' (SPT) states  \cite{Chen,Chen2,Senthil}, which are states that are topologically nontrivial when some global symmetry
is taken into  account but  become 
trivial if that global symmetry is explicitly broken.   Such a system can have an anomalous boundary state in one dimension less.   A much-studied special case
is a fermionic system in spacetime dimension $D=3$ with a global $\Z_2$ symmetry \cite{Qi,YR,GL}.    Those systems have a $\Z_8$ classification.      An
example associated to a nonzero element  $k\in  \Z_8$ has a boundary state in dimension $d=2$ that carries the anomaly that one would find in the GOS
calculation if the number of chiral  fermions considered were $k$ (or any integer congruent to $k$ mod 8) rather than 8.

To put this  problem in the context of gauge theory, we consider a $\Z_2$  gauge theory in two dimensions, with $k$ positive chirality real fermions that are invariant
under $\Z_2$, and $k$ negative chirality real fermions that transform as $-1$ under the nontrivial element of $\Z_2$.
 This system
has no perturbative anomaly,  but an explicit genus 1 calculation shows that if $k$ is not divisible by 8, it has a global anomaly.
We will explore from the vantage point of the present paper the absence of anomalies when $k$ is a multiple of 8.  (For a previous analysis, 
see the discussion of eqn. (24) in  \cite{GlobalString}. An explicit computation of the $\eta$-invariant when $k$ is not a multiple of 8 to show the $\Z_8$ classification was 
done in \cite{Tachikawa:2018njr}.)

Let $W$ be a two-manifold with spin structure  $\alpha$ and some background $\Z_2$ bundle, which we can think of as a real line bundle $L$ with
structure group $\Z_2=\{\pm 1\}$.     In the model just introduced, positive chirality  fermions  (being invariant under  $\Z_2$) are coupled
only to the spin structure $\alpha$.   But negative chirality fermions are coupled to the spin structure $\alpha$ and also to $L$.
Effectively the negative chirality fermions are coupled to a 
 new spin structure $\beta=\alpha\otimes L$.  Thus for studying anomalies, we can forget about the $\Z_2$ gauge field and just say
that fermions of positive or negative chirality are coupled to different spin structures $\alpha$ or $\beta$.

The spin cobordism problem for a two-manifold with two spin structures $\alpha,\beta$ (or even just with one spin structure) is not trivial.
This is related to the fact that in string theory, the sign of the GOS projection in the Ramond sector is not uniquely determined and could be reversed.  It is
also related to the existence of two different Type II superstring theories.\footnote{The cobordism invariant of a Riemann surface with one spin structure is
 $(-1)^\zeta$, which was already discussed,
with references to various applications, at the end of section \ref{d1}. With two spin structures $\alpha,\beta$, one has $(-1)^{\zeta_\alpha}$ and $(-1)^{\zeta_\beta}$.}   
We will pass over such issues here and just ask if there is any consistent way to define the theory with 8 positive chirality fermions coupled to spin structure $\alpha$
and 8 negative chirality fermions coupled to $\beta$.   As usual, the potential obstruction is a global anomaly that can be measured by an $\eta$-invariant.

In detail, let  $\bY$ be a closed three-manifold with spin structures $\alpha,\beta$.   The global anomaly is then measured by $\exp\left(-\frac{\pi\i}{2}8(\eta_{\bY,\alpha}-\eta_{\bY,\beta})\right)$
where $\eta_{\bY,\alpha}$ and $\eta_{\bY,\beta}$ are $\eta$-invariants on $\bY$ for a Majorana fermion coupled to spin structure $\alpha$ or $\beta$.
We note that this is trivial if and only if one always has
\be\label{notfor} \exp\left(-\frac{\pi\i}{2}8\eta_{\bY,\alpha}\right)=\exp\left(-\frac{\pi\i}{2}8\eta_{\bY,\beta}\right), \ee
or in other words if and only if the anomaly for 8 positive chirality fermions in two dimensions does not depend on the spin structure.
This is how we formulated the question initially. 

In this form, it is not immediately obvious how to answer the question.   But a more general  question
 is easier to answer.    Consider a two-dimensional theory with gauge group $\Spin(8)$.   This group has two spinor representations -- spinors of positive
or negative $\Spin(8)$ chirality.  Each of these is a real representation of dimension 8.   Let us call the two representations $S_+$ and $S_-$.   We
consider a two-dimensional theory with gauge group $\Spin(8)$, with  8  positive  chirality real fermions in  the representation $S_+$, and 8 negative chirality fermions
in the representation $S_-$.    This theory is free of perturbative anomalies, because the representations $S_+$ and $S_-$ have the same dimension and quadratic Casimir operator.
We will show that the theory is also free of global anomalies.

To answer this question, we consider a  closed three-manifold $\bY$ with spin structure $\alpha$  and some $\Spin(8)$ bundle.   The global anomaly of the $\Spin(8)$ theory described
in the last paragraph
is measured by $\Upsilon= \exp\left(-\frac{\pi\i}{2}\left(\eta_{\bY,\alpha,S_+}-\eta_{\bY,\alpha,S_-}\right)\right)$.   The notation is hopefully clear; $\eta_{\bY,\alpha,S_\pm}$
is the $\eta$-invariant on $\bY$ for a Majorana fermion in the representation $S_\pm$ coupled to some background $\Spin(8)$ gauge field, as well as to the spin structure
$\alpha$.

Since this theory has no perturbative anomaly, $\Upsilon$ is a cobordism invariant and in particular it is a topological invariant.   But any $\Spin(8) $ bundle on
a three-manifold is topologically trivial (since $\pi_i(\Spin(8)) =0$ for $i\leq 2$).   So we can continuously deform to the case that the background $\Spin(8)$ gauge
field is trivial, in which case trivially $\Upsilon=1$.   Thus the $\Spin(8)$ theory under consideration has no global anomaly, and
for any background $\Spin(8)$ gauge field,
\be\label{bac} \exp\left(-\frac{\pi\i}{2}\eta_{\bY,\alpha,S_+}\right) = \exp\left(-\frac{\pi\i}{2}\eta_{\bY,\alpha,S_-}\right).\ee

  It follows from this, together with a judicious embedding of $\Z_2$ in $\Spin(8)$, that the $\Z_2$ theory with 8 positive
chirality neutral fermions and 8 negative chirality charged fermions is also free of global anomaly.   For this, we embed $\Z_2$ in $\SO(8)$ so that the nontrivial
element  $x\in \Z_2$ maps  to the central element $-1\in \SO(8)$.   The element $-1\in  \SO(8)$ can be lifted to  $\Spin(8)$  in  two ways.   We can pick a lift
so that $x$ acts as $+1$ on $S_+$ and as $-1$ on $S_-$.  (With the other lift, these signs are reversed.)
 Now we consider the identity (\ref{bac}), specialized to the case that the background $\Spin(8)$
gauge field actually has structure group $\Z_2$, embedded in $\Spin(8)$ as just described.    With this choice, the $\Spin(8)$ identity (\ref{bac}) reduces to the
identity (\ref{notfor}), which says that the anomaly of 8 chiral fermions does not depend on the spin structure.  Indeed, for a background $\Spin(8)$ gauge field
that is induced from a $\Z_2$ bundle $L$ by embedding $\Z_2$ in $\Spin(8)$ in the way that we have described, the vector bundle over $\bY$ corresponding to $S_+$ is a rank 8 trivial bundle,
and the vector bundle over $\bY$ corresponding to $S_-$ is the direct sum of 8 copies of $L$.  On the left hand side of eqn. (\ref{bac}), $\eta_{\bY,\alpha,S_+}$ reduces
in this example to $8\eta_{\bY,\alpha}$ on the left hand side of  eqn. (\ref{notfor}), while on the right hand side of eqn. (\ref{bac}), $\eta_{\bY,\alpha,S_-}$  similarly
reduces to
$8\eta_{\bY,\beta}$ on the right hand side of eqn. (\ref{notfor}). So eqn. (\ref{bac}) does reduce to eqn. (\ref{notfor}).

The only property of $\Spin(8)$ that we used was that it  is  connected and simply-connected; the only important property of the fermion representation was
that it has no perturbative anomaly.   So a two-dimensional theory with a connected and simply-connected gauge group and no perturbative anomaly also has
no global anomaly.

\subsection{$d=4$}\label{d4}

For an example in $d=4$ with nontrivial cancellation of perturbative anomalies, we can take the Standard Model of particle physics.
Does it have any global anomaly?

The Standard Model can actually be embedded in the $\SU(5)$ grand unified theory \cite{GG} which  is also free of perturbative
anomalies.  It turns out that the $\SU(5)$ grand unified model has no global anomaly.  
Thus, the phase of the fermion path integral of the $\SU(5)$ grand unified theory, coupled to an arbitrary background metric and $\SU(5)$ gauge field,
can be defined in a consistent way.   Specializing to the case that the structure group of the background gauge field reduces to the gauge group of the
Standard Model, it follows that the Standard Model is also free of global anomalies.
 This was originally shown in \cite{Freed}, Example 3.4. 
 Here we will explain how to establish the result in the framework of the present paper.
 See also \cite{Garcia-Etxebarria:2018ajm} for another discussion.
 
 Since the Standard Model does not have a time-reversal or reflection symmetry, we formulate it only on oriented four-manifolds $W$, and in the anomaly
 inflow problem, we consider only oriented manifolds $Y$.   Since the Standard Model has fermions whose definition requires a spin structure,
 both  $W$ and $Y$ are endowed with spin structures.
 
 We are not going to be able to get a unique answer for the phase of the path integral of  the $\SU(5)$ model in an arbitrary background gauge and gravitational
 field.  The reason is that a four-dimensional spin manifold $W$ with an $\SU(5)$ bundle is not necessarily the boundary of a five-manifold $Y$
 over which the spin structure and $\SU(5)$ bundle of $W$ can be extended.   The relevant cobordism group is $\Z\times \Z$.   
 The two integer-valued invariants  are $\sigma(W)/16$ (where $\sigma$ is the signature) and  the $\SU(5)$ instanton number.
 
 For generators of the cobordism group, we can use the following two manifolds $W_1$ and $W_2$.   For $W_1$, we take a K3 surface, with some chosen metric and orientation,
 and with the background gauge field being $A=0$.   For $W_2$, we  take a four-sphere $S^4$ with some chosen metric and orientation 
 and with some chosen gauge field $A_0$ of
 instanton number 1.
 
 We have no way to determine the phases of the path integral measure for those two examples, so we make arbitrary choices.  One  can think of those choices
 as representing a precise definition of what is meant by the gravitational $\theta$-angle and the $\SU(5)$ $\theta$-angle including quantum effects of the fermions.
 
 Any other $W$ is cobordant, by some manifold  $Y$ over which the relevant structures extend,
  to a  linear combination  $n_1W_1+n_2W_2$ for some integers $n_1$ and $n_2$.  (By $n_1W_1$ or $n_2W_2$ with $n_1$ or $n_2$  negative,
 one means  $|n_1|$ or $|n_2|$ copies of $W_1$ or $W_2$ with  orientation reversed.)
  So once the gauge and gravitational $\theta$-angles have been fixed, the procedure of section \ref{defphase} gives a definition of the  path
  integral measure for any $W$.  {\it A priori}, this  definition might depend on $Y$.
  
  To know that the phase does not depend on $Y$ so that the 
  $\SU(5)$ grand unified theory has  no global anomaly, we need to know that $\Upsilon_{\bY}=\exp\left(-\i\pi\eta_{\bY}/2\right)$ is trivial for any  closed five-manifold
  $\bY$ with $\SU(5)$ gauge field.   Here $\eta_{\bY}$ is the $\eta$-invariant of the operator  $\D_{\bY}$, acting on a five-dimensional field $\Psi$ that  is obtained  by  combining
  the Standard Model fermions $\chi$ with dual fields $\t\chi$. 
     We will use cobordism invariance, plus an examination  of some  special cases, to show that  $\Upsilon$
  is always trivial.
  
  We can proceed  as  follows.   First of all, spin cobordism  is  trivial in five dimensions, so in case the $\SU(5)$ gauge field is trivial, $\bY$ is the boundary of a spin manifold
  and  $\Upsilon_{\bY}=1$. Actually any fermion coupled only to gravity  can have a bare 
  mass  in $d=4$ and hence does not have any pure gravitational anomaly.\footnote{Since a bare mass is possible, one suspects that there will be a simple direct proof, without knowing about the cobordism group,
  that, for a $d=4$ Majorana fermion $\chi$ coupled only to gravity,  in five dimensions
  one has $\Upsilon_{\bY}=0$.  Such a proof  may be constructed as follows.   A rank five Clifford algebra
   has an irreducible representation of dimension four, so there exists on a five-dimensional spin manifold $\bY$
  a self-adjoint Dirac operator $\D_{\bY}^0=\i \slashed{D}$ acting on a four-component fermion field.   It is not possible for the five gamma matrices
  to be all real, but it is possible for three of them to be real while  two, say $\gamma_1$ and $\gamma_2$, are imaginary.  Using this fact in
  a locally Euclidean frame, one can define an antilinear operator $\sC=*\gamma_1\gamma_2$ that commutes with all gamma matrices and anticommutes
  with $\i\slashed{D}$.   Therefore, the $\eta$-invariant of $\D_{\bY}^0$ reduces to the number of its zero-modes.   That number is always even, since $\sC^2=-1$,
  so $\eta(\D^0_{\bY})$ is an even integer.
  The doubling procedure applied to the four-dimensional Majorana fermion $\chi$ produces an operator $\D_{\bY}$ that is the direct sum of two copies
  of $\D_{\bY}^0$.  ($\chi$ has four components, so after doubling we get an eight component fermion in five dimensions, whose Dirac operator is the direct sum
  of two copies of $\D_{\bY}^0$.)   So $\eta_{\bY}=2\eta(\D^0_{\bY})$
  is a multiple of 4 and $\Upsilon_{\bY}=\exp(-\i\pi\eta_{\bY}/2)=1$.} 
    Second, an $\SU(5)$ bundle $E$ over a five-manifold $\bY$ is completely classified by its second Chern class $c_2(E)$.   (This follows from the fact
  that homotopy groups $\pi_d(\SU(5))$ vanish for $d\leq 4$ except for $\pi_3(\SU(5))=\Z$, which is related  to the second Chern class.)    Moreover,
  in five dimensions, $c_2(E)$ is dual to an embedded circle $C\subset \bY$.    That means that, topologically, the $\SU(5)$ gauge field describes some
  integer number $\nu$  of instantons propagating along the circle $C$.   In other words, we can deform the $\SU(5)$ gauge field so that it is trivial except very near
  some circle $C$, while in the normal plane to $C$ the integral that defines the instanton number  integrates  to  $\nu$.    
  
  In this situation, by an elementary cobordism,\footnote{Instead  of cobordism, one can use a cut and paste argument that is explained
  near the end of this section.}   $\bY$ is cobordant to a disjoint union $Y_1+Y_2$, where $Y_1$ and $Y_2$ are as follows.
  $Y_1$ is a copy of $\bY$, but with $A=0$.  $Y_2$ is a copy of $S^4\times C$ ($S^4$ being a four-sphere) with instanton number $\nu$ on $S^4$.
  
We already know that $Y_1$ has $\Upsilon_{Y_1}=1$.
  We therefore only  have to investigate $Y_2$.
  
  There are two possible spin structures on $Y_2=S^4\times C$, since the spin structure around $C$ may be of R or NS type.   In the NS case (antiperiodic fermions),
  $S^4\times C$ is the boundary of $S^4\times D$, where $D$ is a two-dimensional disc, and the $\SU(5)$ gauge field on $S^4\times C$ extends over
  $S^4\times D$.  Therefore, $\Upsilon_{Y_2}=1$ in this case.
  
  In the R case (periodic fermions), there is no obvious six-dimensional spin manifold $X$  with boundary $S^4\times C$
 over which the instanton bundle on $S^4$ extends.
Such an $X$ actually can be constructed,\footnote{This is slightly technical.   First, by an elementary cobordism or by a cut and paste argument (as described later), one can
replace $S^4$ with another convenient manifold containing an instanton.  A useful choice is $S^2 \times T^2$ where $T^2$ is a two-torus,
which we take with a spin structure periodic in both directions.    Then $Y_2=S^4\times S^1$, with Ramond spin structure on $S^1$, is replaced by
$Y_2'=S^2\times T^2\times S^1$,  where now $T^2\times S^1$ is a three-torus with spin structure periodic in all directions.
$T^2\times S^1$, with this spin structure, is the boundary of a ``half-K3 surface,'' that is, a four-manifold $Q$ that maps
to a disc $D$ with generic fiber an elliptic curve.  In particular, the map $Q\to D$ has a section $s:D\to Q$.   We can use $S^2\times Q$
as the six-manifold of boundary $Y_2'$.   Picking any point $p\in S^2$, the world-volume of the instanton in $S^2\times Q$ can be taken
to be the two-manifold $p\times s(D)$.    We proceed with the argument in the text because it seems illuminating to know how to directly
evaluate the $\eta$-invariant, rather than relying on a technical argument about cobordism.  } but without having to know this, we can proceed as follows.
 We can deform the metric on $S^4\times C$ to be a product and we can choose the $\SU(5)$ gauge field on $S^4\times  C$ to be a pullback
 from $S^4$.    This condition means that if $C$ is parametrized by an angle $\tau$, then  the gauge field on $S^4\times C$ is independent of $\tau$ and
 has no component in the $\tau$ direction.   Concretely then, if the metric of $C$ is chosen to be $\d\tau^2$,  the self-adjoint Dirac operator on $Y_2$ takes the form
 \be\label{ondo}   \D_{Y_2}=\i\gamma^\tau\frac{\partial}{\partial\tau} +\D_{S^4}, \ee
 where $\D_{S^4}$ is the self-adjoint Dirac operator of $S^4$.
 
 The chirality operator $\bar\gamma_{S^4}$  of $S^4$ (the product of the four gamma matrices of $S^4$) anticommutes with $\D_{S^4}$ and commutes with $\gamma^\tau$.
 If we combine $\bar\gamma_{S^4} $ with a reflection $\tau\to -\tau$, we get an operator that anticommutes with $\D_{Y_2}$.   Hence the nonzero eigenvalues
 of  $\D_{Y_2}$ come in pairs $\lambda,-\lambda$.   As usual, such  pairs do not contribute to the $\eta$-invariant.  Hence, $\eta_{\D_{Y_2}}$ is just the number of
 zero-modes of  $\D_{Y_2}$.   
 
 Since
 \be\label{wondo}\D_{Y_2}^2=-\frac{\partial^2}{\partial\tau^2}+\D_{S^4}^2, \ee
 a zero-mode of $\D_{Y_2}$ is independent of $\tau$ and is a zero mode of $\D_{S^4}$.   Conversely, any such  mode is a zero-mode of $\D_{Y_2}$.
 So $\eta_{Y_2}$ is the same as the number of zero-modes of $\D_{S^4}$.
 
 Because of the doubling that is involved  in going from $W$ to $Y$ in this construction, the operator  $\D_{S^4}$ acts on the Standard Model fermions $\chi$
 plus a dual or complex conjugate set of fermions $\t\chi$.   
 Let $n$ be the number of zero modes of the Dirac operator on $S^4$ acting on the original Standard  Model
 fermions  $\chi$.
 Complex  conjugation exchanges zero-modes of $\chi$ with zero-modes of $\t\chi$, so the number  of
 zero-modes of $\D_{S^4}$ is $2n$.  Hence $\eta_{Y_2}=2n$.

 As we will explain in a moment, $n$ is always even.   So $\eta_{Y_2}$ is  a multiple of 4, which implies that $\exp\left(-\i\pi\eta_{Y_2}/2\right)=1$.
 
 To show that $n$ is always even, let $n_+$ and $n_-$ be the number  of positive or negative chirality 
 zero-modes of  $\SU(5)$ model fermions, coupled to some background $\SU(5)$ gauge
 field on $S^4$.   Then $n=n_++n_-$, and this is congruent mod 2 to the index $\I=n_+- n_-$.   The index theorem for $\SU(5)$ gauge theory shows that for the  $\SU(5)$ model fermions, $\I$  is always
 even (in fact always a multiple of 4), so $n$ is always even.
 
 It is inevitable that in one way or another we were going to have to check that in the $\SU(5)$ grand unified model, the number of fermion zero-modes in an instanton
 field on $S^4$ is always even.   If this number could be odd, it really would represent an anomaly \cite{WittenSUtwo,Wang:2018qoy}.

  An alternative to  cobordism in  reducing to the two special cases of $Y_1$ and $Y_2$ is the following.   Let $Y'=S^4\times S^1$
  with trivial gauge field.  Given any closed five-manifold $\bY$, to show $\Upsilon_{\bY}=1$, 
  let $ \h Y=\bY+Y'$.   Since $\Upsilon_{Y'}=1$, we have $\Upsilon_{\h Y}=\Upsilon_{\bY}$
  and we want to show that $\Upsilon_{\h Y}=1$.   We will reduce to the special cases $\Upsilon_{Y_1}=\Upsilon_{Y_2}=1$  by showing
   that $\Upsilon_{\h Y}=\Upsilon_{Y_1}\Upsilon_{Y_2}$.
Since cutting $S^4$ on the equator decomposes it into two copies of a four-dimensional ball $B_4$, $Y'$ can be cut to make two copies of $B_4\times S^1$.
The boundary of $B_4\times S^1$ is $S^3\times S^1$.   
Likewise by cutting along the boundary of a tubular neighborhood of $C\subset \bY$, we can remove from $\bY$ a copy of $B_4\times S^1$, leaving behind
another manifold $\t Y$, also with boundary $S^3\times S^1$.   At this point we have decomposed $ \h Y $ to the disjoint union of $\t Y$ and three copies
of $B_4\times S^1$.  Of these four manifolds, only one copy of $B_4\times S^1$ has a nontrivial gauge field.
   Exchanging that copy of $B_4 \times S_1$ with one of the others and then gluing
the pieces back together, we get $Y_1+Y_2$.   In other words, we have described a procedure of cutting and regluing that converts $\h  Y=\bY+Y'$ into $Y_1+Y_2$.
The gluing law (\ref{eq:glue}) for the $\eta$-invariant implies that $\Upsilon$ is invariant under cutting and regluing, so we conclude that
$\Upsilon_{\h Y}=\Upsilon_{Y_1}\Upsilon_{Y_2}$.  Since invariance under cobordism can essentially be deduced from invariance under cutting and regluing and vice versa
\cite{Freed:2016rqq,Yonekura:2018ufj}, it is inevitable that a cobordism argument can be expressed in terms of cutting and regluing.

  What happens if (still assuming there is no time-reversal or reflection symmetry)
 we replace $\SU(5)$ by another compact gauge group $G$?    Since $\pi_2(G)=0$ for any $G$, we can repeat
 the above argument verbatim if $\pi_0(G)=\pi_1(G)=\pi_4(G)=0$.   For example, the grand unified theories  based on $\Spin(10)$ or $E_6$ satisfy
 those conditions.\footnote{The $\Spin(10)$ theory has a refinement
 in which we  
take  the symmetry  group of the fermions to be $(\Spin(4) \times \Spin(10))/\Z_2$
 rather than $\Spin(4) \times \Spin(10)$.  Absence of a global anomaly in this case was argued in  \cite{Wang:2018cai} based on
 cobordism invariance of the anomaly.   It is possible to modify the following argument for this case.
 See also \cite{Wan:2018bns}.}   
 We conclude that if $G$ satisfies those conditions 
 and moreover the fermion representation is free of perturbative anomaly  and the number of fermion zero-modes in a $G$-bundle on $S^4$ is always even,
 then the model is completely anomaly-free.    With some additional knowledge of Lie group topology, one can omit the assumption that $\pi_4(G)=0$.    First of all,  with $\pi_0(G)=\pi_1(G)=0$ but without assuming
 $\pi_4(G)=0$, the classification of $G$-bundles on a five-manifold $\bY$ is modified only in a very simple way.   The only change is
 that we have to allow for the possibility that
 a $G$-bundle over $\bY$ can be modified in a small neighborhood of a point $p\in \bY$ by twisting by a nontrivial element of $\pi_4(G)$.   This means
 that in the cobordism analysis, we have to allow a third special case $Y_3$, namely a five-sphere $S^5$ with some $G$-bundle.   To proceed farther, we need
 some specific facts about Lie group topology.   For simple $G$, one has $\pi_4(G)=0$ except for $G=\Sp(2k)$, which satisfies $\pi_4(\Sp(2k)))=\Z_2$. 
 If $G$ is semi-simple rather than simple, then $\pi_4(G)$ is a product of copies of $\Z_2$, one for each $\Sp(2k)$ factor in $G$.
 A nontrivial
 $\Sp(2k)$-bundle over $S^5$ associated to the nonzero element of $\pi_4(\Sp(2k))$ can be constructed as follows.  Think of an instanton as a particle in five dimensions.
 Consider an instanton propagating in $S^5$ in such a way that its worldline is an embedded circle  $C$.   To make from this a nontrivial $\Sp(2k)$ bundle,
 the instanton  should undergo a $2\pi$ rotation as it propagates around $C$.   Starting with this
 description of the nontrivial $G$-bundles on $S^5$, 
 one can use a cobordism  argument -- or a cut and paste  argument, as in the last paragraph --  to show that  $\Upsilon_{Y_3}$, for 
 a $G$-bundle on $S^5$, is equal to $\Upsilon_{Y_2}$, for a corresponding $G$-bundle on $Y_2=S^4\times S^1$.   So the condition $\Upsilon_{Y_3}=1$ does
 not add anything new.  In fact,  the mod 2 cobordism invariant associated to $\pi_4(\Sp(2k))=\Z_2$ is the mod 2 index of the Dirac operator with values in the
 fundamental representation of $\Sp(2k)$.   This invariant is nonzero for  $Y_2=S^4\times S^1$ with Ramond spin structure around $S^1$ \cite{Wang:2018qoy},
 so in effect our analysis for $Y_2$ already incorporated the role of $\pi_4(G)$.

 The discussion so far demonstrates the general utility of cobordism invariance and a cut and paste argument. 
 It is possible to reorganize the argument  in a perhaps more elementary way as a reduction from any connected and simply-connected $G$
 to the case $G=\SU(2)$. In doing so, we may assume that $G$ is simple; otherwise, we make the following analysis for each simple factor of $G$.
 In this version of the argument, we de-emphasize the role of cobordism invariance and proceed as much as possible
 using only the fact that when a four-dimensional
 theory has  no perturbative
 anomaly, the global anomaly is a topological invariant in five dimensions.
 To make the argument explicit, let  us first consider gauge groups $G = \Spin(n), \SU(n) $, and $ \Sp(2n)$.
 The bundle associated to the fundamental representation of $G$ is a real, complex, or quaternionic vector bundle of rank $n$, respectively.
 Let $ \alpha$ be $1, 2$ or $4$ for $\Spin$, $\SU$ or $\Sp$, respectively. The real rank of the vector bundle is $\alpha n$.
 If $\alpha n > D$, there is a section of the vector bundle which is nonzero everywhere. 
 This is simply because a sufficiently generic section of a vector bundle of rank $\alpha n$  is always nonzero as a function of $D$ variables $x^1, \ldots, x^D$  if $\alpha n >D$.   (Locally, such a section is a collection of
$\alpha n$ real-valued functions, and generically these functions have no common zero as a function of
$D<\alpha n$ real variables.)
 By taking such a nonzero section, the structure group of a vector bundle can be reduced from  $\Spin(n)$, $\SU(n) $, or $ \Sp(2n)$
  to  $ \Spin(n-1), \SU(n-1)$, and $\Sp(2n-2)$, respectively. 
 By repeating this process, the structure group in $D=5$ dimensions is topologically reduced to  $\Spin(5) , \SU(2),$ and $ \Sp(2)$.
 But we have $\Spin(5) \cong \Sp(4)$ and this can be further reduced to $\Sp(2) \cong \SU(2)$, so we can always reduce the structure group from $G$ to $\SU(2)$. 
 Using ``obstruction theory'' and a knowledge of the homotopy groups $\pi_i(G)$, $i\leq 4$,
  one can show that this is also possible if $G$ is a  connected and simply-connected exceptional Lie group.  
Therefore, for any $G$, the global
anomaly can always be captured  for background fields valued in 
a subgroup $\SU(2) \subset G$.
An irreducible representation of $\SU(2)$ is either strictly real or pseudoreal. A strictly real representation does not contribute to the anomaly because a mass term is possible in $d=4$.
The exponentiated $\eta$-invariant of a pseudoreal representation 
is given by the mod 2 index in $D=5$.   This vanishes if the number of zero-modes in an instanton field on $S^4$ is always even.   But to prove
that this last statement holds for a general five-manifold (and therefore that the anomaly associated to the mod 2 index
in $D=5$ is entirely captured by a counting of zero-modes on $S^4$)
appears to require cobordism or cut and paste arguments such as we have explained above.  

At any rate, the conclusion is that a four-dimensional theory with connected and simply-connected gauge group has no anomalies beyond the familiar ones.
If one drops the requirement for the gauge group to be connected, then, as in $d=2$, there definitely are new anomalies.

\section*{Acknowledgments}  We thank R. Mazzeo for some discussions.
The work of KY is supported by JSPS KAKENHI Grant-in-Aid (Wakate-B), No.17K14265.
Research of EW is supported in part by NSF Grant PHY-1911298. 

\bibliographystyle{unsrt}

\end{document}